\pdfoutput=1

\documentclass[11pt,twoside,a4paper,cmspaper,final,collab]{cms-tdr}
\def\svnVersion{483108P}\def\cmsCernNoTag{CERN-EP-2018-214}\def\cmsCernDate{\today}\def\cmsMessage{Published in Physical Review D as \href{http://dx.doi.org/10.1103/PhysRevD.98.092014}{\doi{10.1103/PhysRevD.98.092014.}}}

\begin{document}\cmsNoteHeader{TOP-17-013}

\hyphenation{had-ron-i-za-tion}
\hyphenation{cal-or-i-me-ter}
\hyphenation{de-vices}
\RCS$HeadURL: svn+ssh://svn.cern.ch/reps/tdr2/papers/TOP-17-013/trunk/TOP-17-013.tex $
\RCS$Id: TOP-17-013.tex 478976 2018-10-23 14:41:30Z mseidel $
\newlength\cmsFigWidth
\ifthenelse{\boolean{cms@external}}{\setlength\cmsFigWidth{0.85\columnwidth}}{\setlength\cmsFigWidth{0.4\textwidth}}
\ifthenelse{\boolean{cms@external}}{\providecommand{\cmsLeft}{top\xspace}}{\providecommand{\cmsLeft}{left\xspace}}
\ifthenelse{\boolean{cms@external}}{\providecommand{\cmsRight}{bottom\xspace}}{\providecommand{\cmsRight}{right\xspace}}

\newlength\cmsTabSkip\setlength{\cmsTabSkip}{1ex}

\newcommand{\DIRE} {{\textsc{Dire}}\xspace}
\newcommand{\asmz}{\ensuremath{\alpha_{S}(m_{\PZ})}\xspace}
\newcommand{\asfsr}{\ensuremath{\alpha_{S}^{\text{FSR}}(m_{\PZ})}\xspace}
\newcommand{\msbar}{\ensuremath{\overline{\text{MS}}}\xspace}

\cmsNoteHeader{TOP-17-013}
\title{Measurement of jet substructure observables in \texorpdfstring{\ttbar}{ttbar} events from proton-proton collisions at \texorpdfstring{$\sqrt{s}=13\TeV$}{sqrt(s) = 13 TeV}}

\date{\today}

\abstract{
A measurement of jet substructure observables is presented using \ttbar events in the lepton+jets channel from proton-proton collisions at $\sqrt{s}=13\TeV$ recorded by the CMS experiment at the LHC, corresponding to an integrated luminosity of 35.9\fbinv. Multiple jet substructure observables are measured for jets identified as bottom, light-quark, and gluon jets, as well as for inclusive jets (no flavor information). The results are unfolded to the particle level and compared to next-to-leading-order predictions from \POWHEG interfaced with the parton shower generators \PYTHIA~8 and \HERWIG~7, as well as from \SHERPA~2 and \DIRE~2. A value of the strong coupling at the \PZ boson mass, $\asmz = 0.115^{+0.015}_{-0.013}$, is extracted from the substructure data at leading-order plus leading-log accuracy.
}

\hypersetup{
pdfauthor={CMS Collaboration},
pdftitle={Measurement of jet substructure observables in ttbar events from proton-proton collisions at sqrt(s) = 13 TeV},
pdfsubject={CMS},
pdfkeywords={CMS, physics, top quark, jet substructure}}

\maketitle

\section{Introduction}

The confinement property of quantum chromodynamics (QCD) renders isolated quarks and gluons unobservable.
Instead, strongly interacting partons produced in high-energy hadron-hadron collisions initiate a cascade of lower-energy quarks and gluons that
eventually hadronize into a jet composed of colorless hadrons. Monte Carlo (MC) event generators~\cite{Buckley:2011ms} describe reasonably well
both the perturbative cascade, dominated by soft gluon emissions and collinear parton splittings, as well as the final hadronization
(via nonperturbative string or cluster models at the end of the parton shower below some cutoff scale of the order of 1\GeV).
The details of the perturbative radiation phase have been studied at previous colliders (Tevatron~\cite{Acosta:2005ix,Aaltonen:2008de,Aaltonen:2011pg}, HERA~\cite{Chekanov:2002ux,Chekanov:2004kz,Chekanov:2009bc}), and the various parameters of the parton fragmentation models have been tuned to match jet data from $\Pep\Pem$ collisions, collected mostly at LEP~\cite{Heister:2003aj,Abreu:1996na,DELPHI:2011aa,Achard:2004sv,Abbiendi:2004qz,Ackerstaff:1998hz} and SLC~\cite{Abe:1998zs,Abe:2002iq}.

Precise measurements of jet properties at the LHC allow improvements in the experimental techniques and theoretical predictions for heavy-quark/light-quark/gluon
discrimination, as well as in the identification of merged jets from Lorentz-boosted heavy particle decays~\cite{Ellis:2009su,Altheimer:2013yza}.
They also give information about the limits and applicability of the current parton shower and fragmentation models in the gluon-dominated environment of proton-proton ($\Pp\Pp$) collisions, rather than the quark-dominated one, as in the $\Pep\Pem$ case~\cite{Anderle:2017qwx}.
In addition, jet substructure studies test QCD in the infrared- and/or collinear-safe limits where recent calculations~\cite{Larkoski:2017jix}
provide analytical predictions with increasingly accurate higher-order corrections, including, \eg, up to next-to-leading-order (NLO) terms~\cite{Marzani:2017mva},
and beyond next-to-leading-logarithmic (NLL) resummations~\cite{Frye:2016aiz} for some observables.

Jet shapes and substructure have been measured in $\Pp\Pp$ collisions at $\sqrt{s} = 7\TeV$ by the ATLAS Collaboration in dijet events~\cite{Aad:2011kq,Aad:2012meb}, and by the CMS Collaboration in dijet and W/Z+jet events~\cite{Chatrchyan:2013vbb}.
Furthermore, jet substructure was measured in dijet events at 8\TeV by CMS~\cite{Sirunyan:2017tyr} and at 13\TeV by ATLAS~\cite{Aaboud:2017qwh} and CMS~\cite{Sirunyan:2018xdh}.
Measurements of jet shapes have also been carried out by ATLAS using events containing top quark-antiquark (\ttbar) pairs at 7\TeV~\cite{Aad:2013fba}, exploiting for the first time the possibility of comparing the properties of bottom and light-quark jets from the top quark decays.
The mass distribution of boosted top quark candidates was measured by CMS at 8\TeV~\cite{Sirunyan:2017yar}.

The analysis presented here uses jet samples obtained from fully resolved \ttbar lepton+jets events, where one of the \PW{} bosons decays to a charged lepton (electron or muon) and the corresponding neutrino, while the other {\PW} boson decays to quarks, yielding two separate jets.
Various jet substructure observables are measured in order to characterize the jet evolution, such as generalized angularities, eccentricity, groomed momentum fraction, $\mathcal{N}$-subjettiness ratios, and energy correlation functions.
For comparison with theory predictions, the measured distributions are corrected for detector effects, unfolding them to the particle level that is defined using stable particles with decay length larger than $10 \mm$.

The measurements are performed using data recorded at $\sqrt{s} = 13\TeV$ by the CMS detector described in Section~\ref{sec:detector}.
Section~\ref{sec:samples} contains details of the data and simulated samples.
Events are reconstructed and selected using the algorithms described in Section~\ref{sec:recosel}.
The unfolding to the particle-level of the observables of interest and their associated systematic uncertainties are described in Section~\ref{sec:unfoldsys}.
The jet substructure variables under investigation are defined and the results presented in Section~\ref{sec:obs}.
The \ttbar lepton+jets topology allows for sorting the jets into samples enriched in bottom quarks, light quarks from the {\PW} boson decays, or gluons stemming from initial-state radiation (ISR), as discussed in Section~\ref{sec:flavor}.
The correlation between jet substructure observables, and their level of agreement to different MC predictions are studied in Section~\ref{sec:chi2}.
Finally, an extraction of the strong coupling from jet substructure observables is presented in Section~\ref{sec:asfsr}.

\section{The CMS detector}
\label{sec:detector}

The central feature of the CMS apparatus is a superconducting solenoid of 6\unit{m} internal diameter, providing a magnetic field of 3.8\unit{T}. Within the solenoid volume are a silicon pixel and strip tracker, a lead tungstate crystal electromagnetic calorimeter (ECAL), and a brass and scintillator hadron calorimeter (HCAL), each composed of a barrel and two endcap sections. Forward calorimeters extend the pseudorapidity ($\eta$) coverage provided by the barrel and endcap detectors. Muons are detected in gas-ionization chambers embedded in the steel flux-return yoke outside the solenoid.
Events of interest are selected using a two-tiered trigger system~\cite{Khachatryan:2016bia}. The first level (L1), composed of custom hardware processors, uses information from the calorimeters and muon detectors to select events at a rate of around 100\unit{kHz} within a time interval of less than 4\mus. The second level, known as the high-level trigger (HLT), consists of a farm of processors running a version of the full event reconstruction software optimized for fast processing, and reduces the event rate to around 1\unit{kHz} before data storage.

A more detailed description of the CMS detector, together with a definition of the coordinate system used and the relevant kinematic variables, can be found in Ref.~\cite{Chatrchyan:2008zzk}.

\section{Data and simulated samples}
\label{sec:samples}

The measurements presented in this paper are based on $\Pp\Pp$ collision data recorded by the CMS experiment during the 2016 run at $\sqrt{s}=13\TeV$, corresponding to an integrated luminosity of 35.9\fbinv.
The average number of $\Pp\Pp$ interactions per bunch crossing is $\left< \mu \right> = 27$.

The \ttbar signal process is simulated with the \POWHEG~v2~\cite{Nason:2004rx,powhegv22,powhegv23,Frixione:2007nw} matrix-element (ME) generator at NLO accuracy with a top quark mass value $m_{\cPqt} = 172.5\GeV$.
The \ttbar samples are normalized to the cross section calculated at next-to-next-to-leading order~\cite{Czakon:2011xx}.
The \ttbar{}+\PW, \ttbar{}+\PZ, \PW\cPZ{}, \PW+jets, and $\cPZ\cPZ$$\rightarrow$$2\ell2\cPq$ (where $\ell$ denotes a lepton) background processes are generated at NLO using \MGvATNLO v2.2.2~\cite{Alwall:2014hca} with the FxFx merging scheme~\cite{Frederix:2012ps} for the jets from the ME generator and the parton shower.
The Drell--Yan background is computed at leading order (LO) with the MLM merging prescription~\cite{Alwall:2007fs}.
The \PW\PW{}, $\cPZ\cPZ$$\rightarrow$$2\ell2\nu$, and \cPqt\PW{} backgrounds are generated with \POWHEG~v2~\cite{Melia:2011tj,Re:2010bp}, while single top quark $t$-channel production is simulated using \POWHEG~v2~\cite{Alioli:2009je} complemented with \textsc{MadSpin}~\cite{Frixione:2007zp,Artoisenet:2012st}.
QCD multijet background events are generated with \PYTHIA~v8.219~\cite{Sjostrand:2014zea}.
The NNPDF3.0 NLO~\cite{Ball:2014uwa} set of parton distribution functions (PDFs), and the strong coupling $\asmz=0.118$ are used in the ME calculations.

The ME generators are interfaced with \PYTHIA~8 for parton shower, hadronization, and underlying multiparton interactions (MPI).
\PYTHIA~8 implements a dipole shower ordered in transverse momentum (\pt), with ME corrections~\cite{Norrbin:2000uu} for the leading emissions in the top quark and {\PW} boson decays.
The hadronization of quarks and gluons into final hadrons is described by the Lund string model~\cite{Andersson:1983ia,Sjostrand:1984ic}, with the Bowler--Lund fragmentation function for heavy quarks~\cite{Bowler:1981sb}.
The CUETP8M2 tune, taking into account \ttbar jet multiplicity data~\cite{CMS-PAS-TOP-16-021}, is used for the \ttbar signal and the single top quark background, while the CUETP8M1 tune~\cite{Khachatryan:2015pea} is used for the remaining processes.
Additional \ttbar samples were generated with parameter variations to estimate systematic uncertainties (Section~\ref{sec:sys}), as well as with \POWHEG interfaced with \HERWIGpp v2.7.1~\cite{Bahr:2008pv}. In \HERWIGpp, the parton shower follows angular-ordered radiation~\cite{Gieseke:2003rz}, and the hadronization is described by the cluster model~\cite{Webber:1983if}.

The generated events are processed with the CMS detector simulation based on \GEANTfour~\cite{Agostinelli:2002hh}.
Additional $\Pp\Pp$ interactions in the same bunch crossing (pileup) are taken into account by adding detector hits of simulated minimum-bias events before event reconstruction.
The simulation is weighted to reproduce the pileup conditions observed in the data.
The simulated events are also corrected for the difference in performance between data and simulation of the trigger paths as well as in lepton identification and isolation efficiencies with scale factors depending on \pt and $\eta$.
The simulated tracking efficiency is corrected with scale factors that depend on the track $\eta$.

Additional predictions are generated without detector simulation for comparisons at the particle level.
\POWHEG~v2 is interfaced with \HERWIG~v7.1.1~\cite{Bellm:2015jjp} using the angular-ordered shower.
In addition, a prediction from \SHERPA~v2.2.4~\cite{Gleisberg:2008ta} with MC@NLO~\cite{Frixione:2002ik} corrections is included.
The parton shower in \SHERPA~2 is based on the Catani--Seymour dipole factorization~\cite{Schumann:2007mg}, and hadrons are formed by a modified cluster hadronization model~\cite{Winter:2003tt}.
The parton shower predictions from \PYTHIA~8, \HERWIG~7 and \SHERPA~2 have leading-log (LL) accuracy, with the option to use Catani--Marchesini--Webber (CMW) rescaling of \alpS to account for next-to-leading corrections to soft gluon emissions~\cite{Catani:1990rr}.
Events are also generated with \DIRE~v2.002~\cite{Hoche:2015sya}, a dipole-like parton shower ordered in (soft) \pt available as a plugin for \PYTHIA~8.
\DIRE~2 includes two- and three-loop cusp effects for soft emissions and partial NLO collinear evolution~\cite{Hoche:2017iem,Hoche:2017hno}, denoted nLL accuracy hereafter.
The values of the QCD coupling in the final-state radiation (FSR) showers, \asfsr, are summarized in Table~\ref{tab:alphas_gen}.
They are obtained from tuning the generator to LEP data using its default settings, with the exception of \SHERPA~2, where the \asmz is chosen to be consistent between ME calculation and parton shower.
The \PYTHIA~8 and \SHERPA~2 generators apply a model where the MPIs are interleaved with parton showering~\cite{Sjostrand:1987su}, while \HERWIG~7 models the overlap between the colliding protons through a Fourier transform of the electromagnetic form factor, which plays the role of an effective inverse proton radius~\cite{Butterworth:1996zw,Borozan:2002fk,Bahr:2008dy,Gieseke:2012ft}.
Depending on the amount of proton overlap, the contribution of generated MPIs varies in the simulation.
The MPI parameters of all generators are tuned to measurements in $\Pp\Pp$ collisions at the LHC~\cite{Khachatryan:2015pea}.

\begin{table*}
{
\centering
\topcaption{Overview of the theoretical accuracy and \asfsr settings of the generator setups used for predicting the jet substructure.
The acronym ``nLL'' stands for approximate next-to-leading-log accuracy.}
\label{tab:alphas_gen}
\begin{scotch}{lrrrrrr}
 &  \multicolumn{3}{c}{\POWHEG + \PYTHIA~8} & \POWHEG + & \SHERPA~2 & \DIRE~2 \\
 &  FSR-down & Nominal & FSR-up & \HERWIG~7 &  &  \\
\hline
\ttbar production & NLO & NLO & NLO & NLO & NLO & LO \\
\cPqt/{\PW} decay & NLO & NLO & NLO & NLO & LO & LO \\
Decay emission & LO & LO & LO & LO & LL & nLL \\
Shower accuracy & LL & LL & LL & LL & LL & nLL \\ [\cmsTabSkip]
\asfsr & 0.1224 & 0.1365 & 0.1543 & 0.1262 & 0.118 & 0.1201 \\
Evolution & One-loop & One-loop & One-loop & Two-loop & Two-loop & Two-loop \\
Scheme & \msbar & \msbar & \msbar & \msbar & CMW & \msbar \\ [\cmsTabSkip]
\end{scotch}
}
\end{table*}

\section{Event reconstruction and selection}
\label{sec:recosel}

The particle-flow (PF) event algorithm~\cite{Sirunyan:2017ulk} aims to reconstruct and identify each individual particle in an event with an optimized combination of information from the various elements of the CMS detector. The energy of photons is directly obtained from the ECAL measurement, corrected for zero-suppression effects. The energy of electrons is determined from a combination of the electron momentum at the primary interaction vertex as determined by the tracker, the energy of the corresponding ECAL cluster, and the energy sum of all bremsstrahlung photons spatially compatible with originating from the electron track. The energy of muons is obtained from the curvature of the corresponding track. The energy of charged hadrons is determined from a combination of their momentum measured in the tracker and the matching ECAL and HCAL energy deposits, corrected for zero-suppression effects and for the response function of the calorimeters to hadronic showers. Finally, the energy of neutral hadrons is obtained from the corresponding corrected ECAL and HCAL energies.

For each event, hadronic jets are clustered from these reconstructed particles using the infrared- and collinear-safe anti-\kt algorithm~\cite{Cacciari:2008gp} with a distance parameter $R=0.4$, as implemented in \FASTJET~3.1~\cite{Cacciari:2011ma}. The jet momentum is determined as the vectorial sum of all particle momenta in this jet, and is found in the simulation to agree with the true jet momentum within 5 to 10\% over the whole \pt spectrum and detector acceptance. Jet energy corrections are derived from the simulation, and are confirmed with in situ measurements of the energy balance in dijet, multijet, photon+jet, and leptonically-decaying \PZ{}+jet events. The jet energy resolution amounts typically to 15\% at 10\GeV, 8\% at 100\GeV, and 4\% at 1\TeV~\cite{Khachatryan:2016kdb}.

\begin{figure*}
  \centering
  \includegraphics[width=0.48\textwidth]{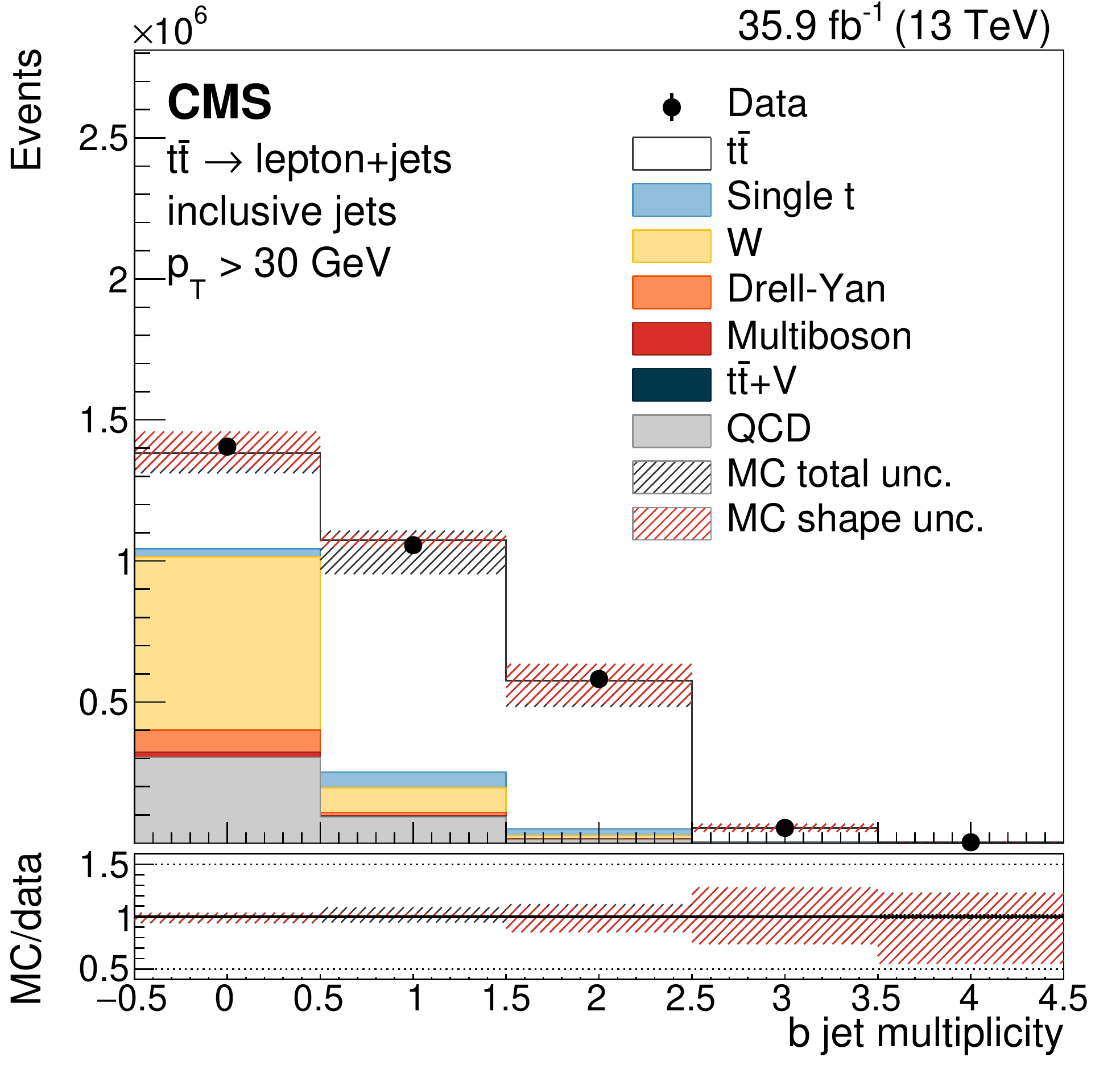}
  \includegraphics[width=0.48\textwidth]{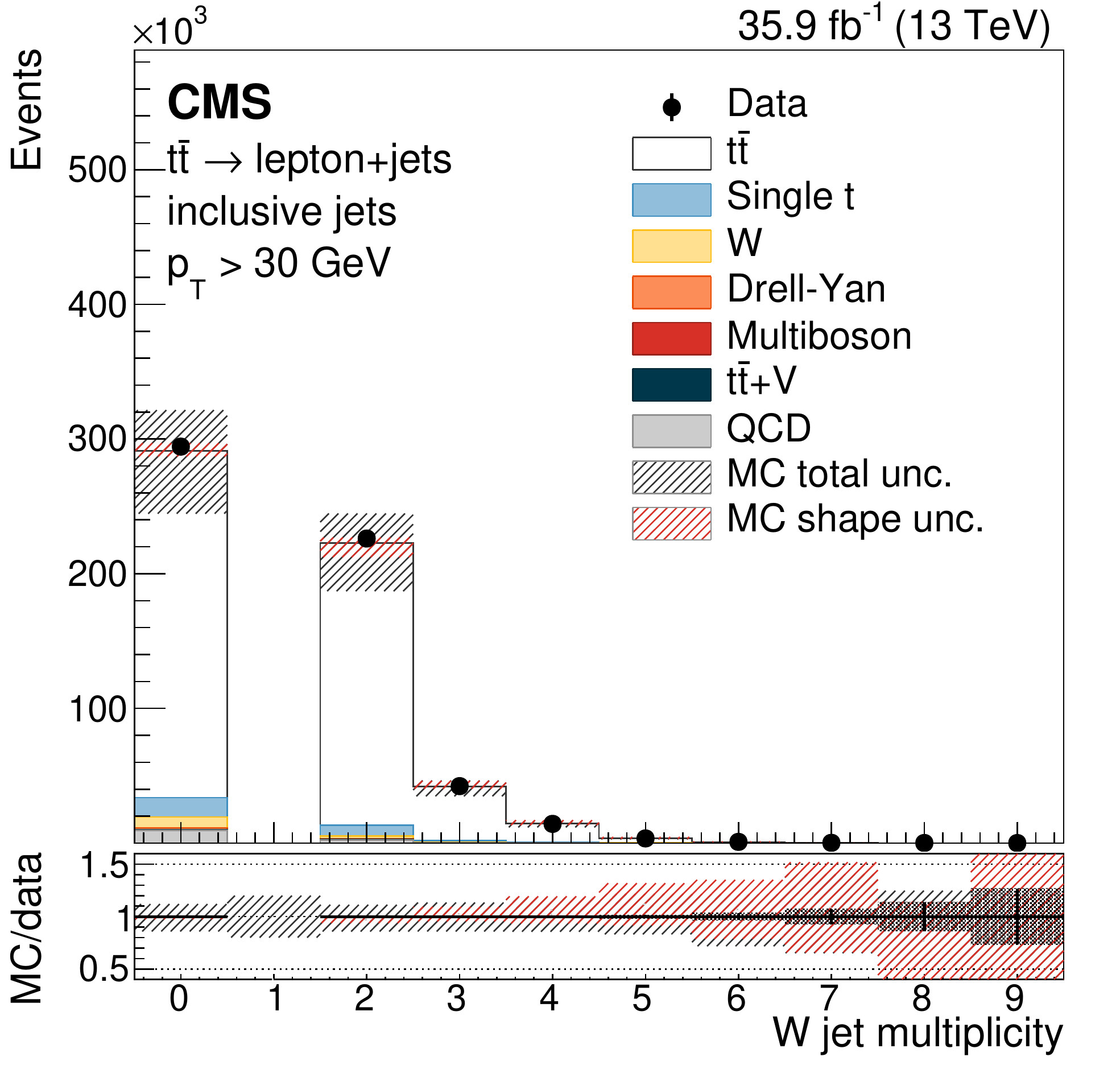}
  \caption{Multiplicity of \cPqb-tagged jets in events with exactly one isolated lepton and four jets (left), and multiplicity of untagged jets yielding a \PW{} boson candidate with $\abs{m_{jj}-80.4\GeV}<15\GeV$ after requiring two \cPqb-tagged jets (right).
These reconstruction-level plots show the sum of the expected contributions from each process (stacked histograms) compared to the data points (upper panels), and the ratio of the MC prediction (\POWHEG + \PYTHIA~8) to the data (lower panels) where the black shaded band represents the statistical uncertainty on the data. The systematic uncertainties on the MC prediction are represented by hatched areas, taking into account either the total uncertainty or shape variations only.
  }
  \label{fig:control_flavor}
\end{figure*}

The event selection is based on the \ttbar lepton+jets decay topology, where data samples are collected using electron or muon triggers with a \pt threshold of 32 or 24\GeV, respectively.
In the offline selection, the relative isolation of electrons (muons) is defined as the scalar sum of PF candidates \pt within a cone of $\Delta R = \sqrt{\smash[b]{(\Delta\eta)^2+(\Delta\phi)^2}} = 0.3 \left(0.4\right)$ (where $\Delta\eta$ and $\Delta\phi$ are the separations in pseudorapidity and azimuth (in radians) of lepton and PF candidate) around the lepton direction, divided by the lepton \pt, and is required to be smaller than 0.06 (0.15).
Leptons have to fulfill tight identification criteria, taking into account track properties and energy deposits, based on their expected signature in the detector.
Exactly one isolated lepton (electron or muon) is required, having $\pt > 34\,(26)\GeV$ and $\abs{\eta} < 2.1\,(2.4)$ for electrons (muons)~\cite{Khachatryan:2015hwa,Sirunyan:2018fpa}.
The event is not selected in the presence of a second loosely-identified lepton with $\pt > 15\GeV$ and $\abs{\eta} < 2.4$, in order to suppress Drell--Yan and \ttbar dilepton events.
Furthermore, the events are required to contain at least four jets with $\pt>30\GeV$ and $\abs{\eta}<2.5$, of which at least two are required to be \cPqb-tagged.
The Combined Secondary Vertex (CSVv2) \cPqb{} tagging algorithm is used at a working point, which has a mean efficiency of 63\% for the correct identification of a bottom jet and a probability of 0.9\% for misidentifying light-flavor (uds or gluon) jets, and 12\% for charm jets in a \ttbar sample~\cite{Chatrchyan:2012jua,Sirunyan:2017ezt}.
Finally, at least two untagged jets are required to yield a \PW{} boson candidate with an invariant mass satisfying $\abs{m_{jj}-80.4\GeV}<15\GeV$, and these jets are composing the light-quark-enriched jet sample.
Events with no (one) \PW{} boson candidate contain no (two) light-quark-enriched jets.
Events are allowed to contain more than one \PW{} boson candidate, leading to more than two jets associated to the light-quark-enriched sample.
The number of events selected in data is 287\,239, with $285\,000 \pm 38\,000$ expected.
The selected sample is composed of 93.8\% \ttbar events as estimated from simulation.
The multiplicities of bottom-quark jets and untagged jets compatible with \PW{} boson candidates at the reconstructed level are presented in Fig.~\ref{fig:control_flavor} and show good agreement between data and MC prediction.

\begin{figure*}
  \centering
  \includegraphics[width=0.48\textwidth]{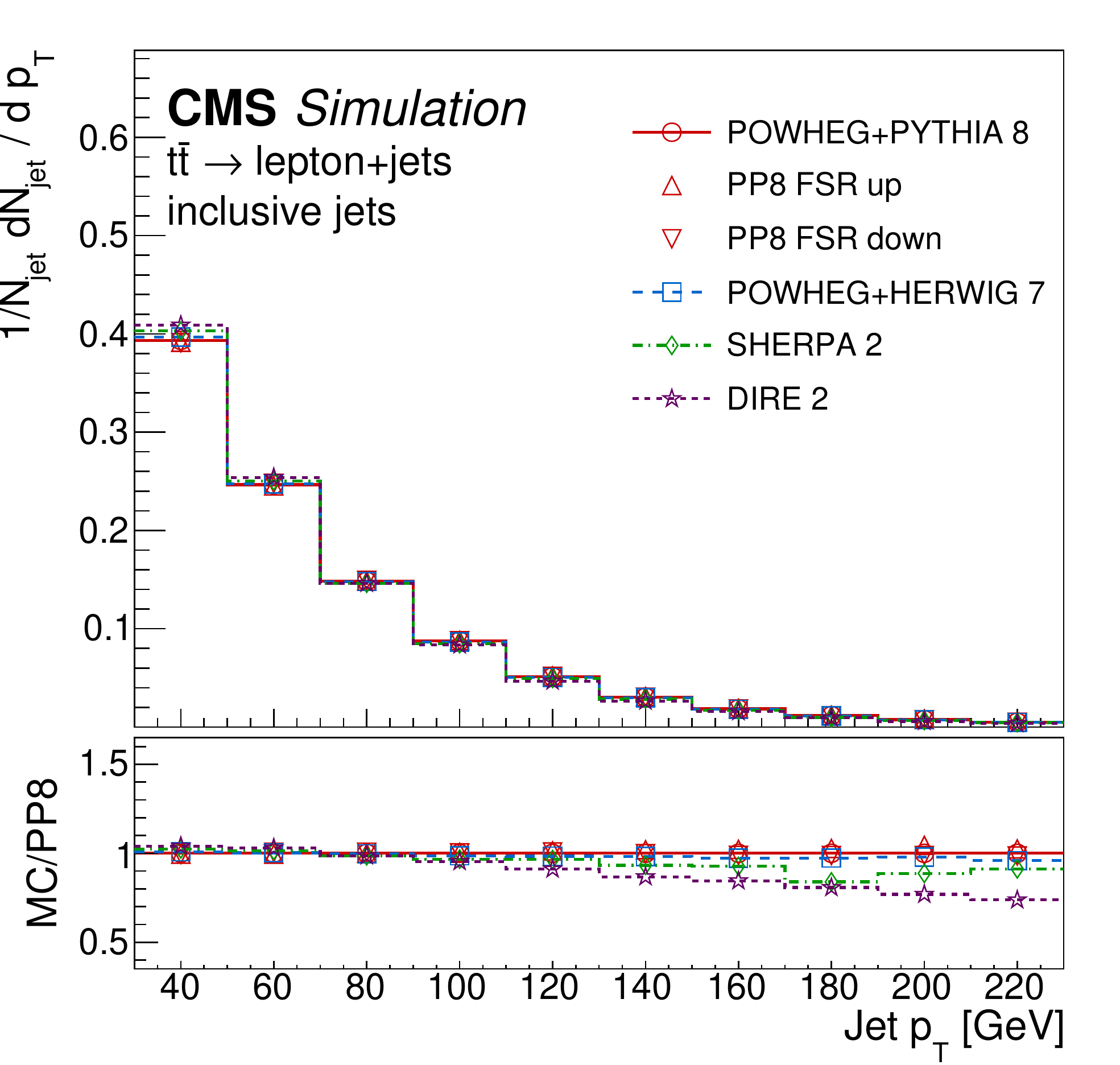}
  \includegraphics[width=0.48\textwidth]{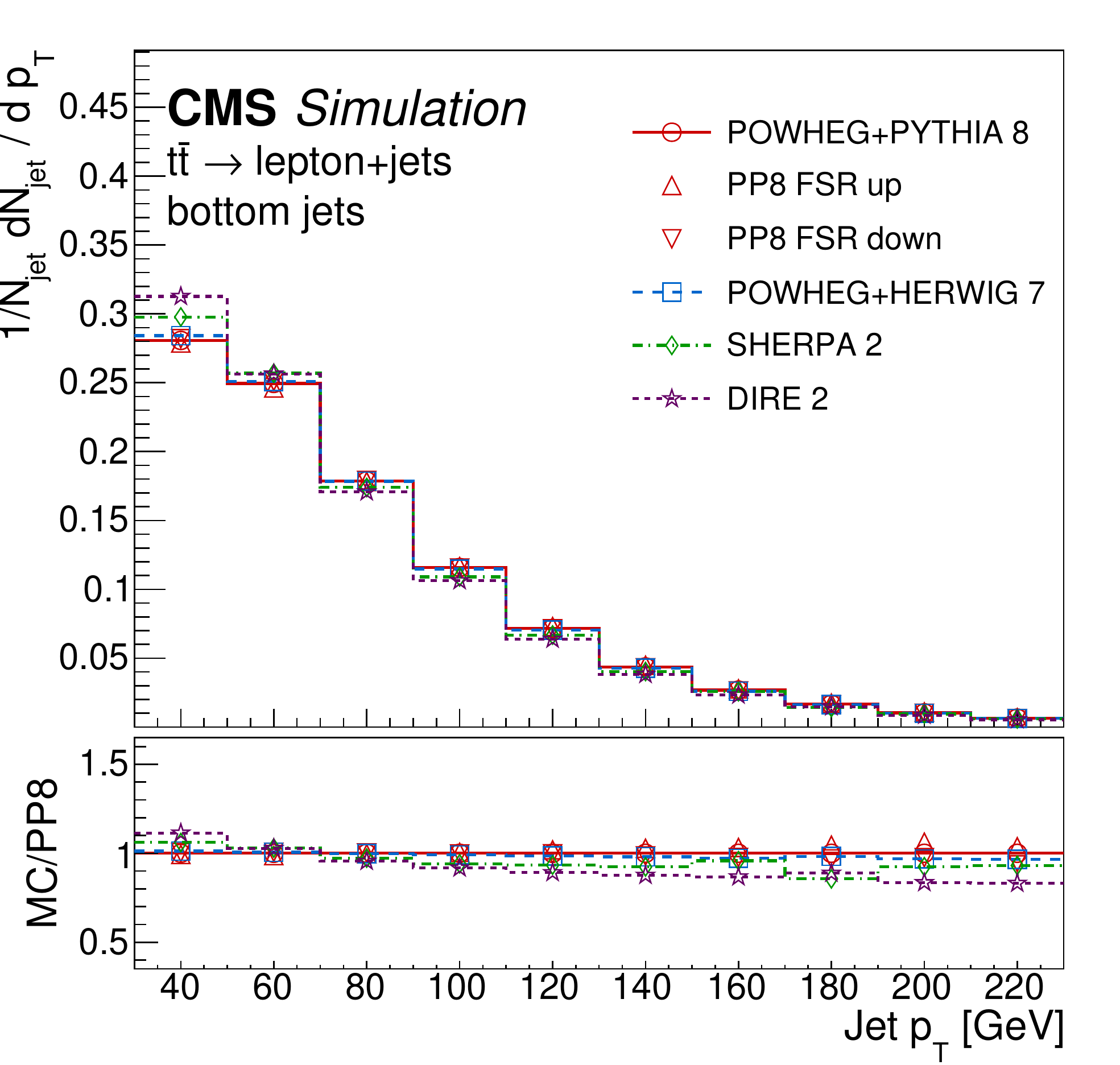}
  \includegraphics[width=0.48\textwidth]{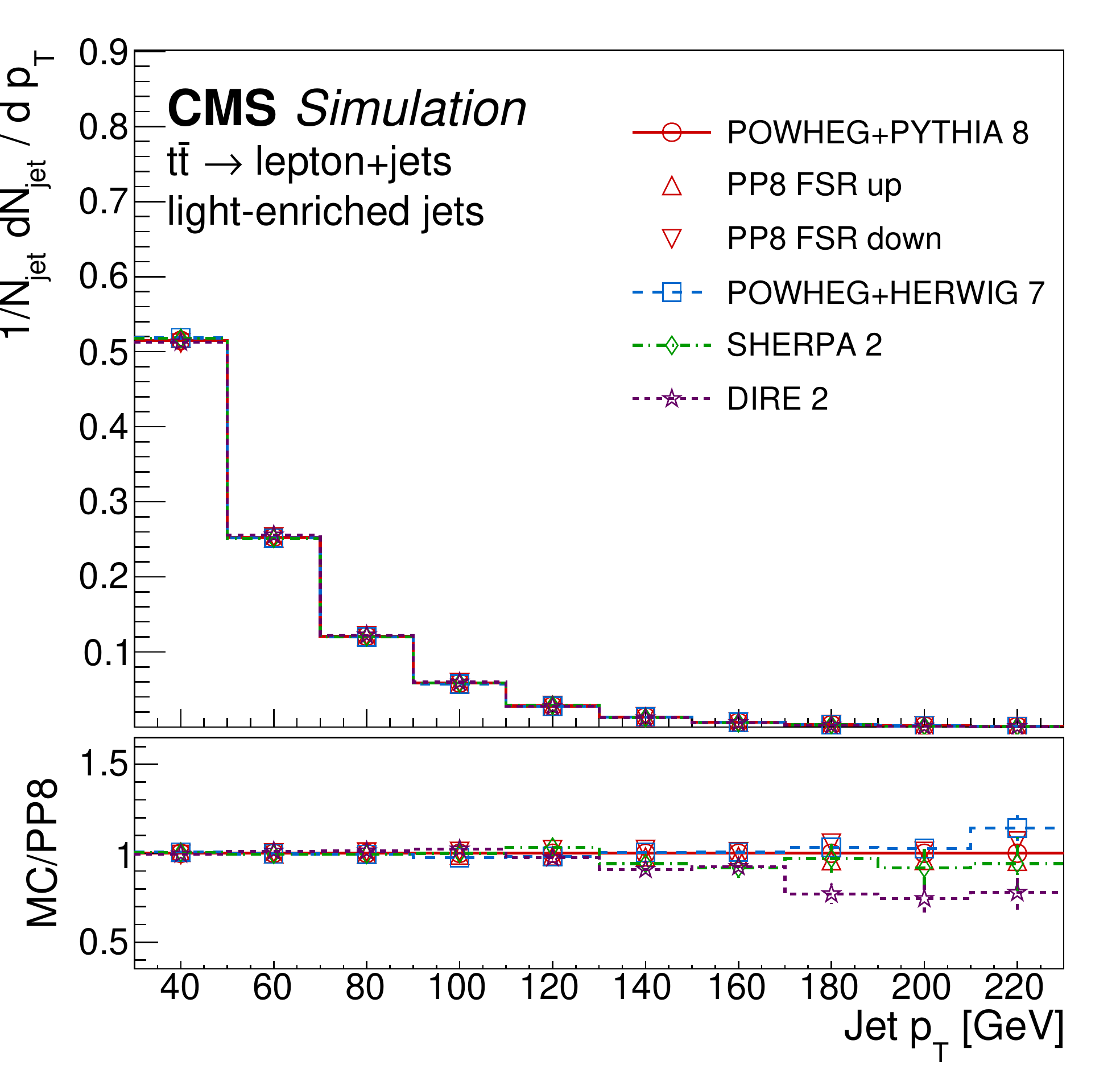}
  \includegraphics[width=0.48\textwidth]{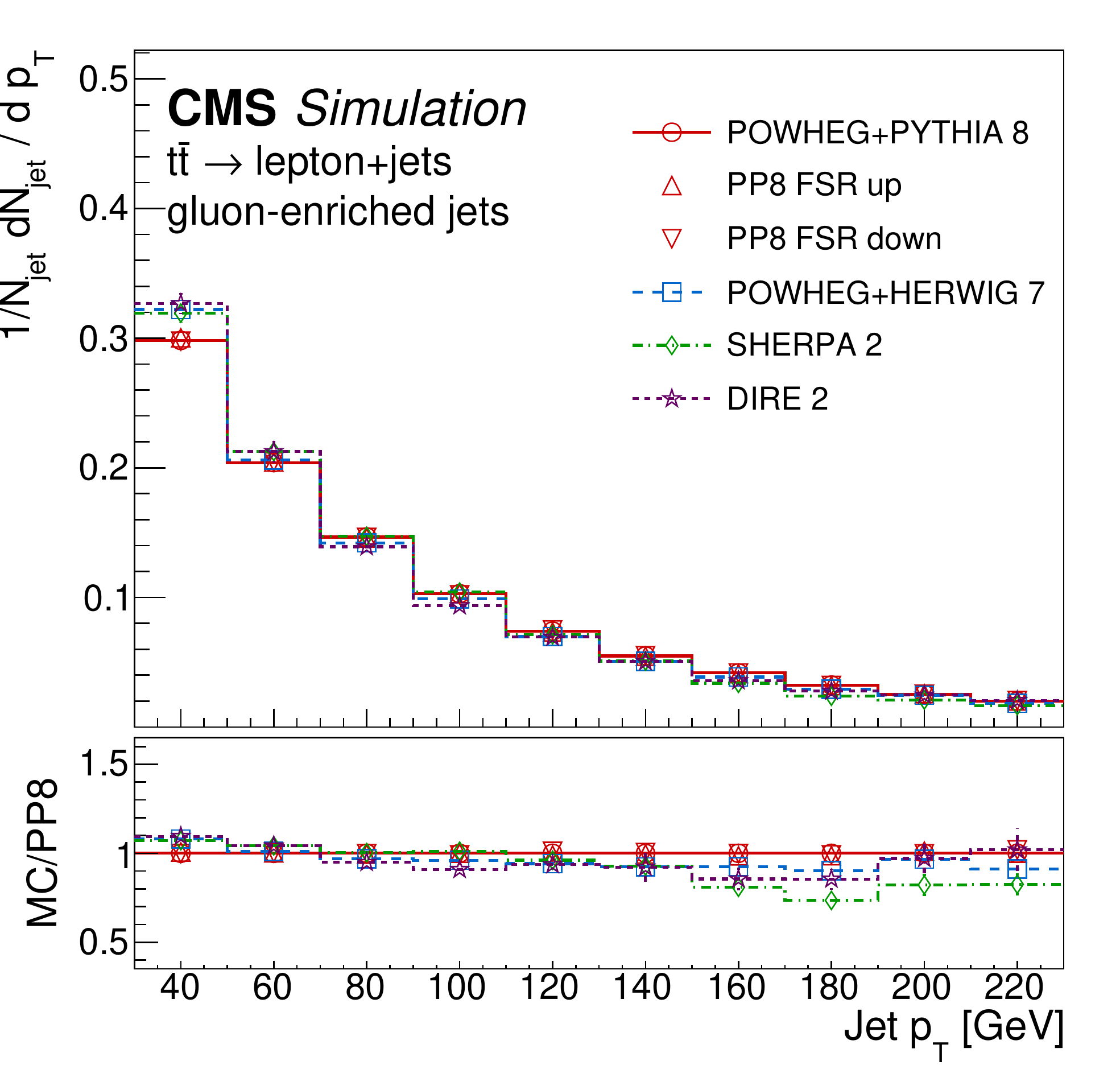}
  \caption{Transverse momentum distribution at the particle level for inclusive jets (upper left), bottom-quark jets (upper right), light-quark-enriched jets (lower left), and gluon-enriched jets (lower right).
  The sub-panels show the corresponding ratios of the different MC predictions over \POWHEG + \PYTHIA~8 (PP8).
  }
  \label{fig:flavor_pt}
\end{figure*}

At the particle level in simulated events, the unfolded jet observables are defined in a phase space region described hereafter.
More details about the algorithms and relevant studies can be found in Ref.~\cite{Collaboration:2267573}.
Leptons are required to be prompt (\ie, not from hadron decays), and the momenta of prompt photons located within a cone of radius $\Delta R = 0.1$ are added to the lepton momentum to account for FSR, referred to as ``dressing''.
Exactly one lepton with $\pt > 26\GeV$ and $\abs{\eta} < 2.4$ is required, while events containing additional dressed leptons fulfilling looser kinematic criteria ($\pt > 15\GeV$, $\abs{\eta} < 2.4$) are rejected.
Jets are clustered from stable particles excluding neutrinos and the dressed leptons with the anti-\kt algorithm using a distance parameter $R=0.4$.
At least four jets with $\pt>30\GeV$ and $\abs{\eta}<2.5$ are required.
In order to identify the jet flavor at particle level, decayed heavy hadrons are included in the jet clustering after scaling their momenta by $10^{-20}$ (known as ``ghost'' tagging~\cite{Cacciari:2008gn}).
A jet is identified as a bottom jet when it contains at least one bottom hadron, and two \cPqb-tagged jets are required in the event.
At least one pair of untagged jet candidates needs to fulfill the \PW{} boson mass constraint $\abs{m_{jj}-80.4\GeV}<15\GeV$.
The \pt distributions at the particle level are shown in Fig.~\ref{fig:flavor_pt} for different MC generators and different jet flavor samples (cf. Section~\ref{sec:flavor} for the flavor definitions).

\section{Unfolding and systematic uncertainties}
\label{sec:unfolding}
\label{sec:sys}
\label{sec:unfoldsys}

All jet substructure distributions described in the following sections are unfolded to the particle level.
Unregularized unfolding as implemented in the \textsc{TUnfold} package~\cite{Schmitt:2012kp} is used to correct the background-subtracted data distributions to the particle level by minimizing $\chi^2=(y-K\lambda)^T V_{yy}^{-1} (y-K\lambda)$,
where $K$ is the particle-to-reconstructed phase space migration matrix,
$V_{yy}$ is an estimate of the covariance of the observations $y$,
and $\lambda$ is the particle level expectation.
The binning of the migration matrix takes into account the resolution of the observables.
We define purity as the fraction of reconstructed events that are generated in the same bin, and stability as the fraction of generated events that are reconstructed in the same bin, divided by the overall reconstruction efficiency per bin.
Both quantities are $\geq 50$\% in most bins.
In each bin the fractional contribution of jets from \ttbar events that pass selection criteria at detector- but not at particle-level is subtracted.
The unfolded distributions are normalized to unity within the chosen axis range, \ie, the overflow is discarded.
Pseudo-experiments are conducted by unfolding pseudo-data distributions sampled from simulated \ttbar events and confirm that the unfolding does not introduce any bias and yields a correct estimate of the statistical uncertainties.

While the central result is unfolded using \POWHEG + \PYTHIA~8 with the nominal data-to-simulation correction factors, systematic uncertainties are assessed by using migration matrices obtained from alternative samples and systematic variations of the correction factors used in this analysis.
The uncertainty in the number of pileup events is estimated by changing the total inelastic pp cross section by $\pm 5$\%~\cite{Sirunyan:2018nqx}.
The data-to-simulation scale factors for lepton trigger and selection efficiencies are varied within their uncertainties.
The energy scale of jets is varied within its uncertainty, as a function of the jet \pt, $\eta$, and flavor, as well as the jet resolution, depending on its $\eta$.
The {\cPqb} tagging efficiency and misidentification probabilities are varied within their uncertainties.
A data-to-simulation tracking efficiency scale factor is determined as a function of $\eta$ for charged pions.
An uncertainty of 3--6\% is assigned to the tracking scale factor, assumed to be correlated across run periods and detector regions, resulting in a global up or down variation.
The cross sections of the most important backgrounds contributions are scaled within their uncertainties: 5\% for single top quark~\cite{Aliev:2010zk,Kant:2014oha,Kidonakis:2010ux,Kidonakis:2013zqa}, 10\% for $\PW+1$ jet, and 33\% for $\PW+2$ jets~\cite{Alwall:2014hca,Frederix:2012ps} processes.
We assume an uncertainty of 100\% on the QCD multijet background predicted by the MC.

\begin{figure*}[ht]
  \centering
  \includegraphics[width=0.48\textwidth]{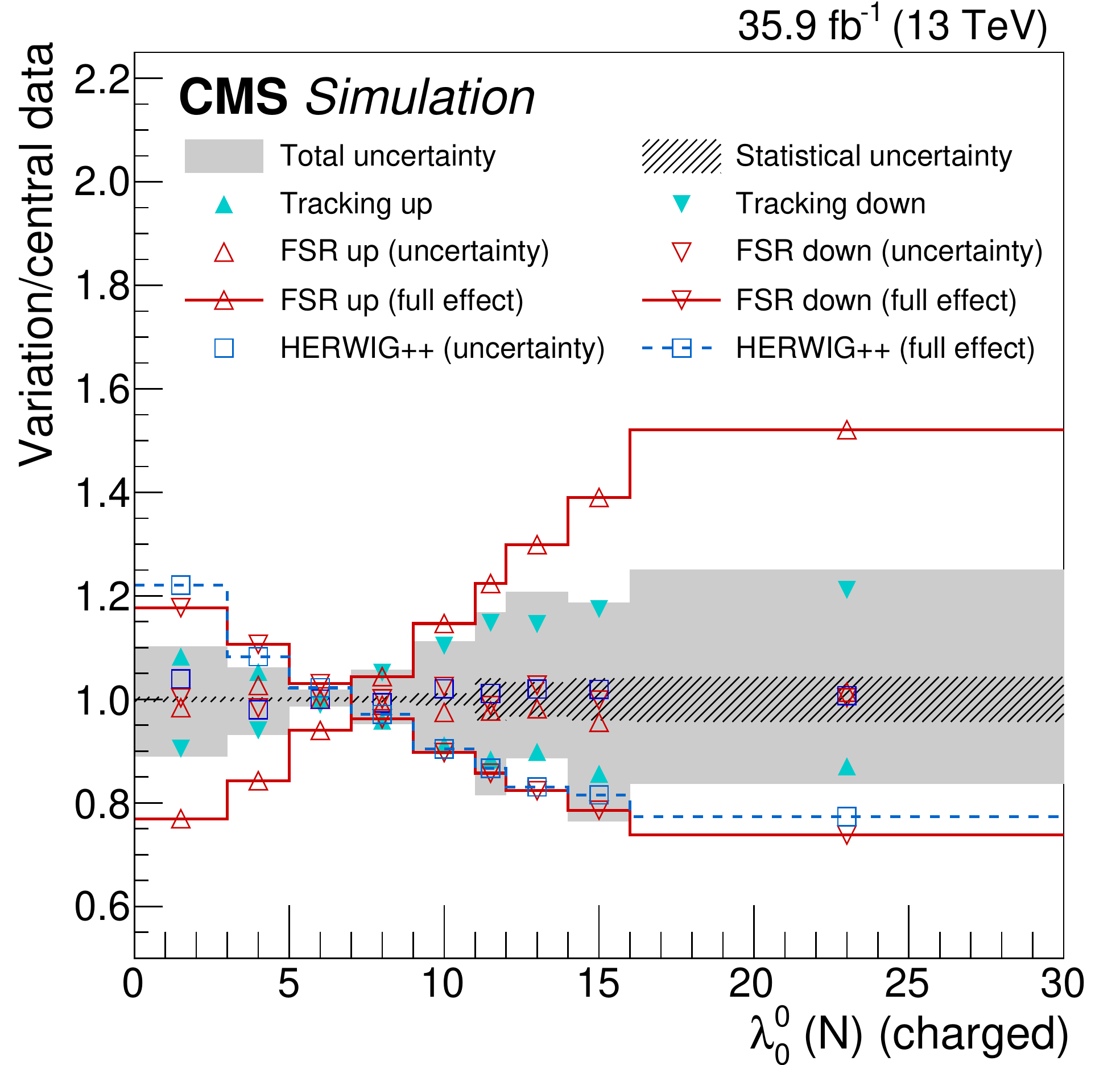}
  \includegraphics[width=0.48\textwidth]{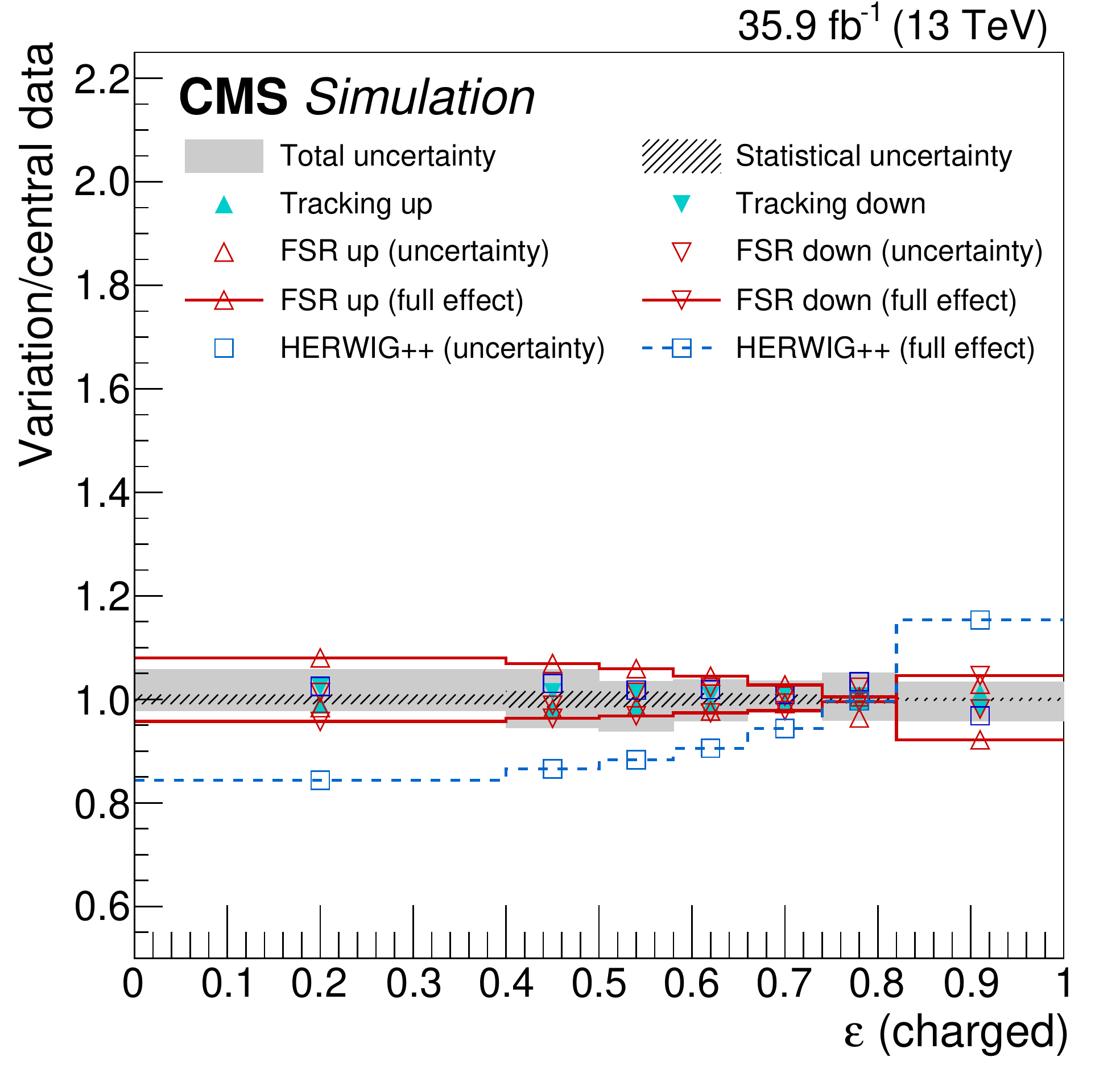}
  \includegraphics[width=0.48\textwidth]{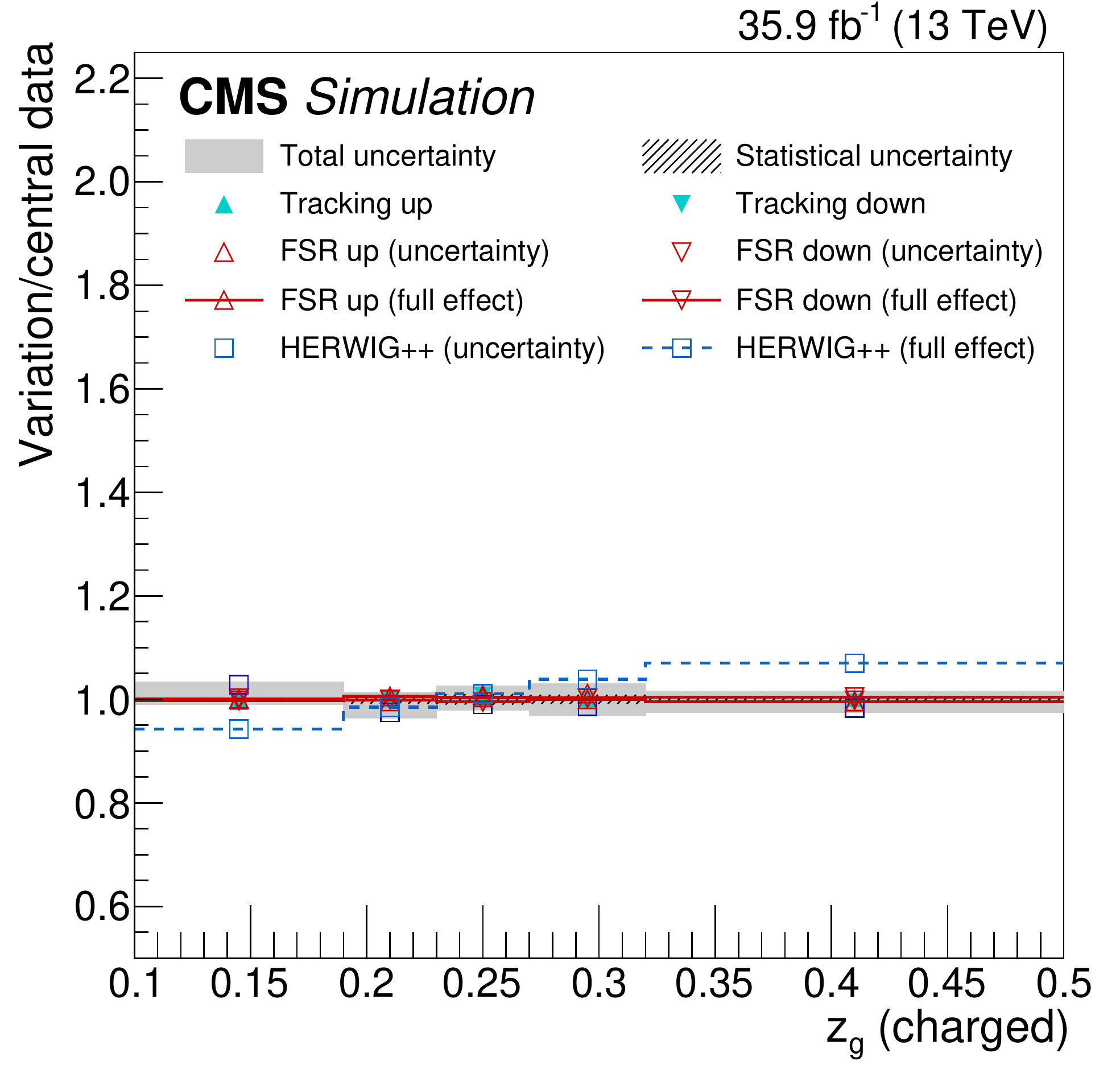}
  \includegraphics[width=0.48\textwidth]{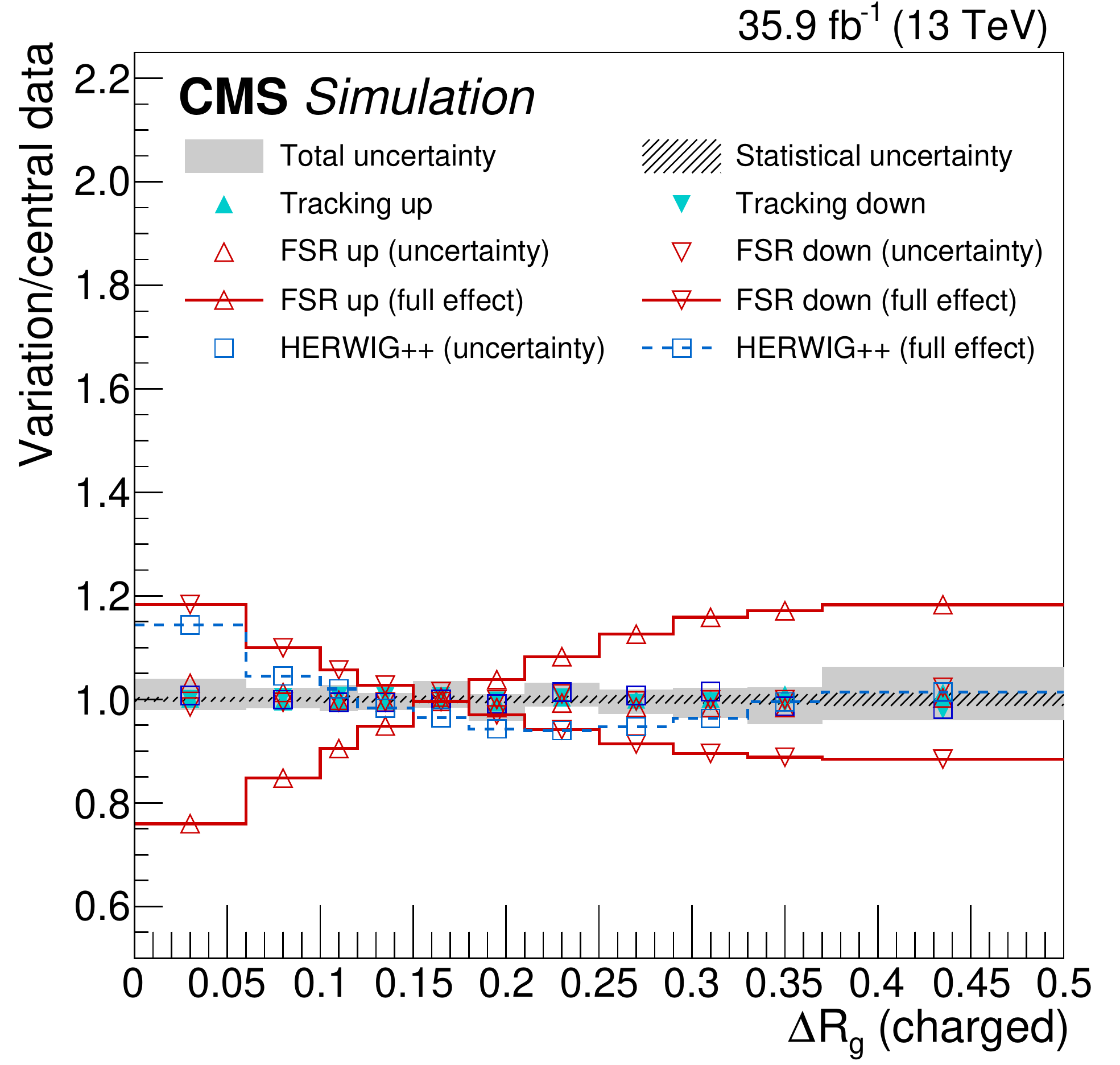}
  \caption{Systematic uncertainties for the charged multiplicity $\lambda_{0}^{0}$ ($N$) (upper left), jet eccentricity $\varepsilon$ (upper right), groomed momentum fraction $z_\mathrm{g}$ (lower left), and angle between the groomed subjets $\Delta R_\mathrm{g}$ (lower right).
  The uncertainties from FSR and \HERWIG (open markers) are compared to the full effect of these variations at the particle level (open markers with lines).
  }
  \label{fig:syst_unc}
\end{figure*}

The uncertainties in the modeling of the \ttbar lepton+jets signal are estimated using migration matrices derived from fully simulated samples with the following variations.
The renormalization and factorization scales in the ME calculation are varied by factors of 0.5 and 2.0 using weights.
CT14 (NLO)~\cite{Dulat:2015mca} and MMHT2014 (NLO)~\cite{Harland-Lang:2014zoa} are used as alternative PDF sets.
The scales for ISR and FSR in the parton shower are varied independently by factors of 0.5 and 2.0 with respect to their default values.
The $h_\text{damp}$ parameter regulating the real emissions in \POWHEG is varied from its central value of $1.58 \, m_{\cPqt}$ using samples with $h_\text{damp}$ set to $0.99\,m_{\cPqt}$ and $2.24\,m_{\cPqt}$, as obtained from tuning to \ttbar data at $\sqrt{s} = 8 \TeV$~\cite{CMS-PAS-TOP-16-021}.
Additional samples are generated with the MPI tune varied within its uncertainties.
For estimating the uncertainty due to color reconnection (CR), we consider the difference between including and excluding (default) the top quark decay products in the default model which fuses the color flow of different systems to minimize the total color string length~\cite{Sjostrand:1987su}.
Two additional models are taken into account, including the top quark decay products: a new model respecting QCD color rules~\cite{Christiansen:2015yqa}, and the gluon move scheme~\cite{Argyropoulos:2014zoa} for minimizing the total string length.
An additional sample is generated using \POWHEG interfaced with \HERWIGpp for testing an alternative model of parton shower, hadronization, MPI, and CR.
The \cPqb{} fragmentation function is varied to cover $\Pep\Pem$ data at the \Z pole~\cite{Heister:2001jg,DELPHI:2011aa,Abbiendi:2002vt,Abe:2002iq} with the Bowler--Lund~\cite{Bowler:1981sb} and the Peterson~\cite{Peterson:1982ak} parametrizations.
Semileptonic branching fractions of b hadrons are varied within their measured values~\cite{Olive:2016xmw}.
The top quark mass is measured by CMS with an uncertainty of $\pm$0.49\GeV~\cite{Khachatryan:2015hba} and samples in this analysis are generated with $\pm$1\GeV in order to estimate its impact on the jet substructure measurements.
The \pt distribution of the top quark was found to be in disagreement with NLO predictions by recent CMS measurements at $\sqrt{s} = 13\TeV$~\cite{Khachatryan:2016mnb,Sirunyan:2017mzl}.
Therefore, the full data-to-simulation difference in the top quark \pt distribution is taken as an uncertainty.
The effects of the most important systematic uncertainties on selected observables (cf. Sections~\ref{sec:obs} and \ref{sec:chi2}) are shown in Fig.~\ref{fig:syst_unc}.
The uncertainties from the FSR modeling are shown to be significantly smaller than the respective full effect of the variations at the particle level, demonstrating the stability of the unfolded measurement against the MC model used for constructing the migration matrices.

\section{Jet substructure observables}
\label{sec:obs}

Jets are selected for further analysis if they satisfy $\pt>30\GeV$, $\abs{\eta} < 2.0$, so that jets with $R=0.4$ are completely contained within the tracker acceptance $\abs{\eta} < 2.5$.
Furthermore, jets are required to be separated in $\eta$--$\phi$ space by $\Delta R(jj)>0.8$ to avoid overlap.
The jet substructure observables are calculated from the jet constituents with $\pt > 1\GeV$, so as to avoid the rapid decrease (increase) in tracking efficiency (misidentification rate) below 1\GeV~\cite{Chatrchyan:2014fea}.
We present our results either with all (charged+neutral) particles, or with only charged particles if the resolution on the variable reconstructed from both charged+neutral particles is poor.
The whole set of jet results obtained from charged and charged+neutral particles is available in the \textsc{HepData} database~\cite{Buckley:2010jn,Maguire:2017ypu}.
Hereafter, a variety of jet substructure observables are presented for the inclusive set of jets.
Individual jet flavor-tagged results are shown in Section~\ref{sec:flavor}.

\subsection{Generalized angularities}

Generalized angularities~\cite{Larkoski:2014pca} are defined as
\begin{linenomath}
\begin{equation}\lambda_{\beta}^{\kappa} = \sum_i{z_i^{\kappa}\left(\frac{\Delta R\left(i,\hat{n}_{r}\right)}{R}\right)^{\beta}}
\end{equation}
\end{linenomath}
where $z_i=\pt^{i}/\sum_{j}\pt^{j}$ is the \pt fraction carried by the particle $i$ inside the jet, $\Delta R\left(i,\hat{n}_{r}\right)$ is its separation in $\eta$--$\phi$ space from the jet axis $\hat{n}_{r}$, $R=0.4$ is the distance parameter used for the jet clustering, and $\kappa$ and $\beta$ are positive real exponents of the energy and angular weighting factors.
The recoil-free jet axis $\hat{n}_{r}$~\cite{Bertolini:2013iqa} is calculated with the ``winner-takes-all'' (WTA) recombination scheme~\cite{Larkoski:2014uqa} mitigating the impact of soft radiation.
Angularities with $\kappa = 1$ are infrared- and collinear- (IRC) safe, while those with $\kappa \neq 1$ are IRC-unsafe (but ``Sudakov'' safe)~\cite{Larkoski:2015lea}.
With the exception of $\lambda_0^0$, at least two selected particles are required in the jet in order to construct these observables.

\begin{figure}
  \centering
  \includegraphics[width=0.48\textwidth]{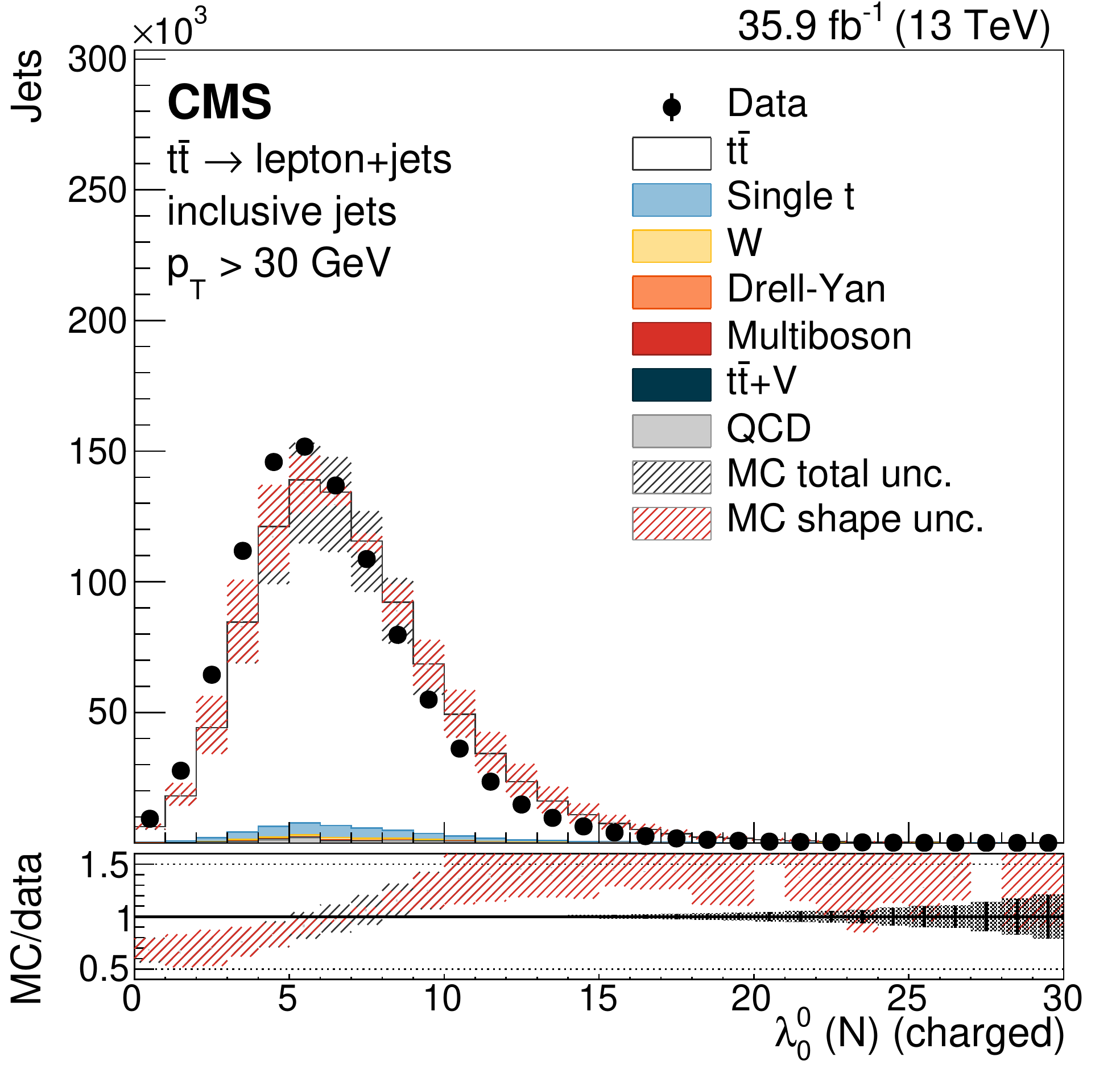}
  \caption{Charged particle multiplicity $\lambda_{0}^{0}$ ($N$) at the reconstructed level after full event selection. The lower panel shows the ratio of the MC prediction (\POWHEG + \PYTHIA~8) to the data (lower panels) where the black shaded band represents the statistical uncertainty on the data. The systematic uncertainties on the MC prediction are represented by hatched areas, taking into account either the total uncertainty or shape variations only.
  }
  \label{fig:control_mult}
\end{figure}

\begin{figure}
  \centering
  \includegraphics[width=0.48\textwidth]{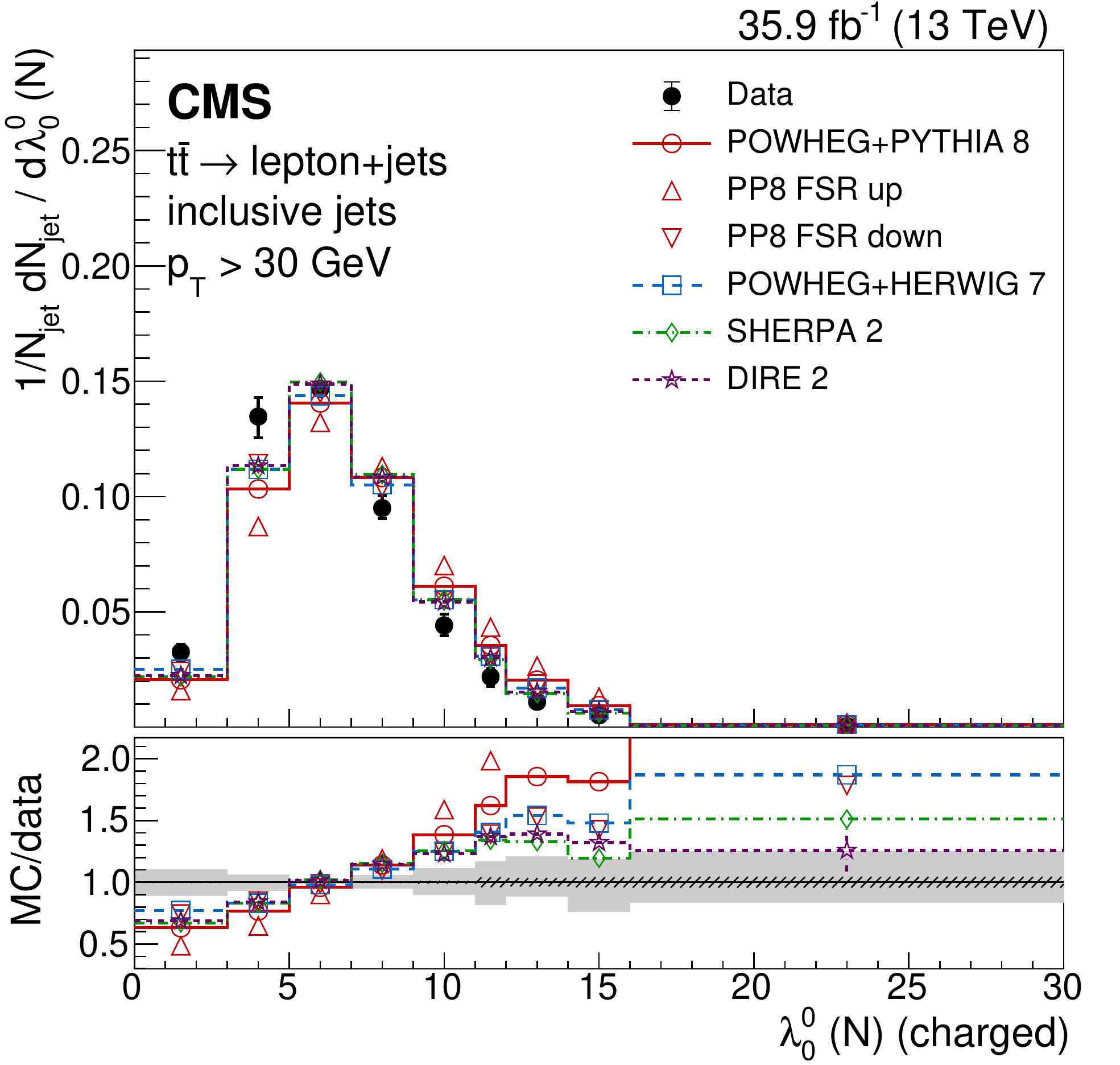}
  \caption{Charged particle multiplicity $\lambda_{0}^{0}$ ($N$) normalized and unfolded to the particle level, for inclusive jets.
  Data (points) are compared to different MC predictions (upper), and as MC/data ratios (lower). The hatched and shaded bands represent the statistical and total uncertainties, respectively.
  }
  \label{fig:mult_charged}
\end{figure}

The particle multiplicity $\lambda_{0}^{0}$ is neither infrared-, nor collinear-safe, as its value is changed by additional soft emissions and/or collinear splitting of partons.
In this analysis, $\lambda_{0}^{0} = N$ (charged) is the number of charged jet constituents passing the particle \pt threshold of 1\GeV and is shown at the reconstructed level in Fig.~\ref{fig:control_mult} and normalized and unfolded to the particle level in Fig.~\ref{fig:mult_charged}.
In general, the MC generators predict a higher (integrated) charged particle multiplicity than seen in the data but the \SHERPA~2 and \DIRE~2 predictions achieve a fair agreement.
An improved agreement could be achieved by including this or similar data in the tuning of the parton showering and hadronization~\cite{ATL-PHYS-PUB-2014-021,Aad:2016oit}.

\begin{figure*}
  \centering
  \includegraphics[width=0.48\textwidth]{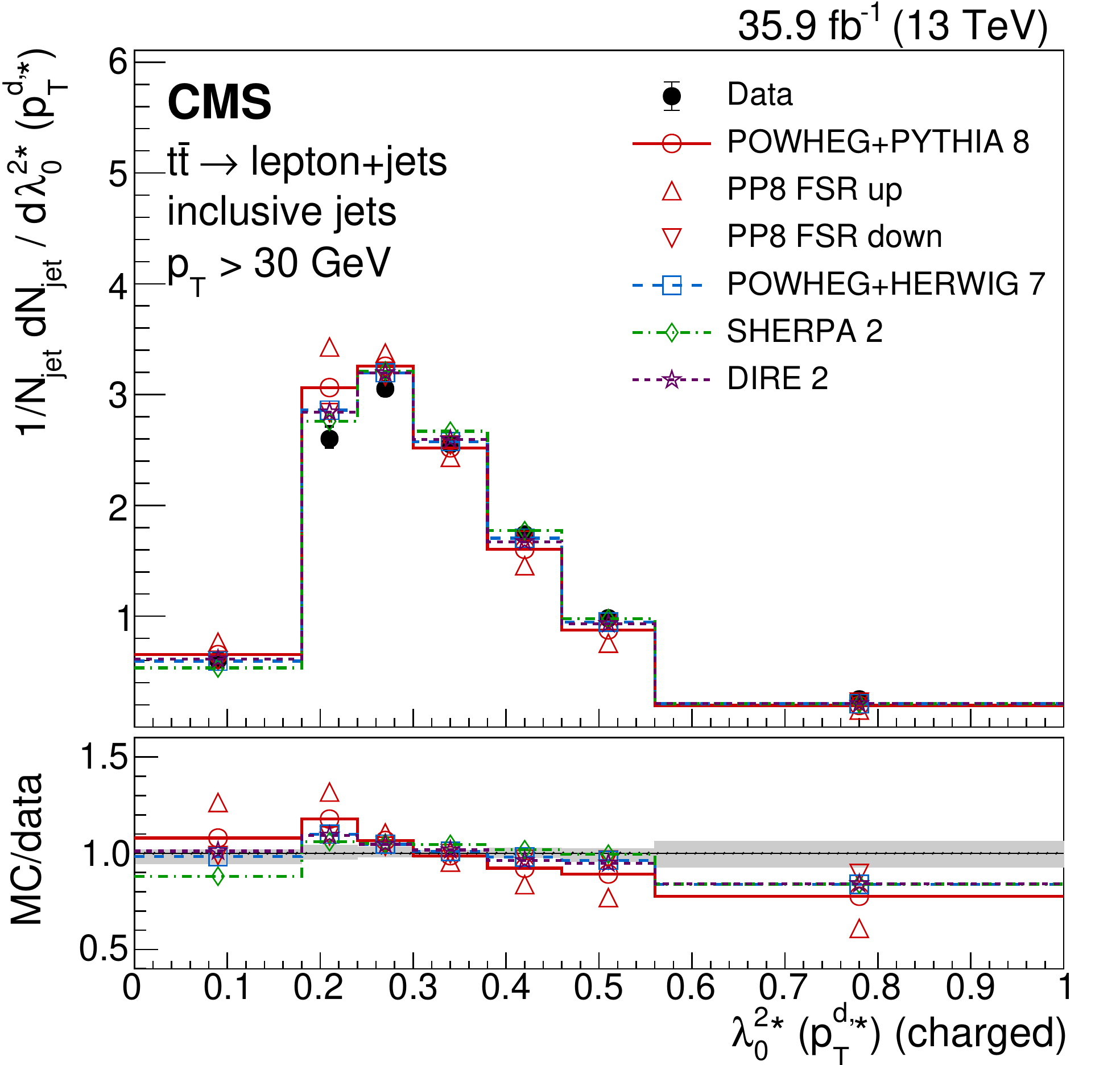}
  \includegraphics[width=0.48\textwidth]{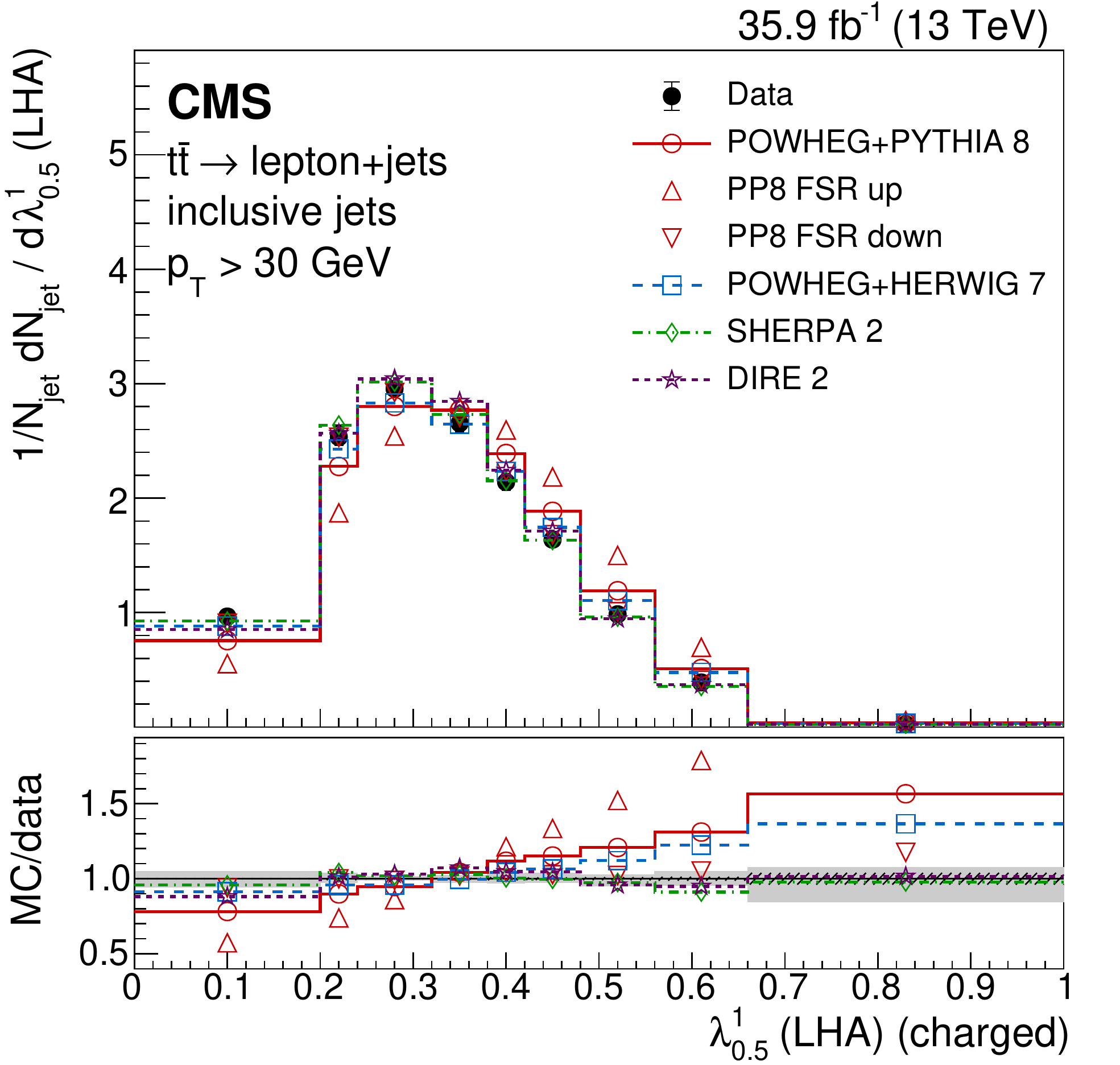}
  \caption{Distributions of the scaled \pt dispersion ($\lambda_0^{2*}$, left) and Les Houches angularity ($\lambda_{0.5}^1$, right), unfolded to the particle level, for inclusive jets reconstructed with charged particles. Data (points) are compared to different MC predictions (upper), and as MC/data ratios (lower). The hatched and shaded bands represent the statistical and total uncertainties, respectively.
  }
  \label{fig:lambda1_charged}
\end{figure*}

The jet \pt dispersion $\lambda_{0}^{2} = \pt^\mathrm{d} = {\sum_i (\pt^i)^2}/{\left(\sum_i \pt^i\right)^2}$~\cite{Chatrchyan:2012sn} is an infrared-, but not collinear-safe quantity, highly correlated with the particle multiplicity.
A scaled \pt dispersion is thus defined as
\begin{linenomath}
\begin{equation}
\lambda_{0}^{2*} = \pt^\mathrm{d,*} = \sqrt{\left(\pt^\mathrm{d} - \frac{1}{N}\right) \frac{N}{N-1}};
\end{equation}
\end{linenomath}
that ensures $\pt^\mathrm{d,*} \to 0$ when the \pt is equally distributed over all jet constituents, irrespective of their number, and $\pt^\mathrm{d,*} \to 1$ when most of the jet momentum is carried by a single particle.
The scaled \pt dispersion is shown in Fig.~\ref{fig:lambda1_charged} (left) compared to the MC predictions.

The ``Les Houches angularity'' (LHA) $\lambda_{0.5}^{1}$ variable, a quantity proposed for quark-gluon discrimination~\cite{Badger:2016bpw}, is well described by most available MC programs (Fig.~\ref{fig:lambda1_charged}, right).
The high \asfsr value associated with the \PYTHIA~8 FSR-up setting is disfavored.

\begin{figure*}
  \centering
  \includegraphics[width=0.48\textwidth]{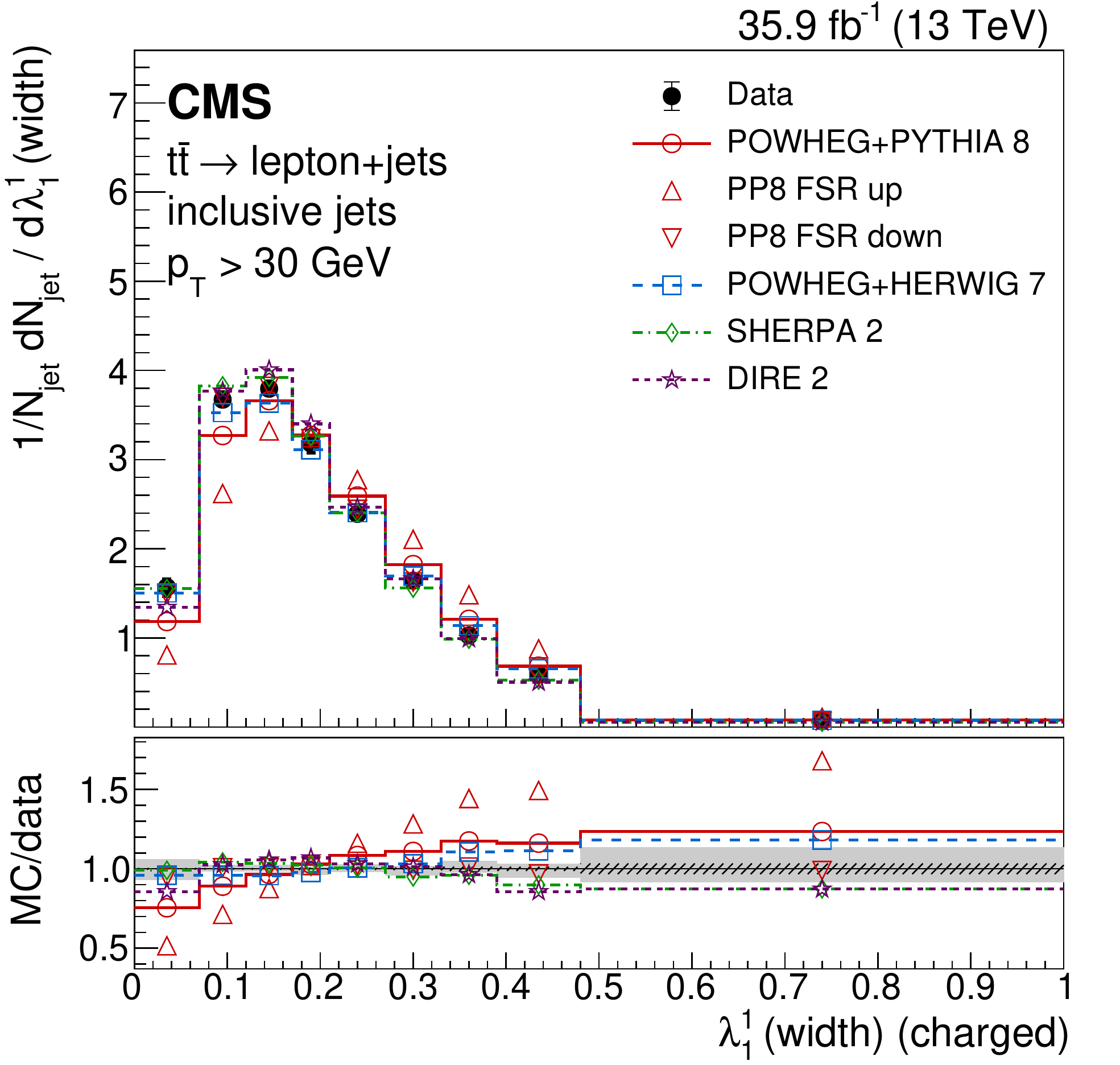}
  \includegraphics[width=0.48\textwidth]{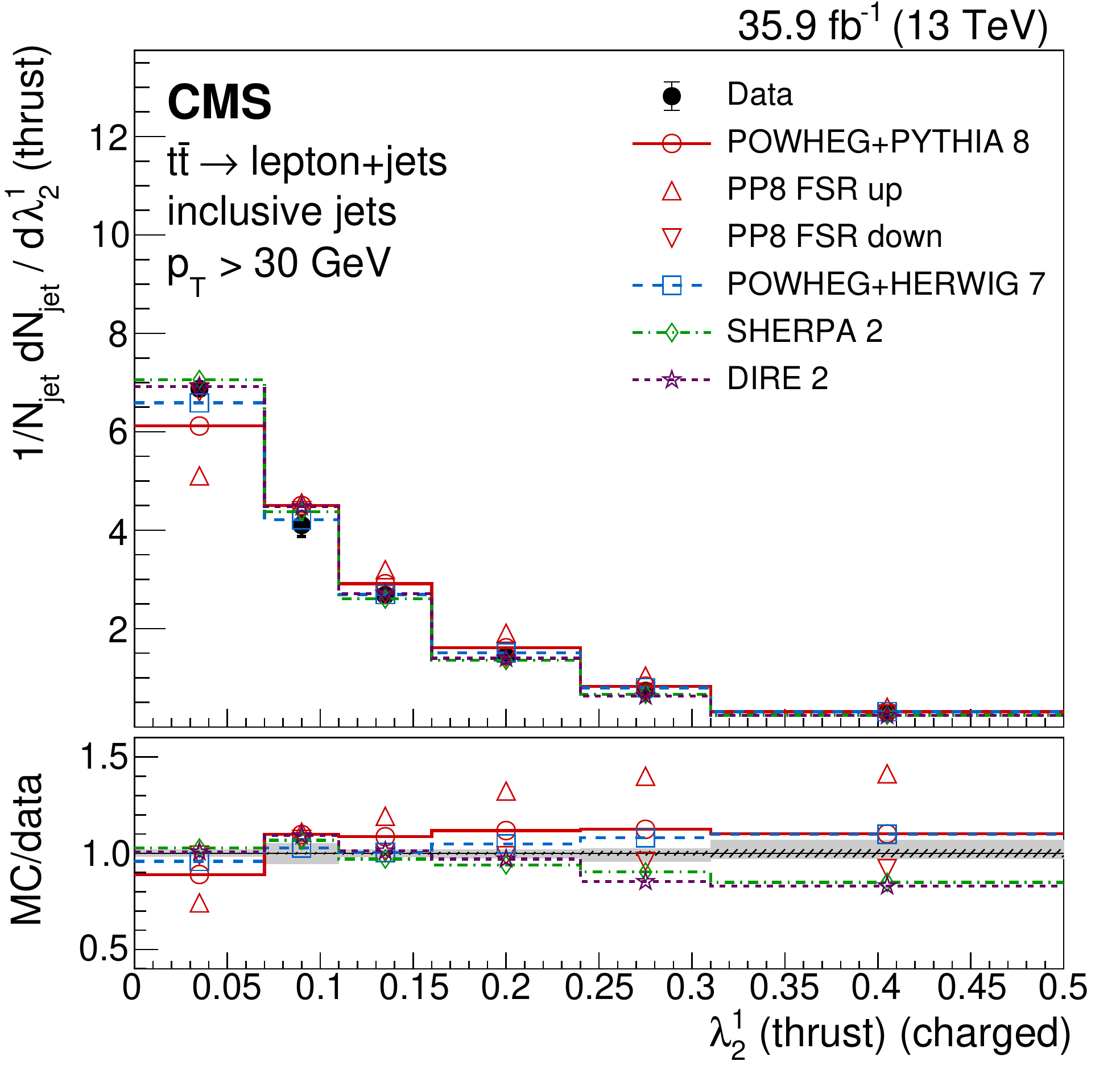}
  \caption{Distributions of the  jet width ($\lambda_1^1$, left) and thrust ($\lambda_2^1$, right), unfolded to the particle level, for inclusive jets reconstructed with charged particles. Data (points) are compared to different MC predictions (upper), and as MC/data ratios (lower). The hatched and shaded bands represent the statistical and total uncertainties, respectively.
  }
  \label{fig:lambda2_charged}
\end{figure*}

The jet width $\lambda_{1}^{1}$, closely related to the jet broadening~\cite{CATANI1992269,RAKOW198163,Ellis:1986ig}, is shown in Fig.~\ref{fig:lambda2_charged} (left).
The data favors the FSR-down variation ($\asfsr = 0.1224$) for \PYTHIA~8.
The jet thrust $\lambda_{2}^{1} \simeq m^2/E^2$~\cite{PhysRevLett.39.1587} is shown in Fig.~\ref{fig:lambda2_charged} (right).
The nominal settings of \POWHEG + \PYTHIA~8 and \POWHEG + \HERWIG~7 provide a good reproduction of the data.

\begin{figure*}
  \centering
  \includegraphics[width=0.48\textwidth]{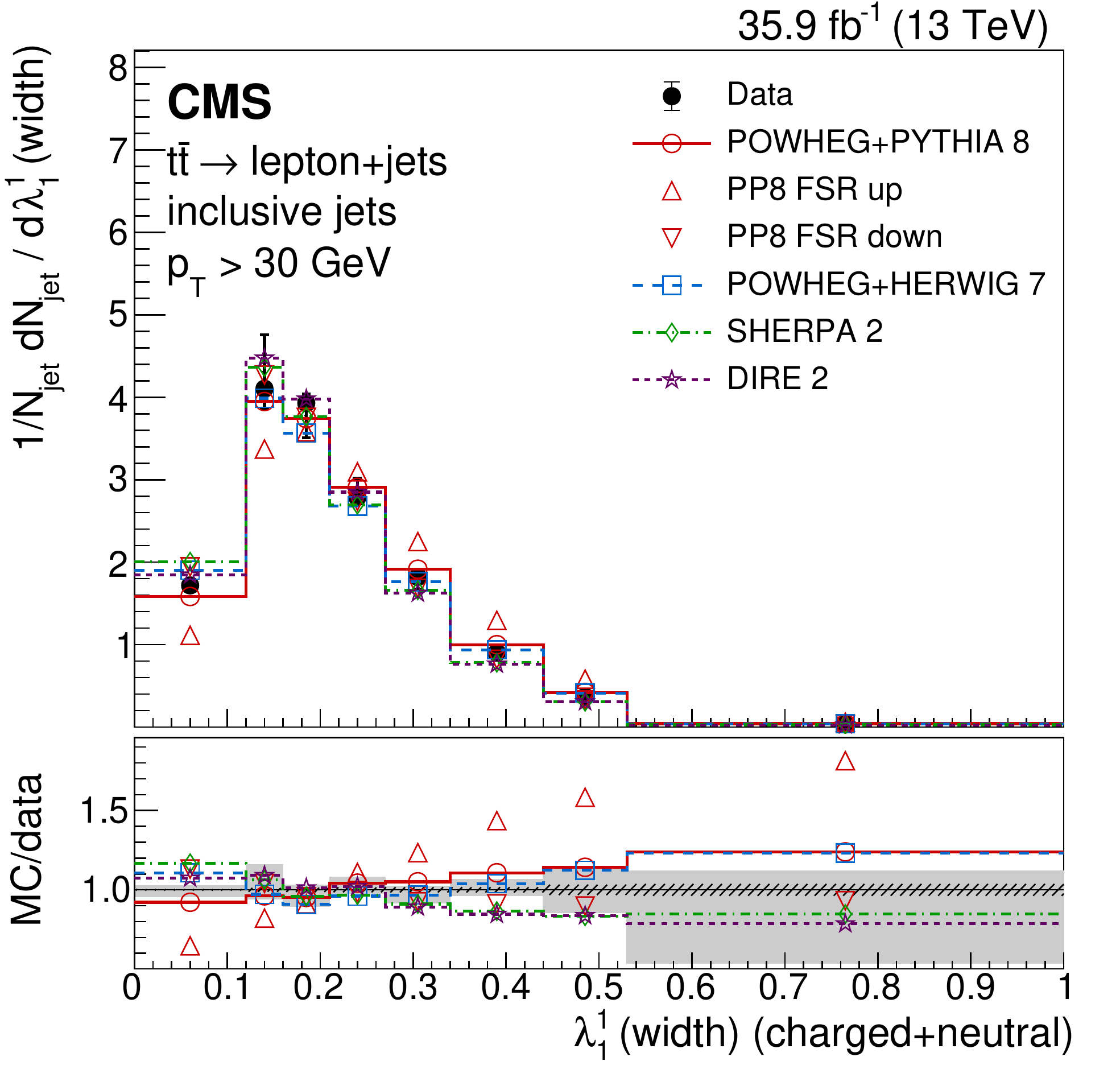}
  \includegraphics[width=0.48\textwidth]{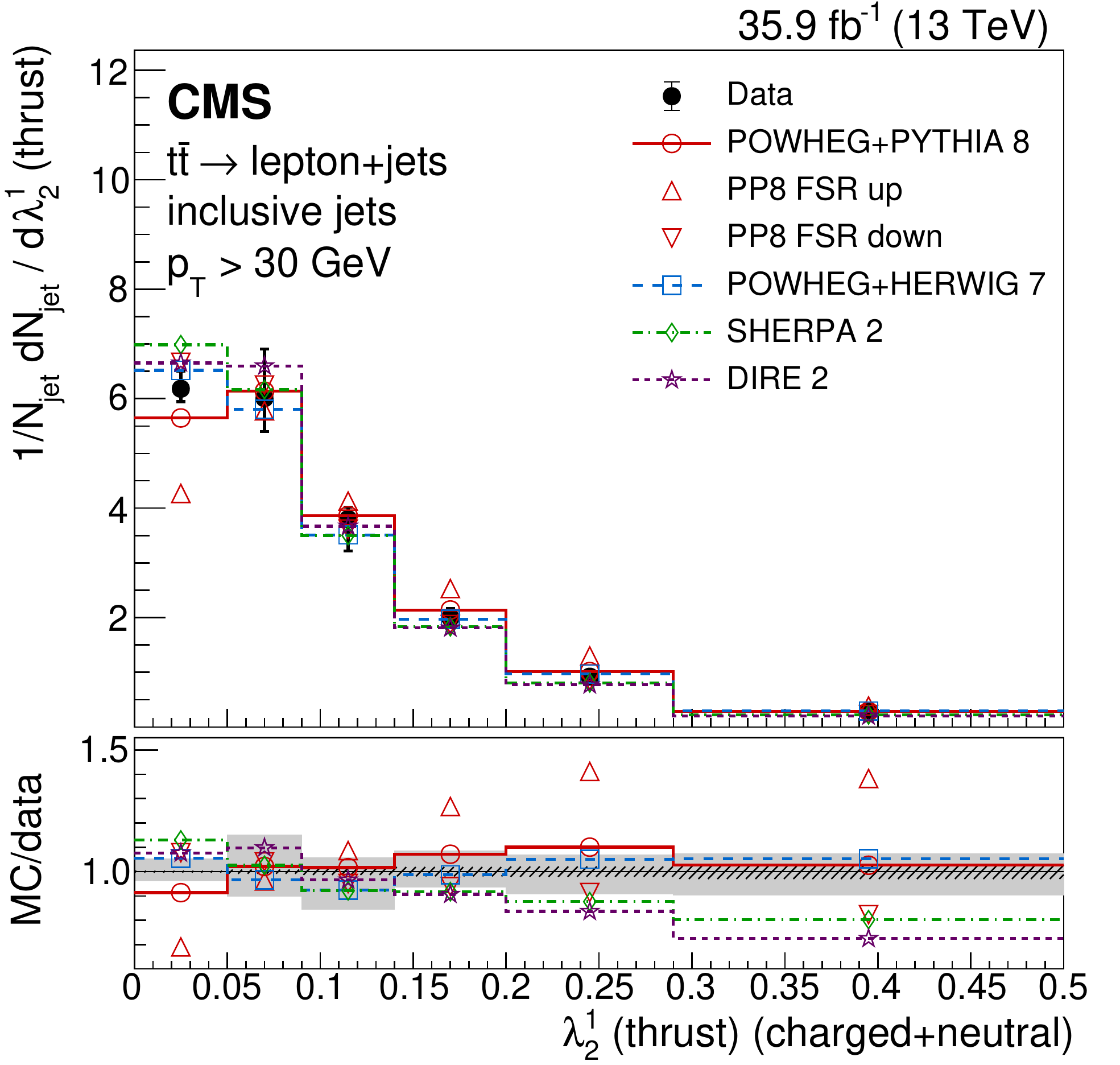}
  \caption{Distributions of the jet width ($\lambda_1^1$, left) and thrust ($\lambda_2^1$, right), unfolded to the particle level, for inclusive jets reconstructed with charged+neutral particles. Data (points) are compared to different MC predictions (upper), and as MC/data ratios (lower). The hatched and shaded bands represent the statistical and total uncertainties, respectively.
}
  \label{fig:lambda_all}
\end{figure*}

For completeness, Fig.~\ref{fig:lambda_all} shows the jet width and thrust distributions obtained using charged+neutral particles
in the jet reconstruction. The comparison to the MC confirms the conclusions extracted with the charged particle-only jet reconstruction
seen in Fig.~\ref{fig:lambda2_charged}.

\subsection{Eccentricity}

\begin{figure}
  \centering
  \includegraphics[width=0.48\textwidth]{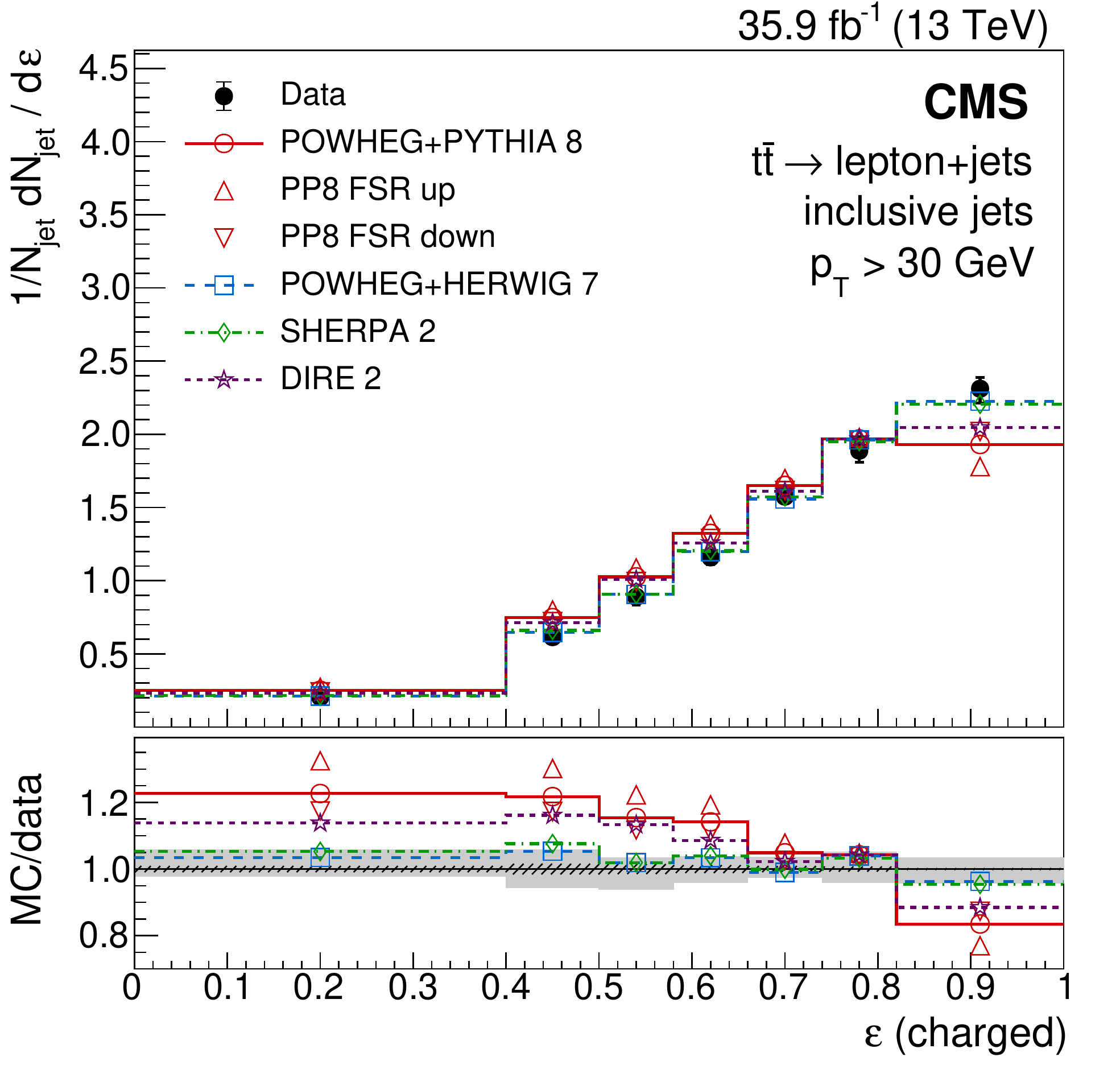}
  \caption{Distribution of the eccentricity $\varepsilon$, unfolded to the particle level, for inclusive jets reconstructed with charged particles. Data (points) are compared to different MC predictions (upper), and as MC/data ratios (lower). The hatched and shaded bands represent the statistical and total uncertainties, respectively.
  }
  \label{fig:ecc_charged}
\end{figure}

The eccentricity~\cite{Chekanov:2010vc} is calculated as $\varepsilon = 1-{v_{\text{min}}}/{v_{\text{max}}}$, where $v$ are the eigenvalues of the energy-weighted covariance matrix $M$ of the $\Delta\eta$ and $\Delta\phi$ distances between the jet constituents $i$ and the WTA jet axis $\hat{n}_{r}$:
\begin{linenomath}
\begin{equation}
M=\sum_{i}E_{i}\,\left(\begin{array}{cc}
\left(\Delta\eta_{i,\hat{n}_{r}}\right)^{2} & \Delta\eta_{i,\hat{n}_{r}}\Delta\phi_{i,\hat{n}_{r}}\\
\Delta\phi_{i,\hat{n}_{r}}\Delta\eta_{i,\hat{n}_{r}} & \left(\Delta\phi_{i,\hat{n}_{r}}\right)^{2}
\end{array}\right).
\end{equation}
\end{linenomath}
A jet perfectly circular in $\eta$--$\phi$ would result in $\varepsilon = 0$, while an elliptical jet gives a value $\varepsilon \to 1$.
At least four particles are required in the jet to calculate the eccentricity.
As shown in Fig.~\ref{fig:ecc_charged}, the \POWHEG + \HERWIG~7 prediction agrees better with the measured distribution than the other MC programs.

\subsection{Soft-drop observables}

The constituents of each individual jet are first reclustered using the Cambridge--Aachen algorithm~\cite{Dokshitzer:1997in,Wobisch:1998wt}.
The ``soft-drop'' (SD) algorithm~\cite{Larkoski:2014wba} is then applied to remove soft, wide-angle radiation from the jet.
Using the angular exponent $\beta = 0$, the soft cutoff threshold $z_{\text{cut}} = 0.1$, and the characteristic parameter $R_{0} = 0.4$, the SD algorithm behaves like the ``modified mass drop tagger''~\cite{Dasgupta:2013ihk}.
At least two particles are required in the jet to perform soft-drop declustering.

\begin{figure*}
  \centering
  \includegraphics[width=0.48\textwidth]{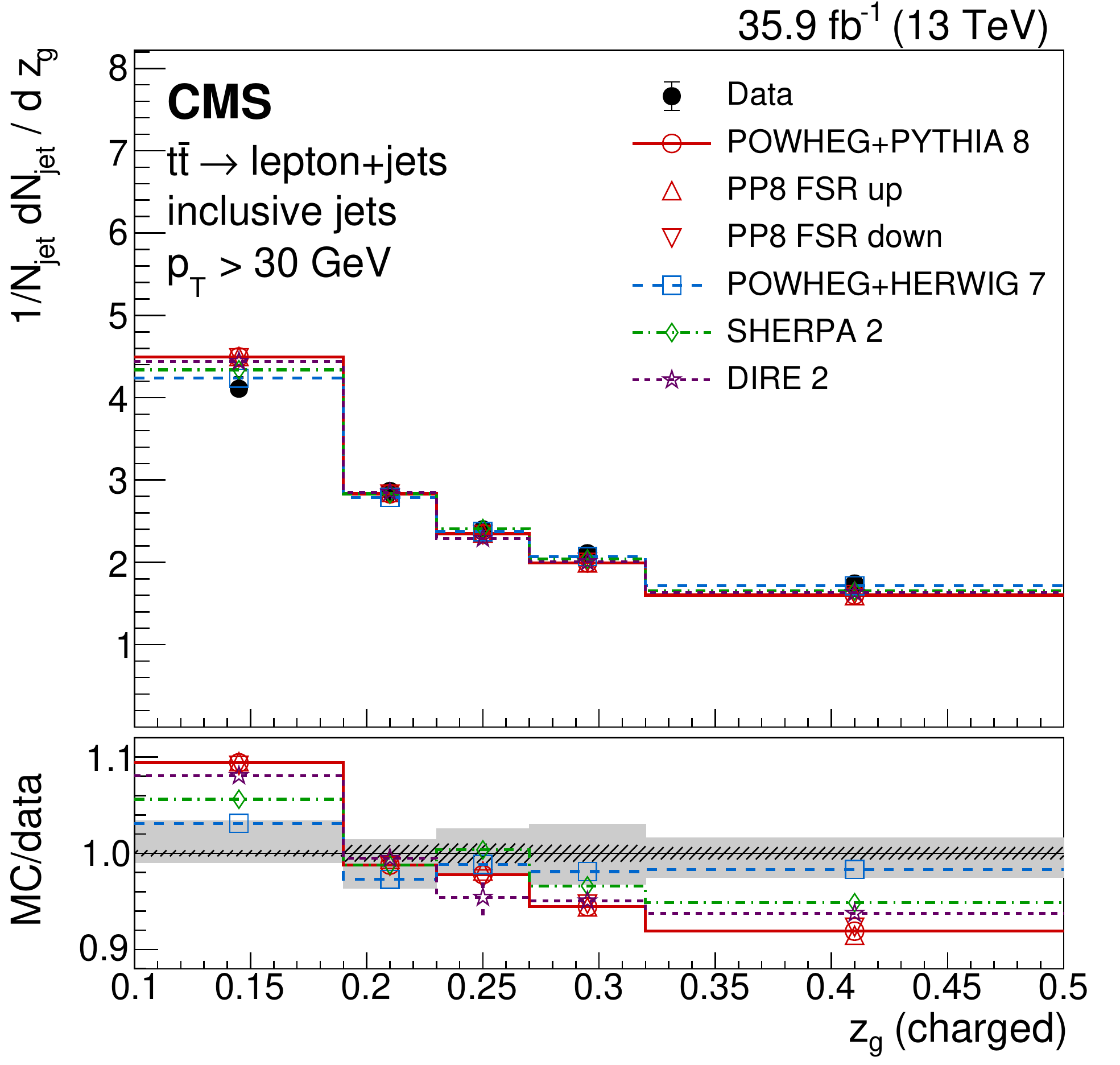}
  \includegraphics[width=0.48\textwidth]{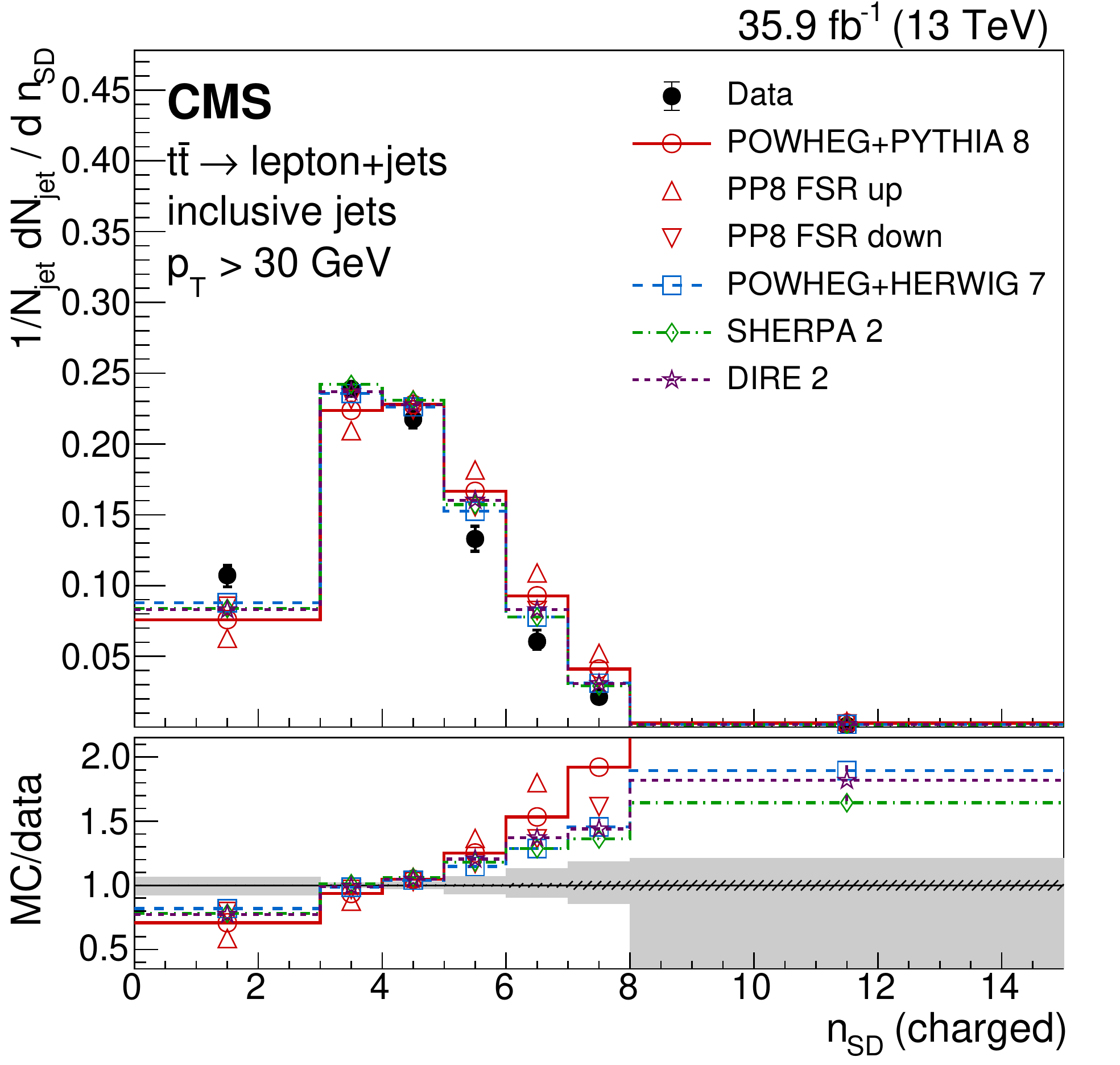}
  \caption{Distributions of the groomed momentum fraction $z_\mathrm{g}$ (left) and the soft-drop multiplicity $n_\mathrm{SD}$ (right), unfolded to the particle level, for inclusive jets reconstructed with charged particles. Data (points) are compared to different MC predictions (upper), and as MC/data ratios (lower). The hatched and shaded bands represent the statistical and total uncertainties, respectively.
  }
  \label{fig:zg_charged}
\end{figure*}

\begin{figure*}
  \centering
  \includegraphics[width=0.48\textwidth]{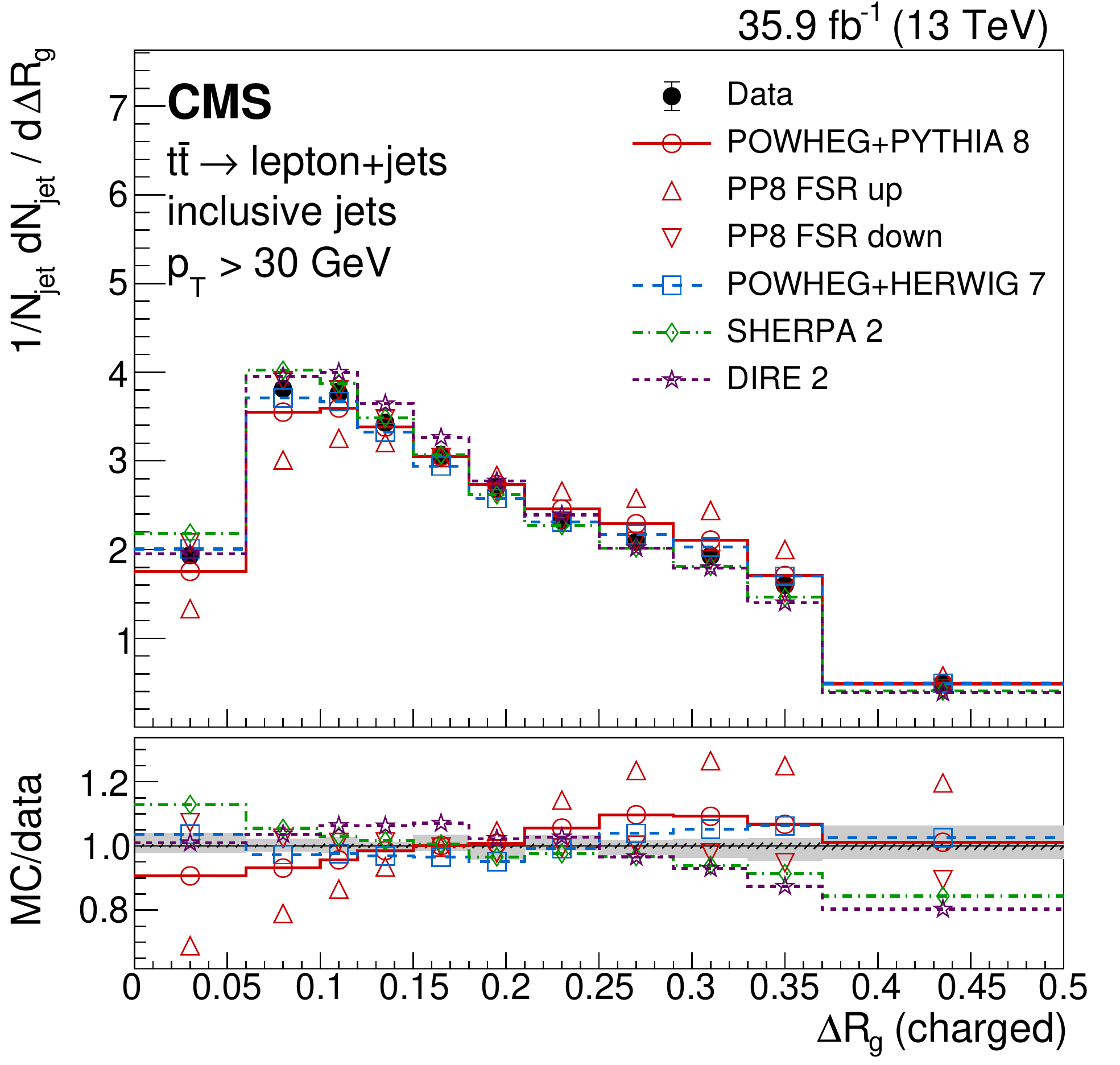}
  \includegraphics[width=0.48\textwidth]{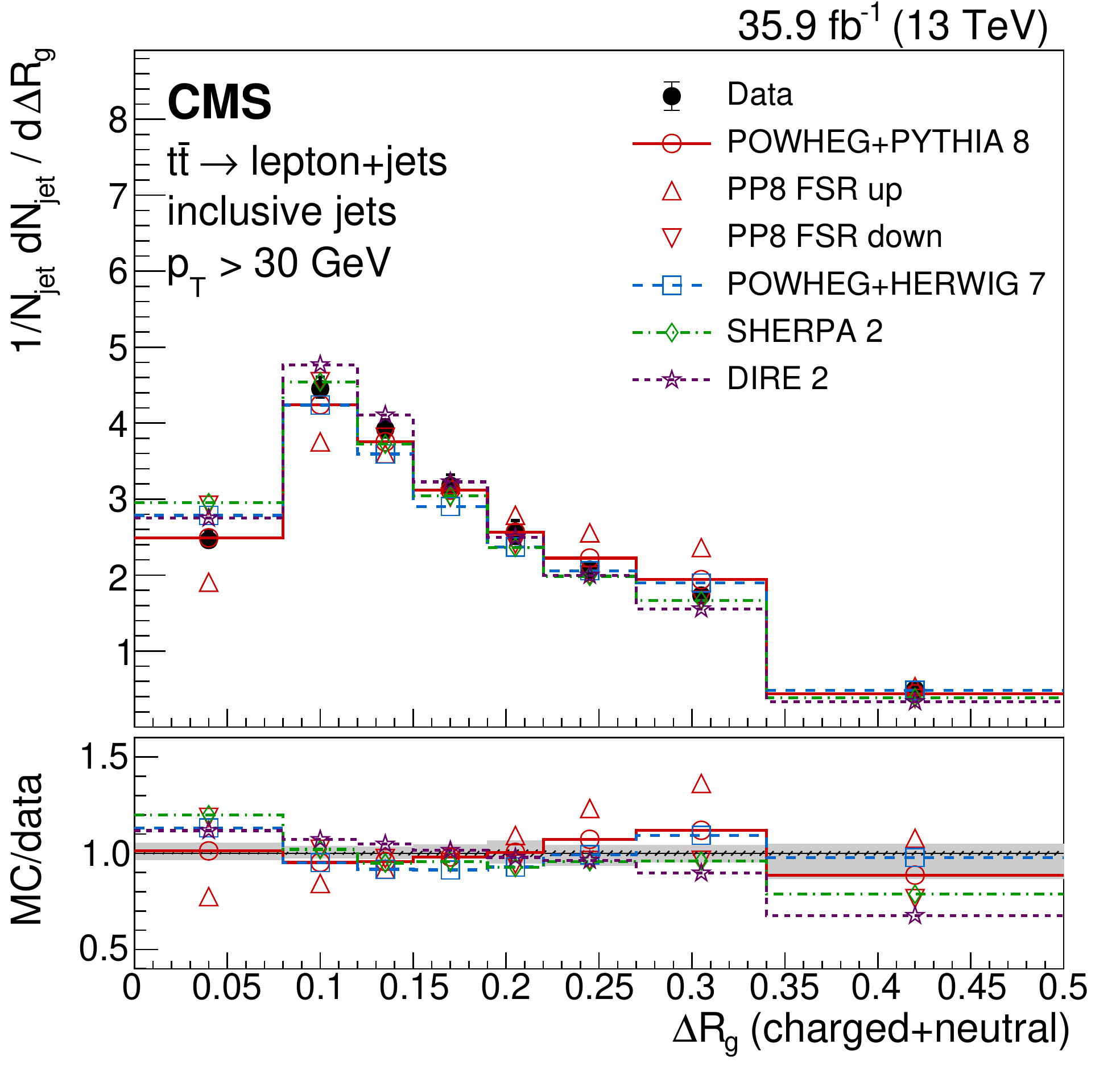}
  \caption{Distributions of the angle between the groomed subjets $\Delta R_\mathrm{g}$, unfolded to the particle level, for inclusive jets reconstructed with charged (left) and charged+neutral particles (right). Data (points) are compared to different MC predictions (upper), and as MC/data ratios (lower). The hatched and shaded bands represent the statistical and total uncertainties, respectively.
  }
  \label{fig:zgdr_charged}
\end{figure*}

After removing soft radiation, the groomed momentum fraction is defined as $z_\mathrm{g}=\pt\left(j_{2}\right)/\pt\left(j_{0}\right)$ of the last declustering iteration $j_{0} \to j_{1}+j_{2}$, where $j_2$ is the softer subjet.
Such a quantity is closely related to the QCD splitting function~\cite{Larkoski:2017bvj}, and does not depend on the value of \alpS.
Recently, uncorrected jet SD measurements were presented for pp collisions at 7~TeV from CMS Open Data~\cite{Larkoski:2017bvj}, as well as in PbPb collisions at 5~TeV~\cite{Sirunyan:2017bsd}.
This analysis presents, for the first time, unfolded $z_\mathrm{g}$ distributions, shown in Fig.~\ref{fig:zg_charged} (left).
The data-model agreement is especially good for the angular-ordered shower of \HERWIG~7.
The angle between two groomed subjets $j_{1}$ and $j_{2}$, $\Delta R_\mathrm{g}$, is related to the jet width but also to the groomed jet area which in turn is relevant for the pileup sensitivity of the algorithm~\cite{Larkoski:2014wba}.
Its measured distribution is shown in Fig.~\ref{fig:zgdr_charged} for both charged and charged+neutral particles, and depends strongly on the amount of FSR.

A soft-drop multiplicity~\cite{Frye:2017yrw}, $n_\mathrm{SD}$, can be defined as the number of branchings in the declustering tree that satisfy the angular cutoff $\Delta R_\mathrm{g} > \theta_{\text{cut}}$ and
\begin{linenomath}
\begin{equation}
z_\mathrm{g} > z_{\text{cut}}\left( \frac{\Delta R_\mathrm{g}}{R_0} \right)^\beta.
\end{equation}
\end{linenomath}
In contrast to the particle multiplicity $N$, $n_\mathrm{SD}$ is IRC-safe for a vast range of parameter settings, \eg, for the one used in this analysis: $z_{\text{cut}} = 0.007$,  $\beta = -1$,  $\theta_{\text{cut}} = 0$.
As shown in Fig.~\ref{fig:zg_charged} (right), the measured data distribution is higher (lower) in the data than in the MC predictions at small (large) $n_\mathrm{SD}$ values, a behavior similar to that observed for the charged multiplicity $\lambda_0^0$ ($N$) in Fig.~\ref{fig:mult_charged}.

\subsection{\texorpdfstring{$\mathcal{N}$}{N}-subjettiness}

The $\mathcal{N}$-subjettiness $\tau_\mathcal{N}$ variable is constructed by first finding exactly $\mathcal{N}$ subjet seed axes using the exclusive \kt clustering algorithm~\cite{Ellis:1993tq} and the WTA recombination scheme.
Starting from these seed axes, a local minimum of $\tau_\mathcal{N}$ is found, where $\tau_\mathcal{N}$ is calculated by summing over all particles belonging to a jet the particle \pt weighted by their radial distance to the nearest of the $\mathcal{N}$ candidate subjet axes:
\begin{linenomath}
\begin{equation}
\tau_\mathcal{N}= \frac{1}{d_0} \sum_i \pt{}_{,i} \min \left\{ \left(\Delta R_{1,i}\right), \left(\Delta R_{2,i}\right), \dots, \left(\Delta R_{\mathcal{N},i}\right) \right\}\;,
\end{equation}
\end{linenomath}
with a normalization factor
\begin{linenomath}
\begin{equation}
d_0 = \sum_i \pt{}_{,i} \left( R_0 \right),
\end{equation}
\end{linenomath}
assuming the original jet distance parameter $R_0 = 0.4$.

\begin{figure*}
  \centering
  \includegraphics[width=0.48\textwidth]{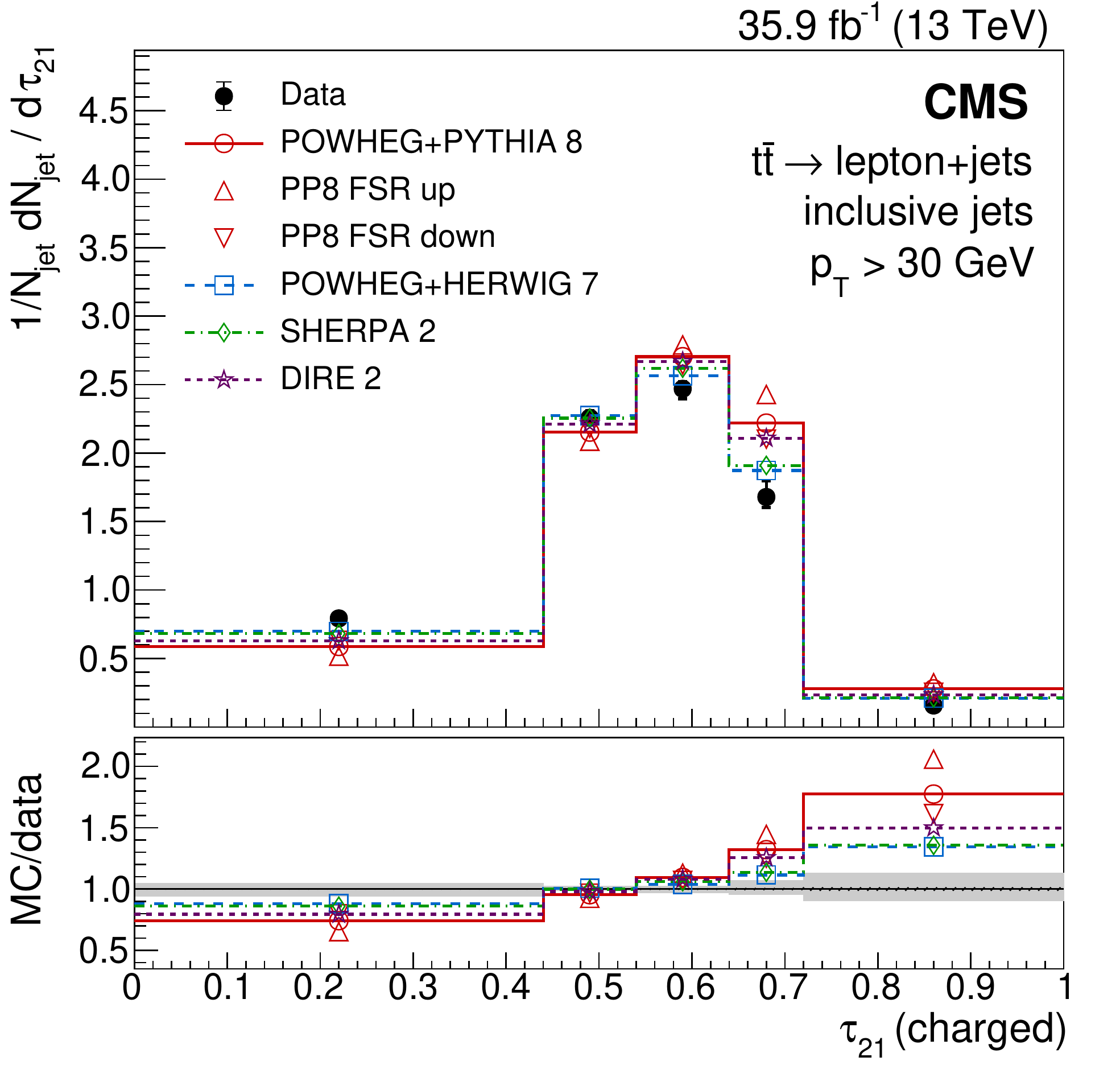}
  \includegraphics[width=0.48\textwidth]{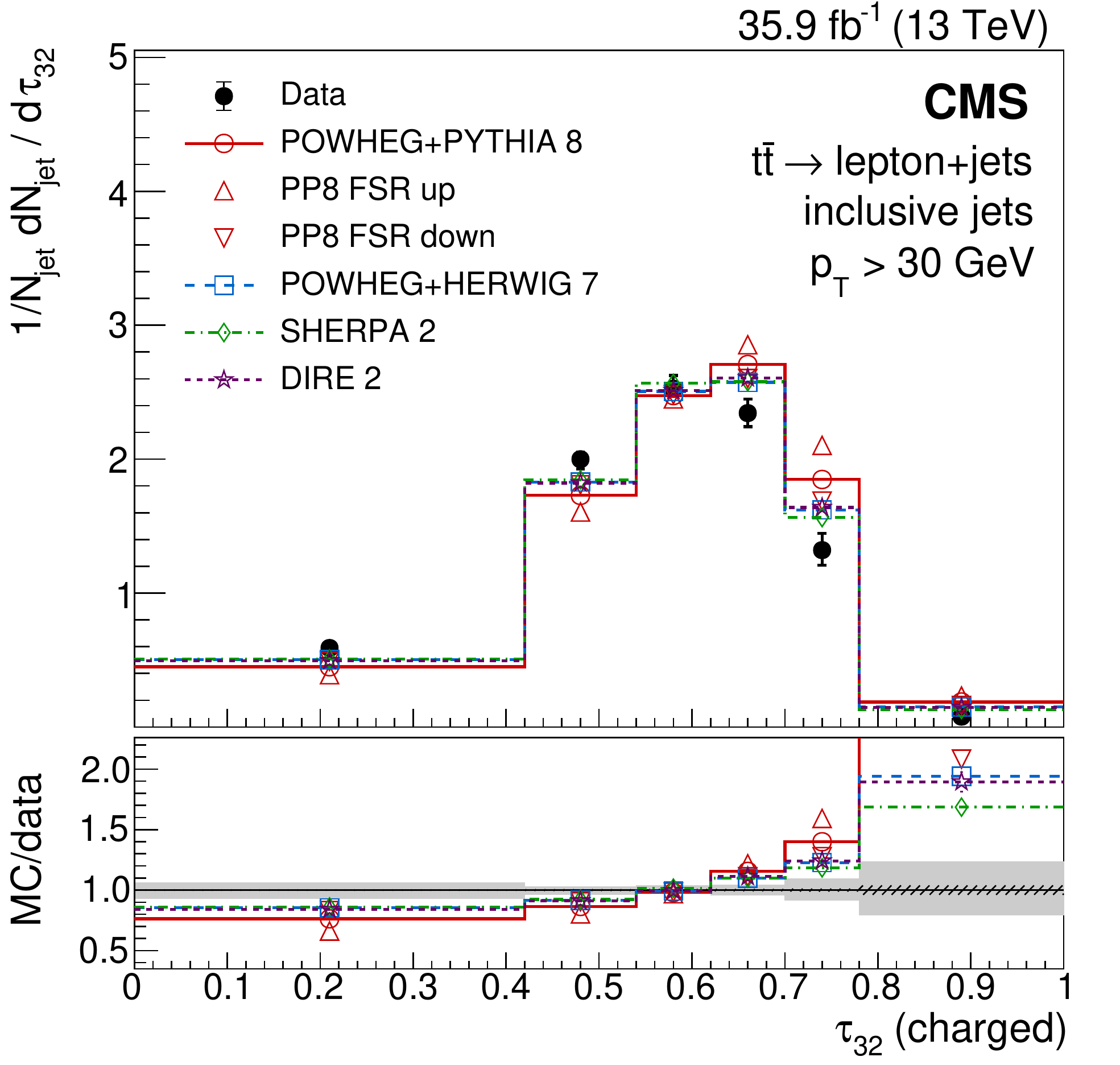}
  \caption{Distributions of the $\mathcal{N}$-subjettiness ratios $\tau_{21}$ (left) and $\tau_{32}$ (right), unfolded to the particle level, for inclusive jets reconstructed with charged particles. Data (points) are compared to different MC predictions (upper), and as MC/data ratios (lower). The hatched and shaded bands represent the statistical and total uncertainties, respectively.
  }
  \label{fig:tau_charged}
\end{figure*}

\begin{figure*}
  \centering
  \includegraphics[width=0.48\textwidth]{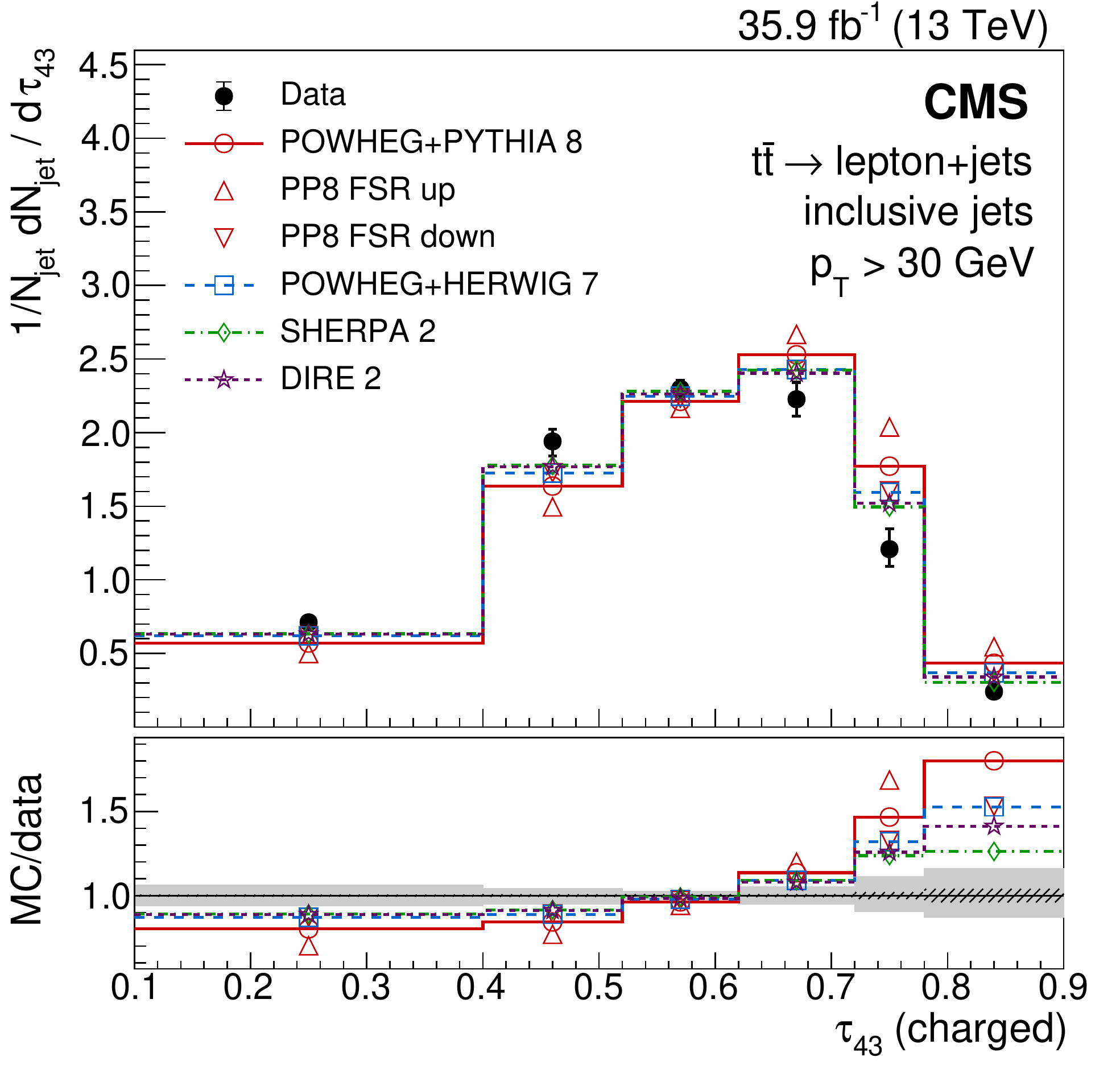}
  \includegraphics[width=0.48\textwidth]{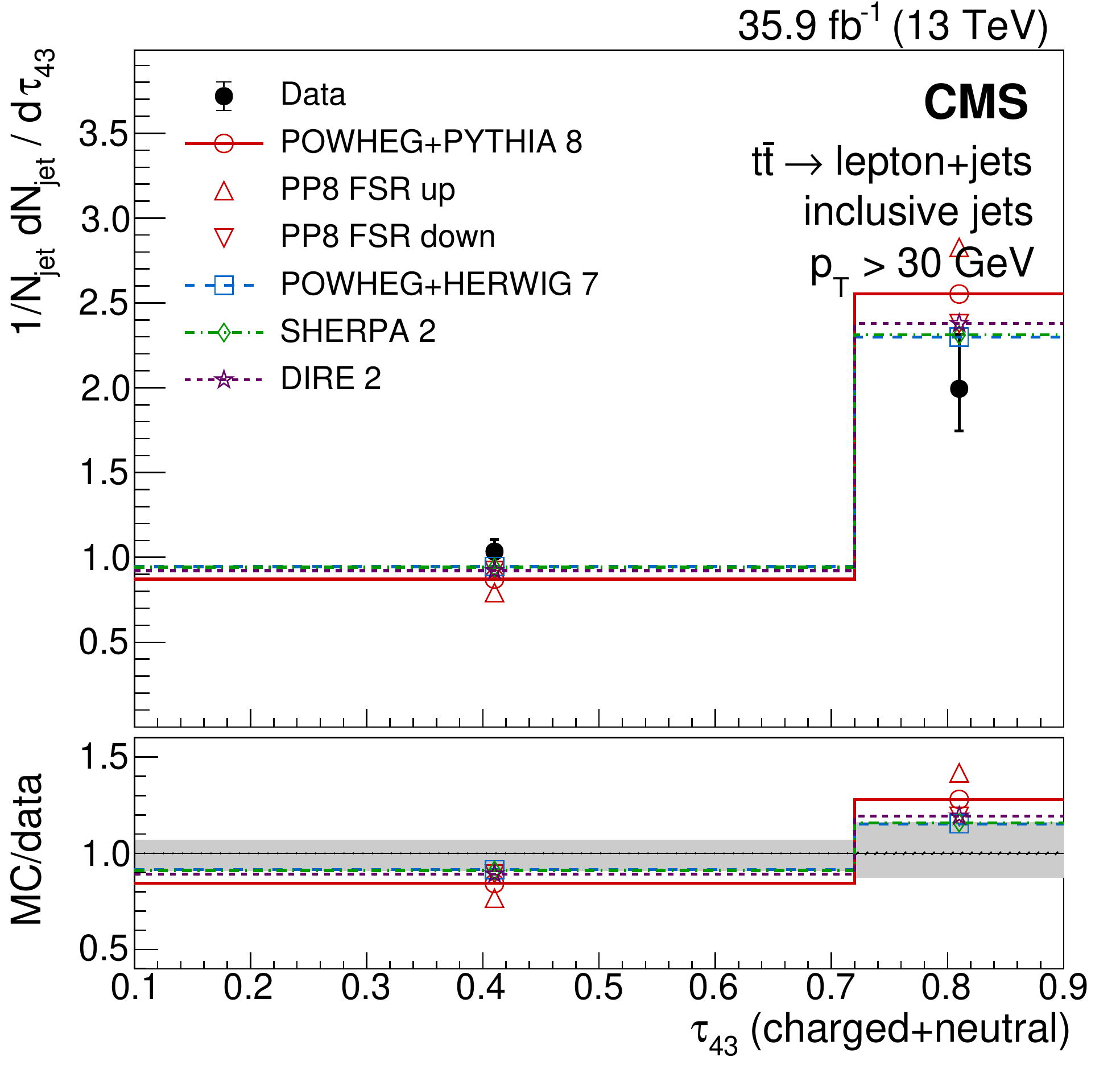}
  \caption{Distributions of the $\mathcal{N}$-subjettiness ratio $\tau_{43}$, unfolded to the particle level, for inclusive jets reconstructed with charged (left) and charged+neutral particles (right). Data (points) are compared to different MC predictions (upper), and as MC/data ratios (lower). The hatched and shaded bands represent the statistical and total uncertainties, respectively.
  }
  \label{fig:tau_charged2}
\end{figure*}

The $\mathcal{N}$-subjettiness ratios $\tau_\mathcal{NM}= \tau_\mathcal{N}/\tau_\mathcal{M}$, defined in~\cite{Thaler:2010tr,Thaler:2011gf}, were shown to be especially useful for distinguishing jets with $\mathcal{N}$ or $\mathcal{M}$ subjets.
In this analysis, $\tau_{21}$, $\tau_{32}$, and $\tau_{43}$ are measured, which are frequently used in the identification of heavy Lorentz-boosted objects.
At least $\mathcal{N}+1$ particles are required in the jet to calculate these observables.
As shown in Figs.~\ref{fig:tau_charged} and \ref{fig:tau_charged2}, the measured $\tau_\mathcal{NM}$ distributions are consistently shifted to lower values than those predicted by the MC programs.
While the expectation from boosted object studies is that $\mathcal{N}$-prong ($\mathcal{M}$-prong) jets acquire a lower (higher) value of $\tau_\mathcal{NM}$, the behavior of $\tau_\mathcal{NM}$ in a resolved topology seems to be mainly driven by the particle multiplicity.

\subsection{Energy correlation functions}

\begin{figure*}
  \centering
  \includegraphics[width=0.48\textwidth]{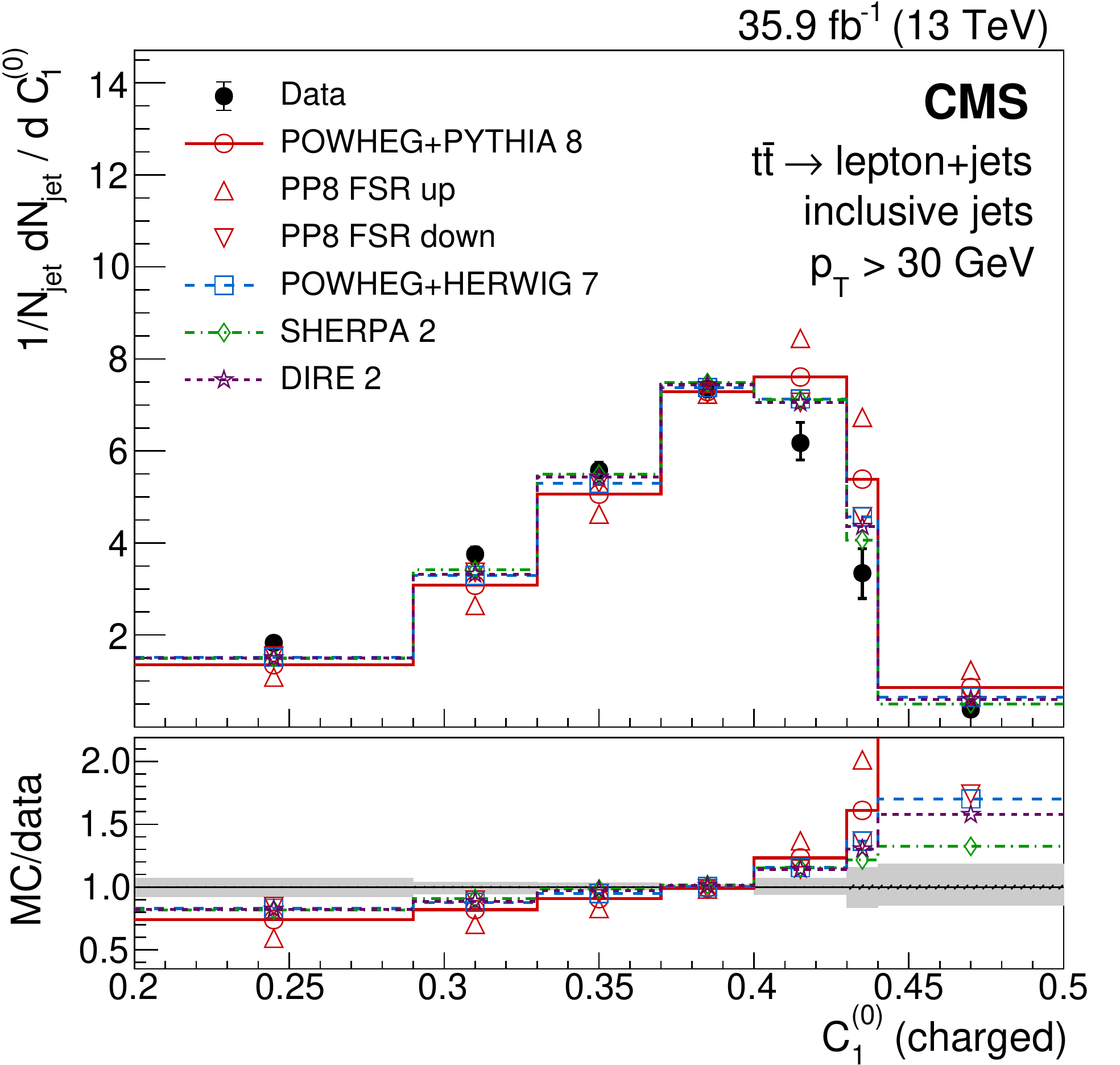}
  \includegraphics[width=0.48\textwidth]{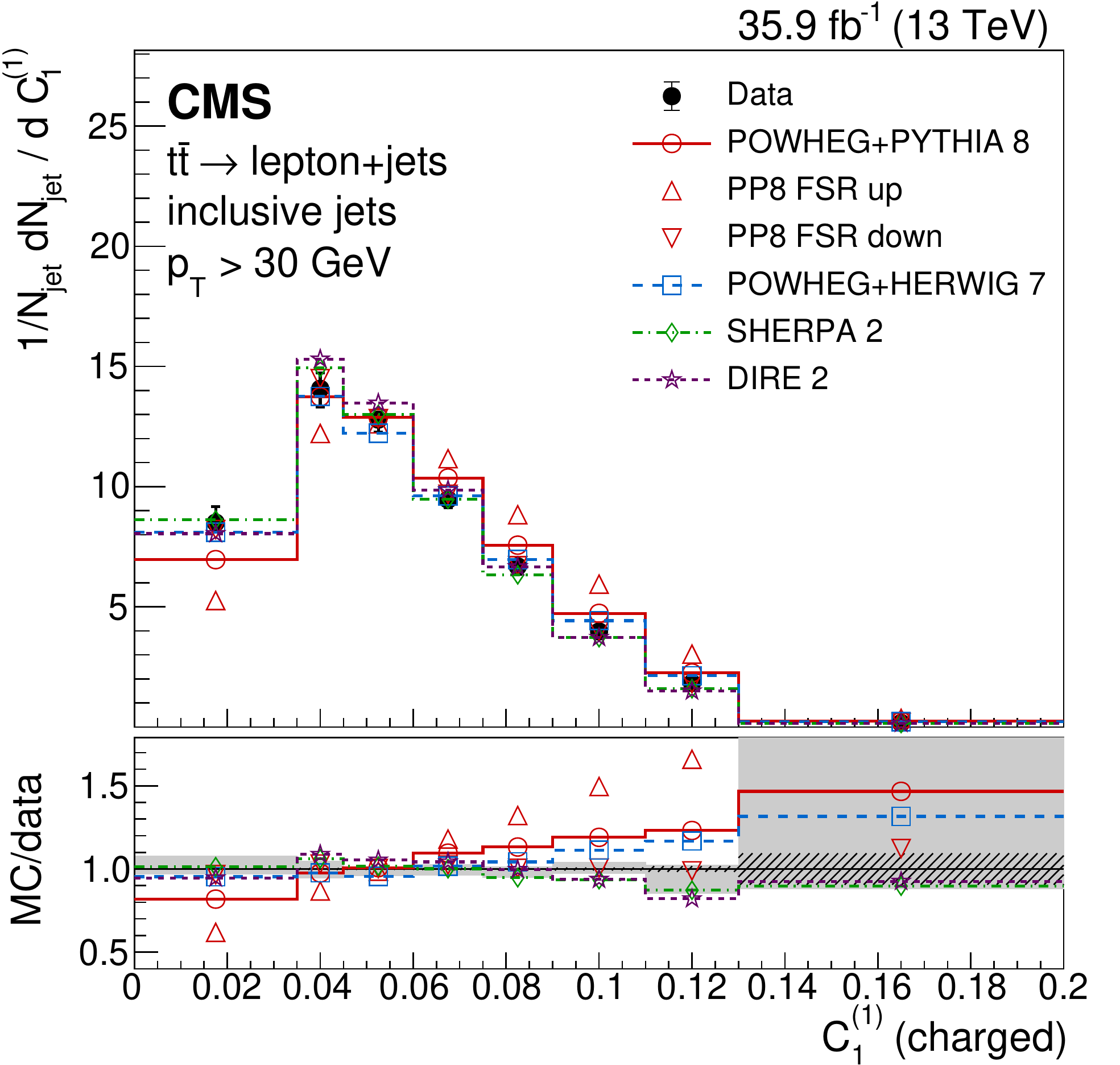}
  \includegraphics[width=0.48\textwidth]{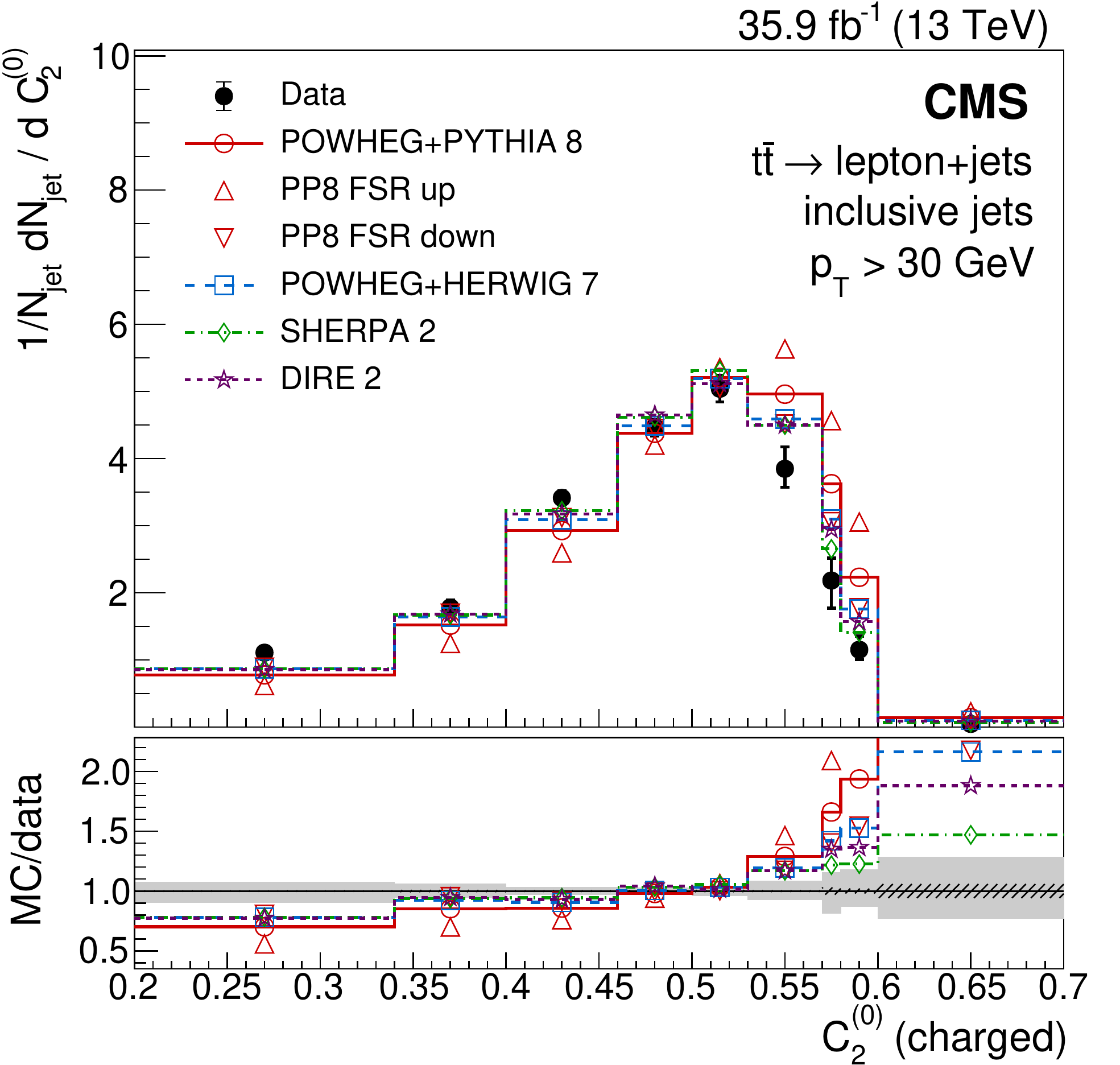}
  \includegraphics[width=0.48\textwidth]{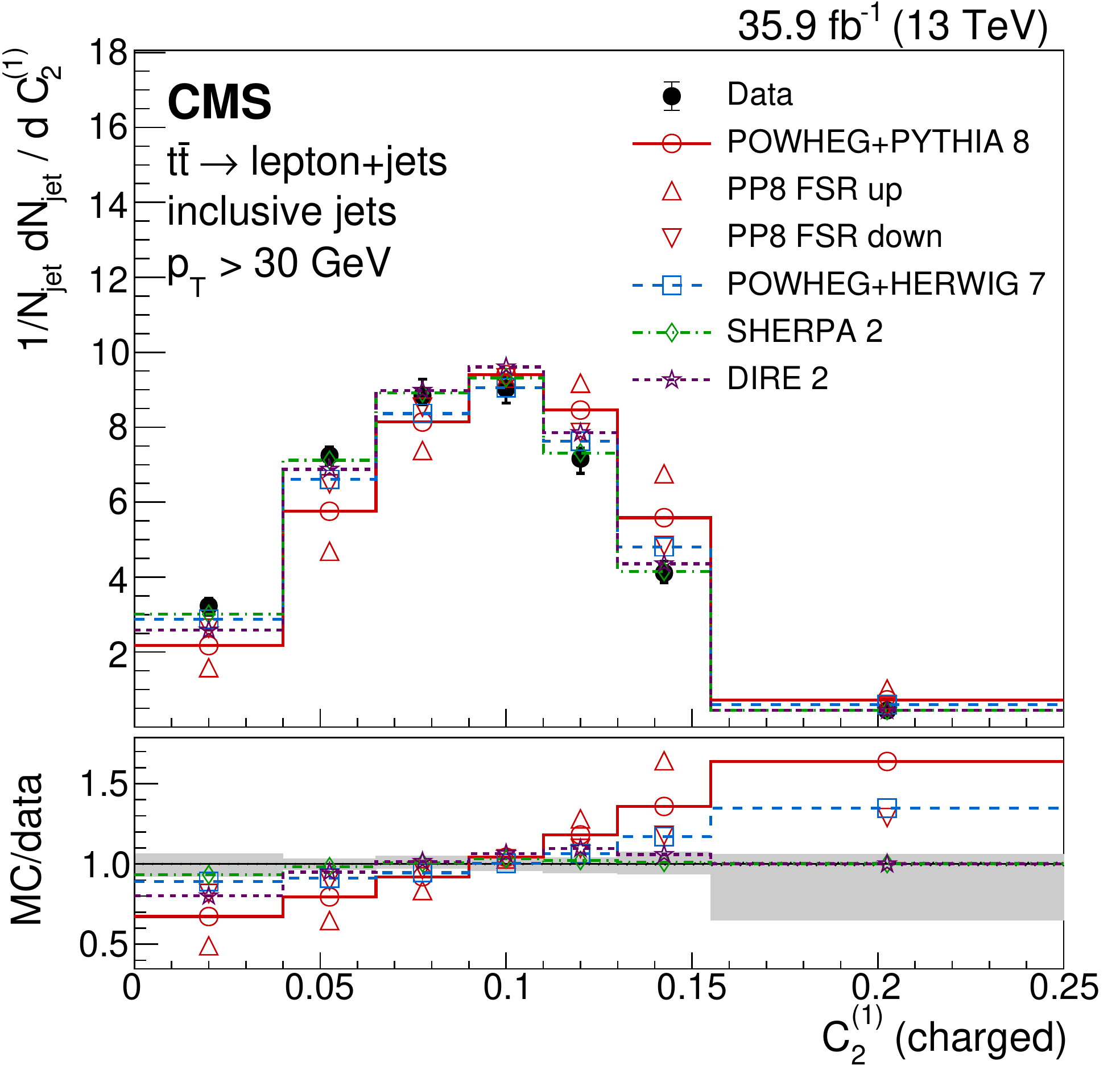}
  \caption{Distributions of energy correlation ratios $C_{1}^{\left(0\right)}$ (upper left), $C_{1}^{\left(1\right)}$ (upper right), $C_{2}^{\left(0\right)}$ (lower left) and $C_{2}^{\left(1\right)}$ (lower right), unfolded to the particle level, for inclusive jets reconstructed with charged particles. Data (points) are compared to different MC predictions (upper), and as MC/data ratios (lower). The hatched and shaded bands represent the statistical and total uncertainties, respectively.
  }
  \label{fig:c12_charged}
\end{figure*}

\begin{figure*}
  \centering
  \includegraphics[width=0.48\textwidth]{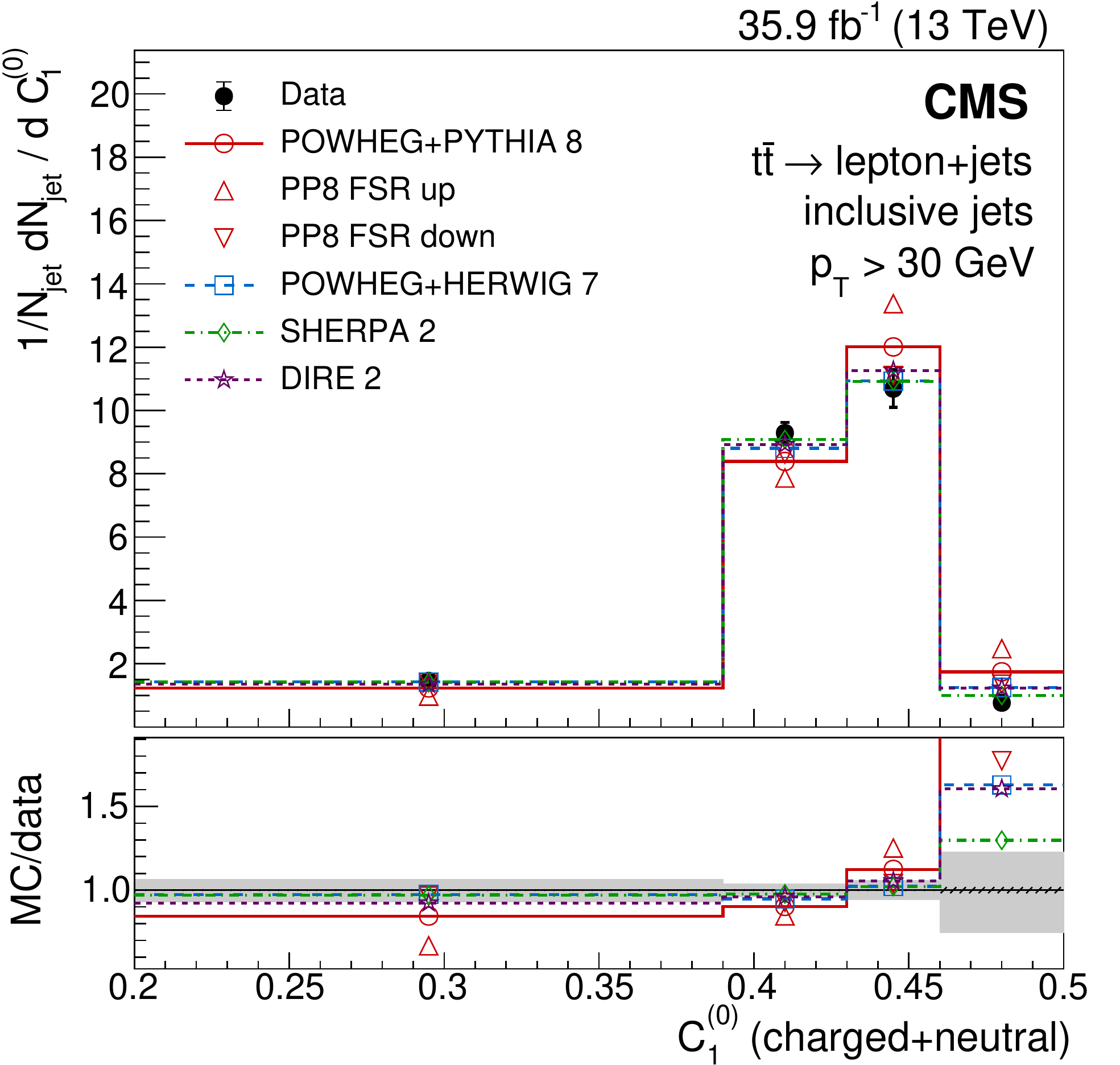}
  \includegraphics[width=0.48\textwidth]{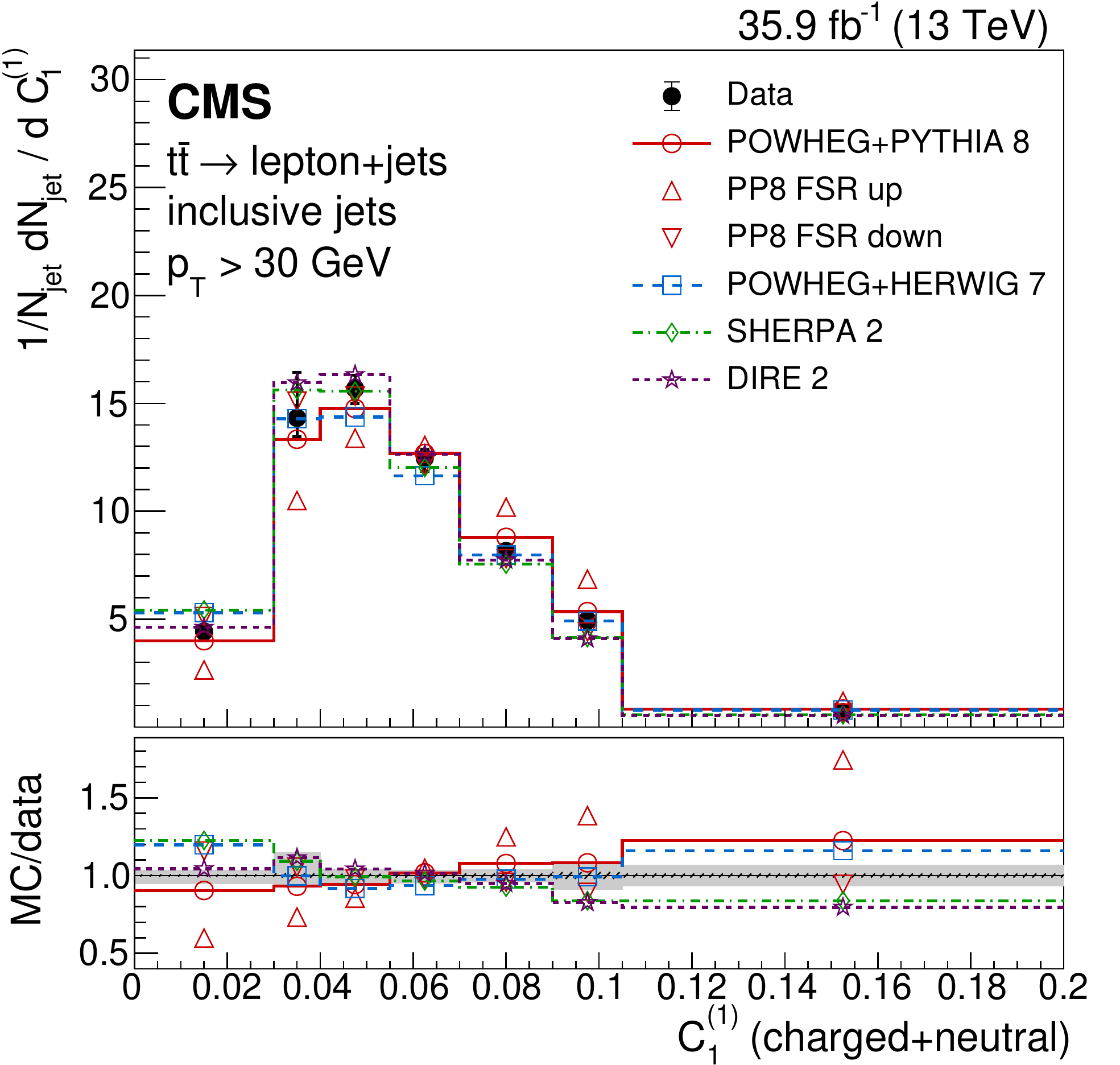}
  \caption{Distributions of energy correlation ratios $C_{1}^{\left(0\right)}$ (left) and $C_{1}^{\left(1\right)}$ (right), unfolded to the particle level, for inclusive jets reconstructed with charged+neutral particles. Data (points) are compared to different MC predictions (upper), and as MC/data ratios (lower). The hatched and shaded bands represent the statistical and total uncertainties, respectively.
  }
  \label{fig:c1_all}
\end{figure*}

\begin{figure*}
  \centering
  \includegraphics[width=0.48\textwidth]{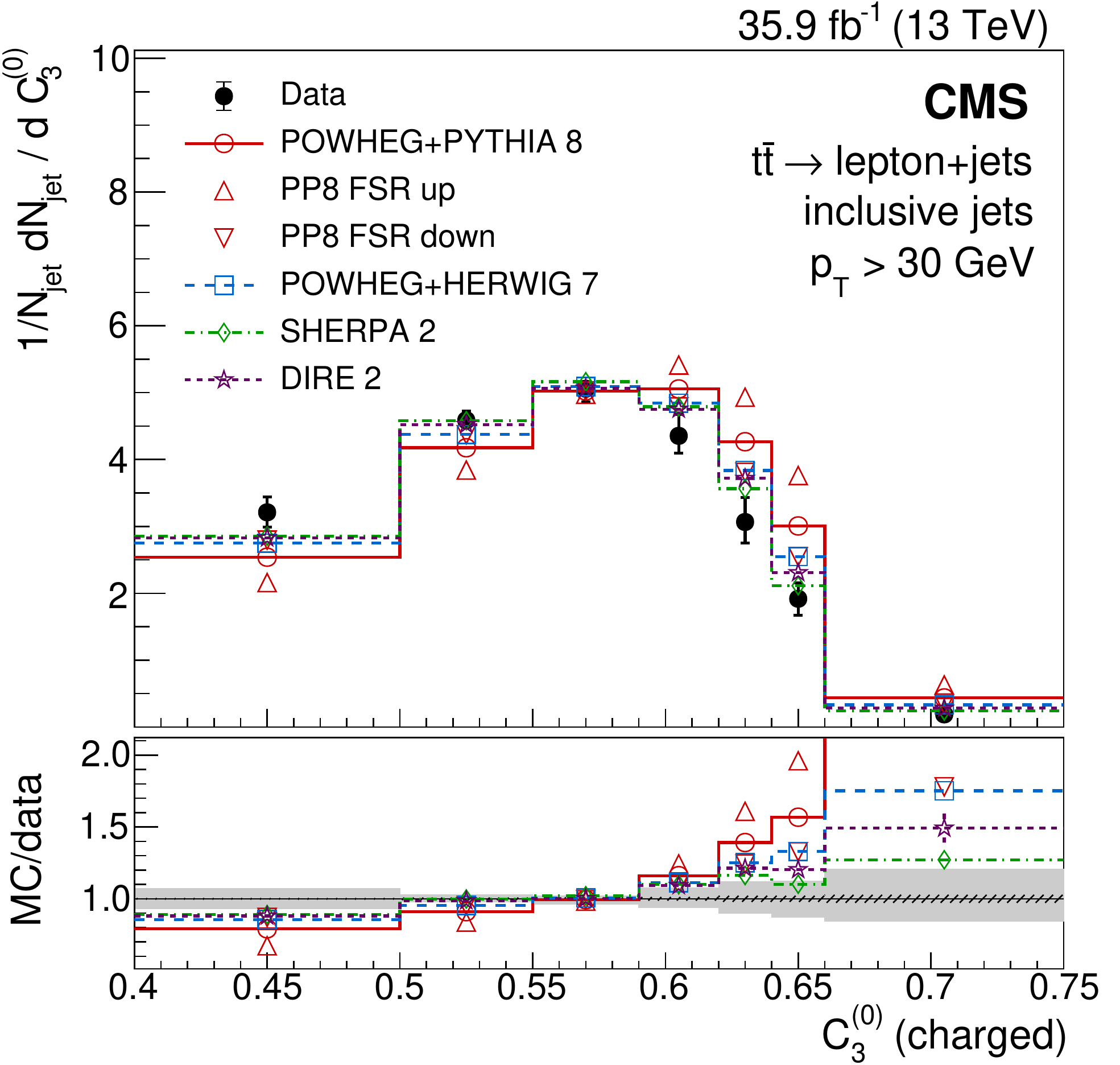}
  \includegraphics[width=0.48\textwidth]{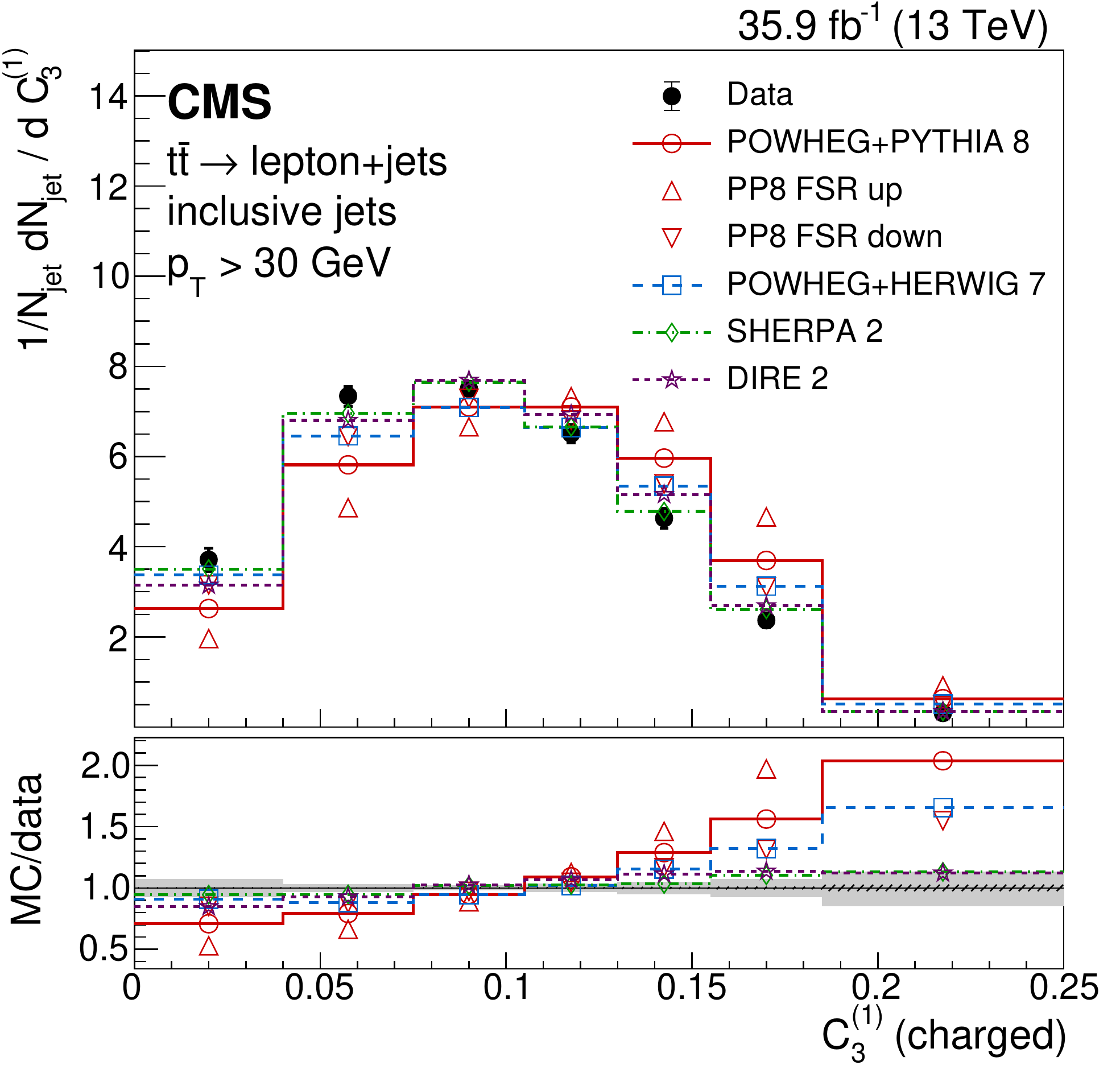}
  \caption{Distributions of energy correlation ratios $C_{3}^{\left(0\right)}$ (left) and $C_{3}^{\left(1\right)}$ (right), unfolded to the particle level, for inclusive jets reconstructed with charged particles. Data (points) are compared to different MC predictions (upper), and as MC/data ratios (lower). The hatched and shaded bands represent the statistical and total uncertainties, respectively.
  }
  \label{fig:c3_charged}
\end{figure*}

The $\mathcal{N}$-point energy correlation double ratios $C_\mathcal{N}^{\left(\beta\right)}$~\cite{Larkoski:2013eya} are defined as
\begin{linenomath}
\begin{equation}
C_\mathcal{N}^{\left(\beta\right)} = \frac{\mathrm{ECF}\left(\mathcal{N}+1,\beta\right)\mathrm{ECF}\left(\mathcal{N}-1,\beta\right)}{\mathrm{ECF}\left(\mathcal{N},\beta\right)^2},
\end{equation}
\end{linenomath}
where
\begin{linenomath}
\ifthenelse{\boolean{cms@external}}
{
\begin{multline*}
\mathrm{ECF}\left(\mathcal{N},\beta\right) =\\ \sum_{i_1 < i_2 < \ldots <  i_\mathcal{N} \in j}  \left( \prod_{a = 1}^\mathcal{N} {p_T}_{i_a} \right) \left(  \prod_{b=1}^{\mathcal{N}-1} \prod_{c = b+1}^\mathcal{N} \Delta R_{i_b i_c} \right)^\beta
\end{multline*}
}
{
\begin{equation*}
\mathrm{ECF}\left(\mathcal{N},\beta\right) = \sum_{i_1 < i_2 < \ldots <  i_\mathcal{N} \in j}  \left( \prod_{a = 1}^\mathcal{N} {p_T}_{i_a} \right)  \left(  \prod_{b=1}^{\mathcal{N}-1} \prod_{c = b+1}^\mathcal{N} \Delta R_{i_b i_c} \right)^\beta
\end{equation*}
}
\end{linenomath}
and the sum runs over the constituents $i$ of the jet $j$ with their \pt product being multiplied with the pairwise distances $\Delta R_{i_b i_c}$ in $\eta$--$\phi$ space.
Each $C_\mathcal{N}$ is sensitive to the $\left(\mathcal{N}-1\right)$-prong substructure of the jet, while the angular exponent $\beta$ adjusts the sensitivity to near-collinear splittings.
At least $\mathcal{N}+1$ particles are required in the jet to calculate these observables.
In this analysis, parameter values $\mathcal{N}=\left\{1,2,3\right\}$ and $\beta=\left\{0,0.2,0.5,1,2\right\}$ are investigated.
The distributions for each $\mathcal{N}$ and $\beta=\left\{0,1\right\}$ are shown in Figs.~\ref{fig:c12_charged}--\ref{fig:c3_charged}.
For $\beta > 0$ the observable is IRC-safe and the data are better described by the MC generators than for $\beta = 0$.
Many observables of this family show significant differences between the jet flavors, as shown later in Fig.~\ref{fig:flavors} (bottom, right).

\begin{figure*}
  \centering
  \includegraphics[width=0.48\textwidth]{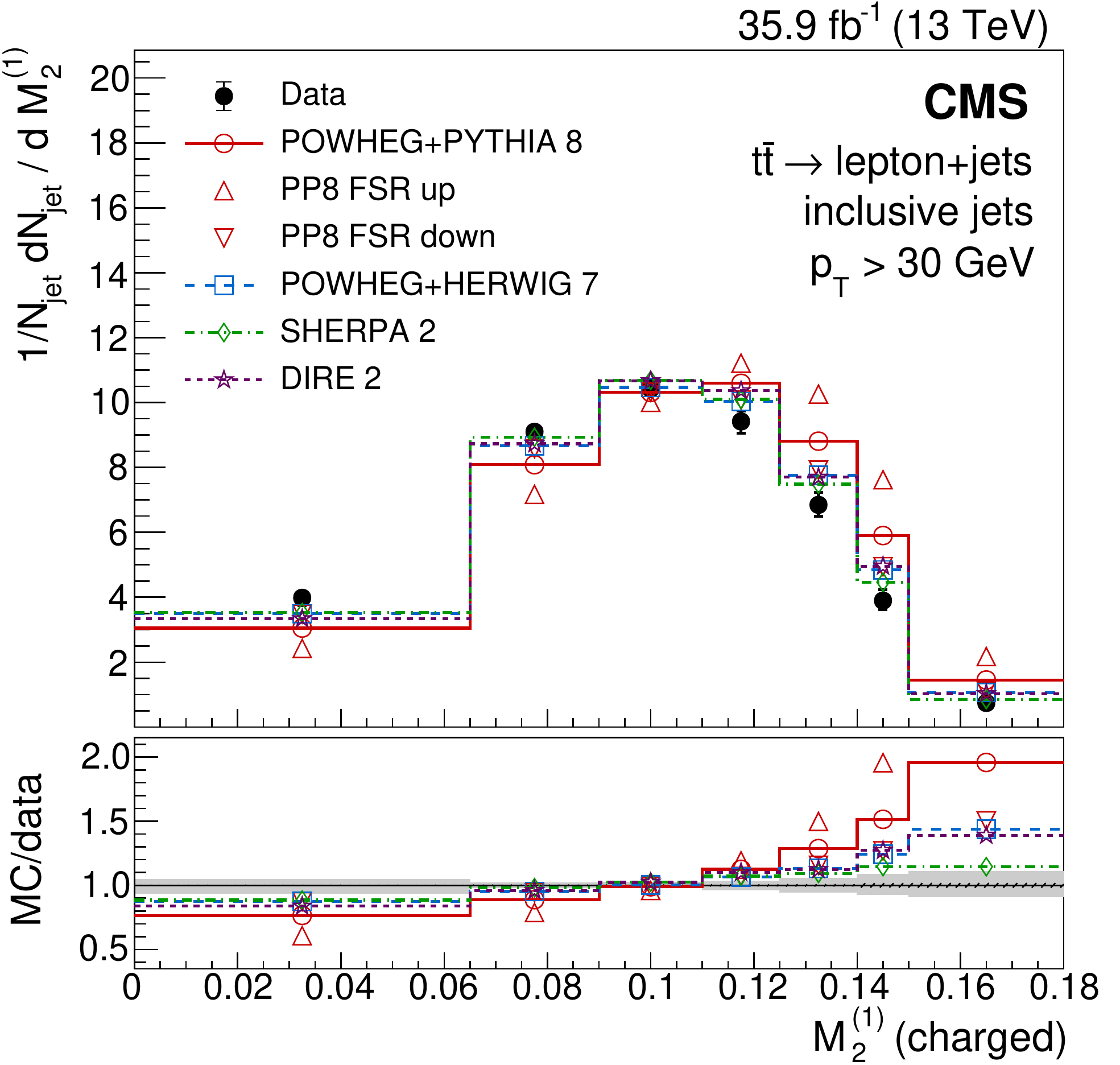}
  \includegraphics[width=0.48\textwidth]{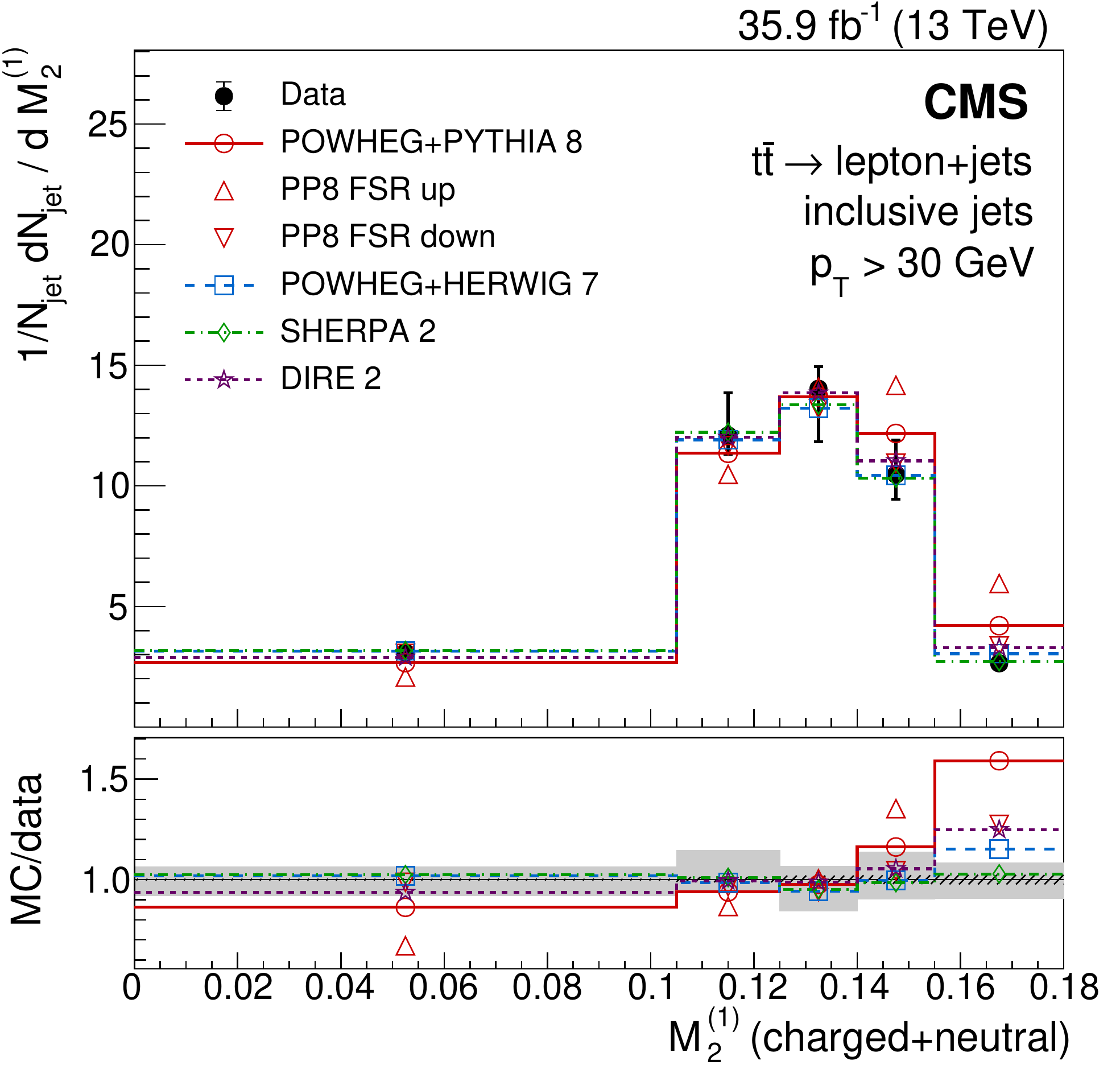}
  \caption{Distributions of the energy correlation ratio $\mathrm{M}_{2}^{\left(1\right)}$, unfolded to the particle level, for inclusive jets reconstructed with charged (left) or charged+neutral particles (right). Data (points) are compared to different MC predictions (upper), and as MC/data ratios (lower). The hatched and shaded bands represent the statistical and total uncertainties, respectively.
  }
  \label{fig:m_i}
\end{figure*}

\begin{figure*}
  \centering
  \includegraphics[width=0.48\textwidth]{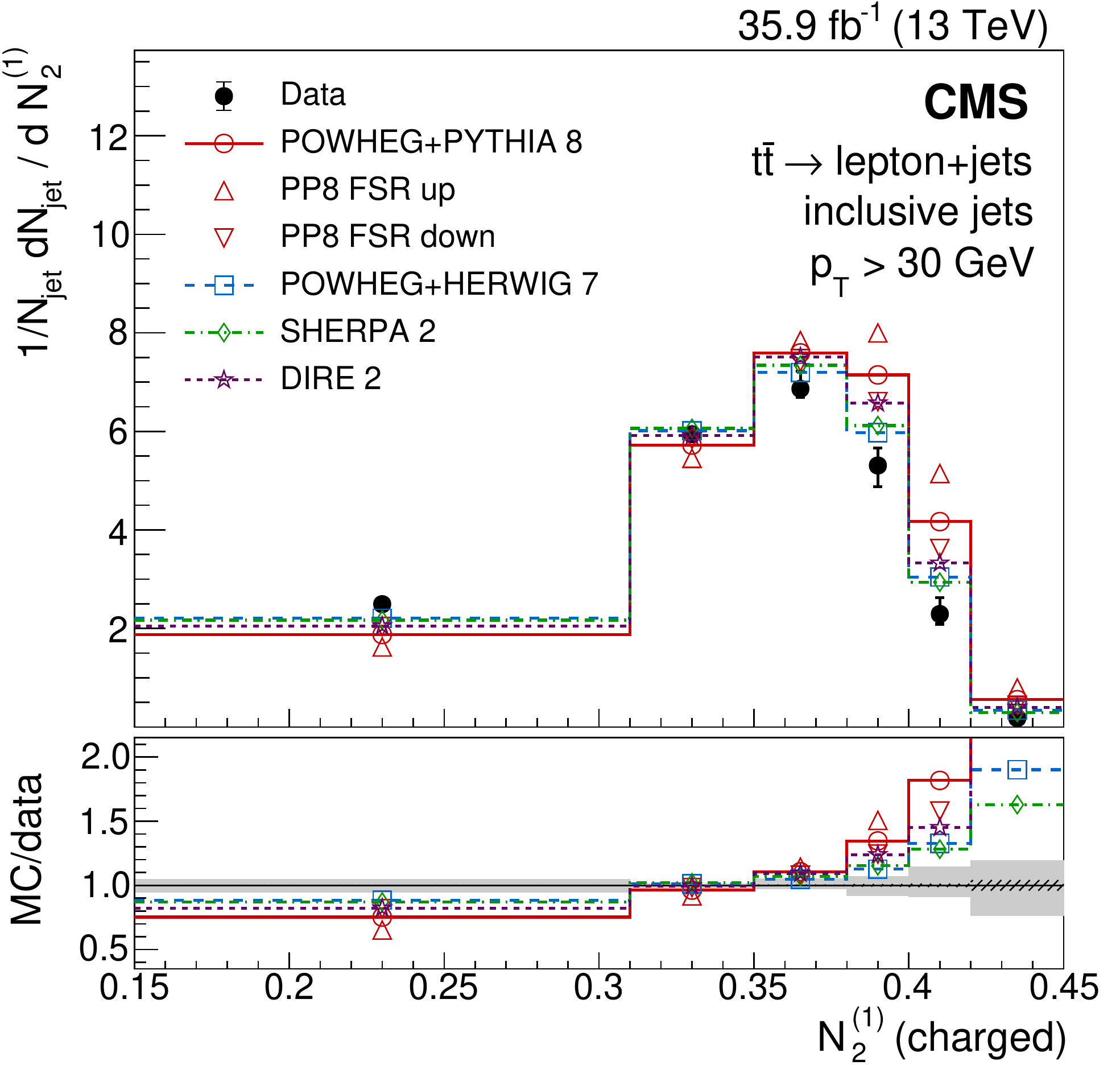}
  \includegraphics[width=0.48\textwidth]{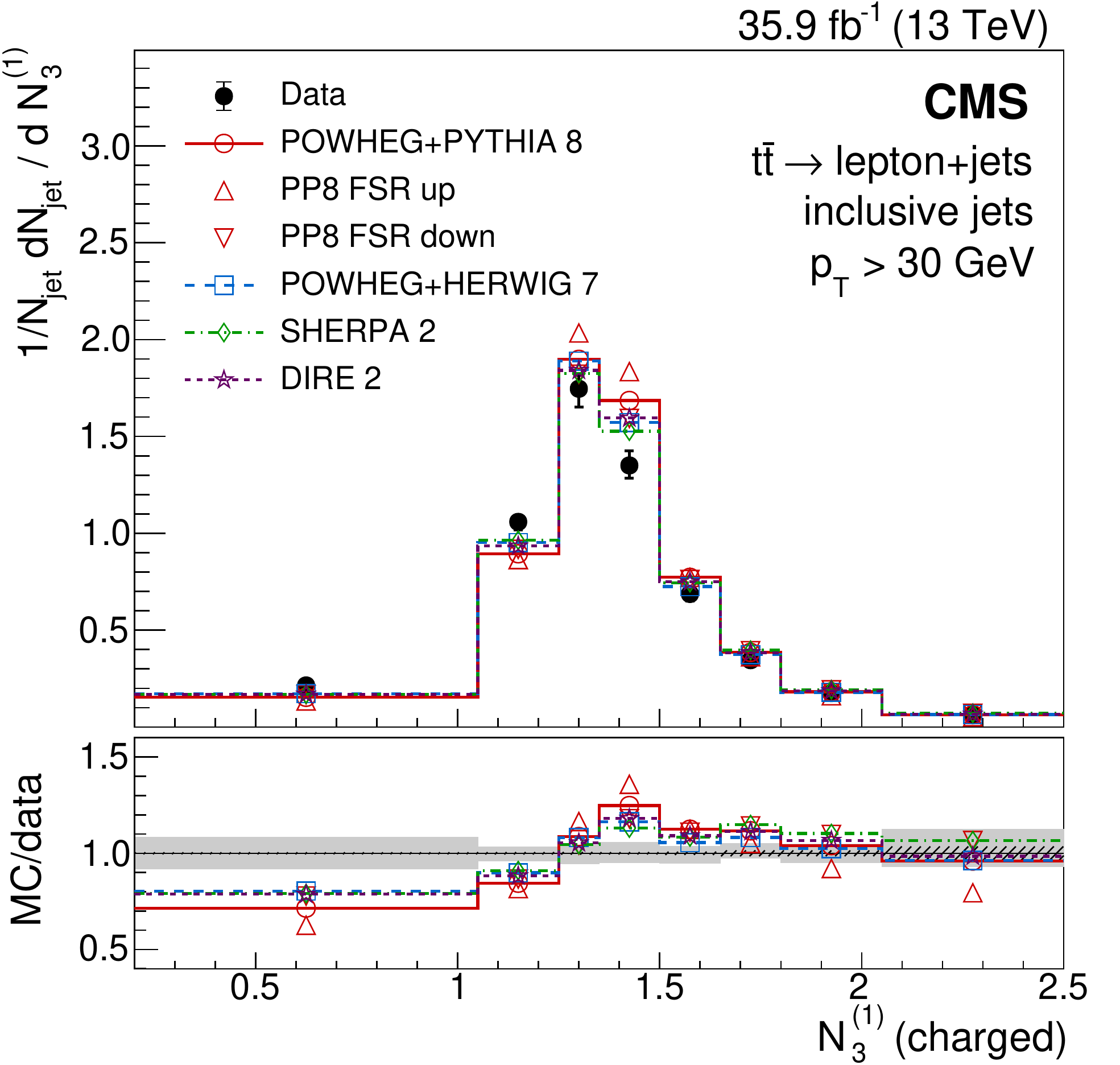}
  \caption{Distribution of the energy correlation ratios $\mathrm{N}_{2}^{\left(1\right)}$ (left) and $\mathrm{N}_{3}^{\left(1\right)}$ (right), unfolded to the particle level, for inclusive jets reconstructed with charged particles. Data (points) are compared to different MC predictions (upper), and as MC/data ratios (lower). The hatched and shaded bands represent the statistical and total uncertainties, respectively.
  }
  \label{fig:n_i}
\end{figure*}

More recently, the $\mathrm{M}_{i}$ and $\mathrm{N}_{i}$ series observables~\cite{Moult:2016cvt} were proposed as the following ratios of generalized energy correlation functions:
\begin{linenomath}
\begin{equation}
\mathrm{M}_{i}^{(\beta)} = \frac{{}_{1} e_{i+1}^{(\beta)}}{ {}_{1} e_{i}^{(\beta)}},\quad
\mathrm{N}_{i}^{(\beta)} = \frac{{}_{2} e_{i+1}^{(\beta)}}{({}_{1} e_{i}^{(\beta)})^2},
\end{equation}
\end{linenomath}
where
\begin{linenomath}
\ifthenelse{\boolean{cms@external}}
{
\begin{multline*}
{}_v e_n^{\left( \beta \right)} = \sum_{1 \leq i_1 < i_2 < \dots < i_n \in j} z_{i_1} z_{i_2} \dots z_{i_n}\\ \prod_{m = 1}^{v} \min^{(m)}_{s < t \in \{i_1, i_2 , \dots, i_n \}} \left\{ \Delta R_{st}^{\beta} \right\}
\end{multline*}
}
{
\begin{equation*}
{}_v e_n^{\left( \beta \right)} = \sum_{1 \leq i_1 < i_2 < \dots < i_n \in j} z_{i_1} z_{i_2} \dots z_{i_n} \prod_{m = 1}^{v} \min^{(m)}_{s < t \in \{i_1, i_2 , \dots, i_n \}} \left\{ \Delta R_{st}^{\beta} \right\}
\end{equation*}
}
\end{linenomath}
and the sum runs over all particles $i$ in the jet $j$ with \pt fractions $z_i$, $\min^{(m)}$ denotes the $m$-th smallest angular distance, $n$ is the number of particles to be correlated, $v$ denotes the number of pairwise angles entering the product, and $\beta$ is the angular exponent.
The observables are Lorentz invariant under boosts along the jet axis and IRC-safe for $\beta > 0$.
The distributions of $\mathrm{M}_{2}$, $\mathrm{N}_{2}$, and $\mathrm{N}_{3}$ have been measured for $\beta=\left\{1,2\right\}$.
Figures~\ref{fig:m_i} and \ref{fig:n_i} show the results for $\beta = 1$.
The \POWHEG + \HERWIG~7 prediction describes these data better than the other MC generators.

\section{Jet substructure for different jet flavors}
\label{sec:flavor}

All jet substructure observables have been measured not only for inclusive jets, but also for b quark jets, and for samples enriched in light-quark or gluon jets, respectively.
The flavor categories are defined as follows below.
The relative contributions to the inclusive jet sample at the particle level are obtained from the default \POWHEG + \PYTHIA~8 simulation with little dependence on the generator.
The parton flavor (quarks and gluons) is determined from the leading \pt parton that can be associated with a jet in \POWHEG + \PYTHIA~8 simulation.
It should be noted that the parton information is very generator-dependent and only serves for illustration of the level of purity of the light- and gluon-enriched samples.

\begin{description}
	\item[Bottom quark jets] (44\% of the inclusive jet sample) \hfill \\
        At detector level, jets are identified as b-tagged by the CSVv2 algorithm.
        At particle level, at least one b hadron is required to be clustered in the jet.

	These jets originate from b quarks in more than 99\% of the cases.
	No distinction is made between b jets from the top quark decay and additional b jets from gluon splitting.
	\item[Light-quark jets] (46\% of the inclusive jet sample) \hfill \\ Jets are assigned to the light-quark-enriched jet sample if they are not b-tagged and are paired with another similar jet to give a \PW{} boson candidate with an invariant mass satisfying $\abs{m_{jj}-80.4\GeV}<15\GeV$.
	Of these jets, 50\% stem from light quarks, 21\% from charm quarks, and 29\% from gluons.
	\item[Gluon jets] (10\% of the inclusive jet sample) \hfill \\ A sample enriched in gluon jets is obtained by selecting jets that are neither b-tagged nor associated to a \PW{} boson candidate, but instead are likely to originate from ISR.
	This sample is composed of jets stemming from  bottom (1\%), charm (11\%), and light quarks (31\%), and gluons (58\%).
\end{description}

\begin{figure*}
  \centering
  \includegraphics[width=0.48\textwidth]{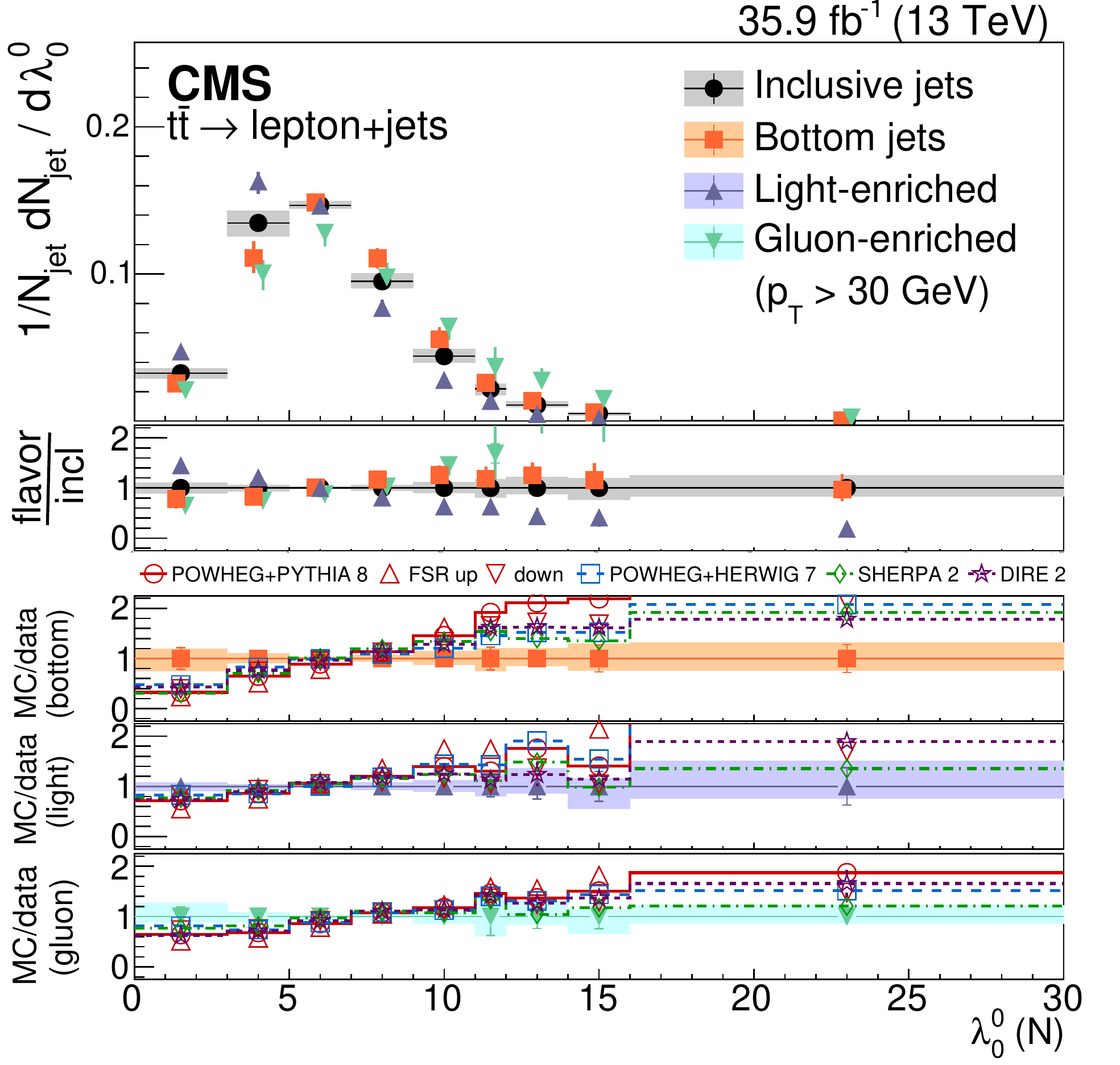}
  \includegraphics[width=0.48\textwidth]{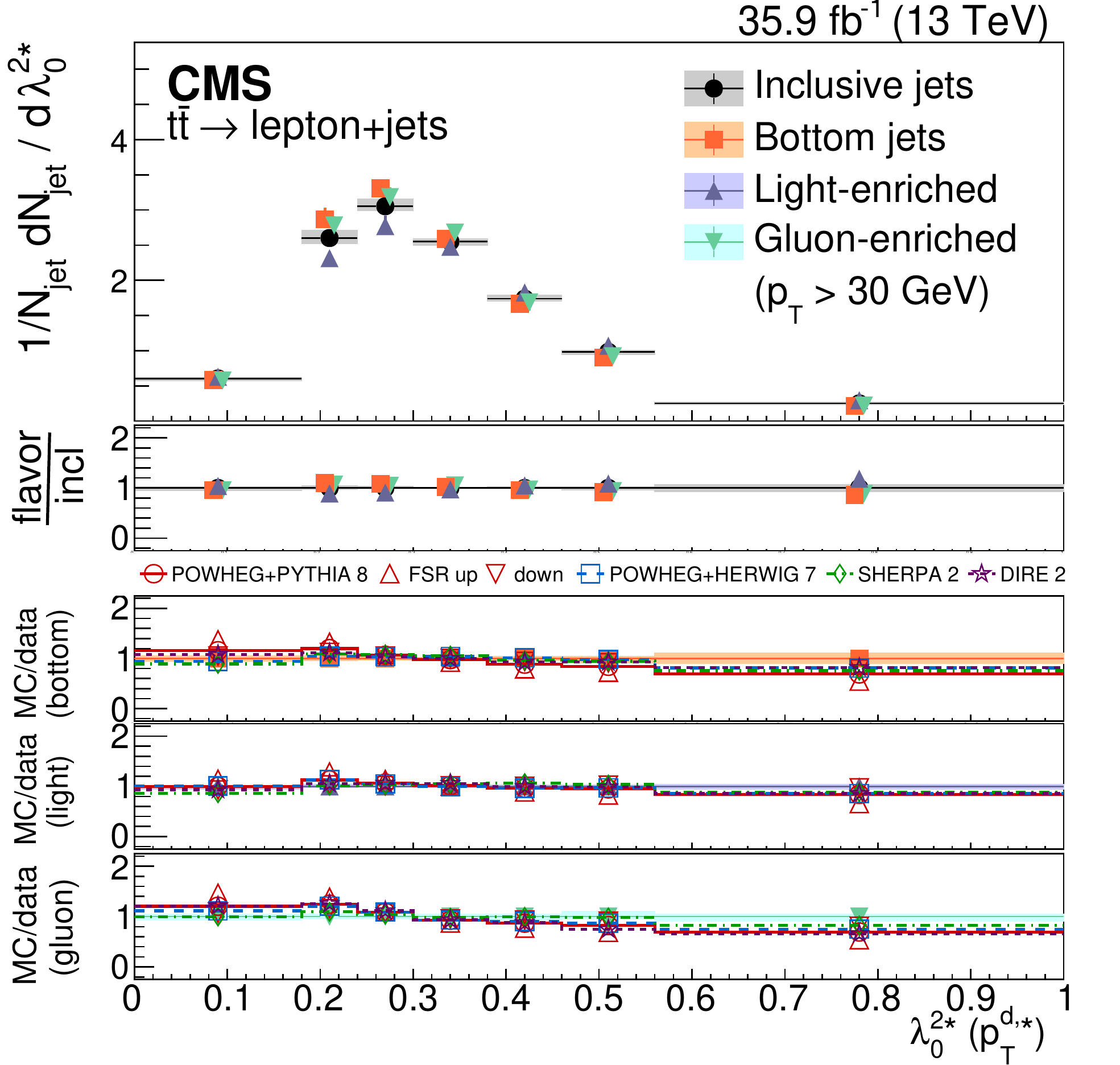}
  \includegraphics[width=0.48\textwidth]{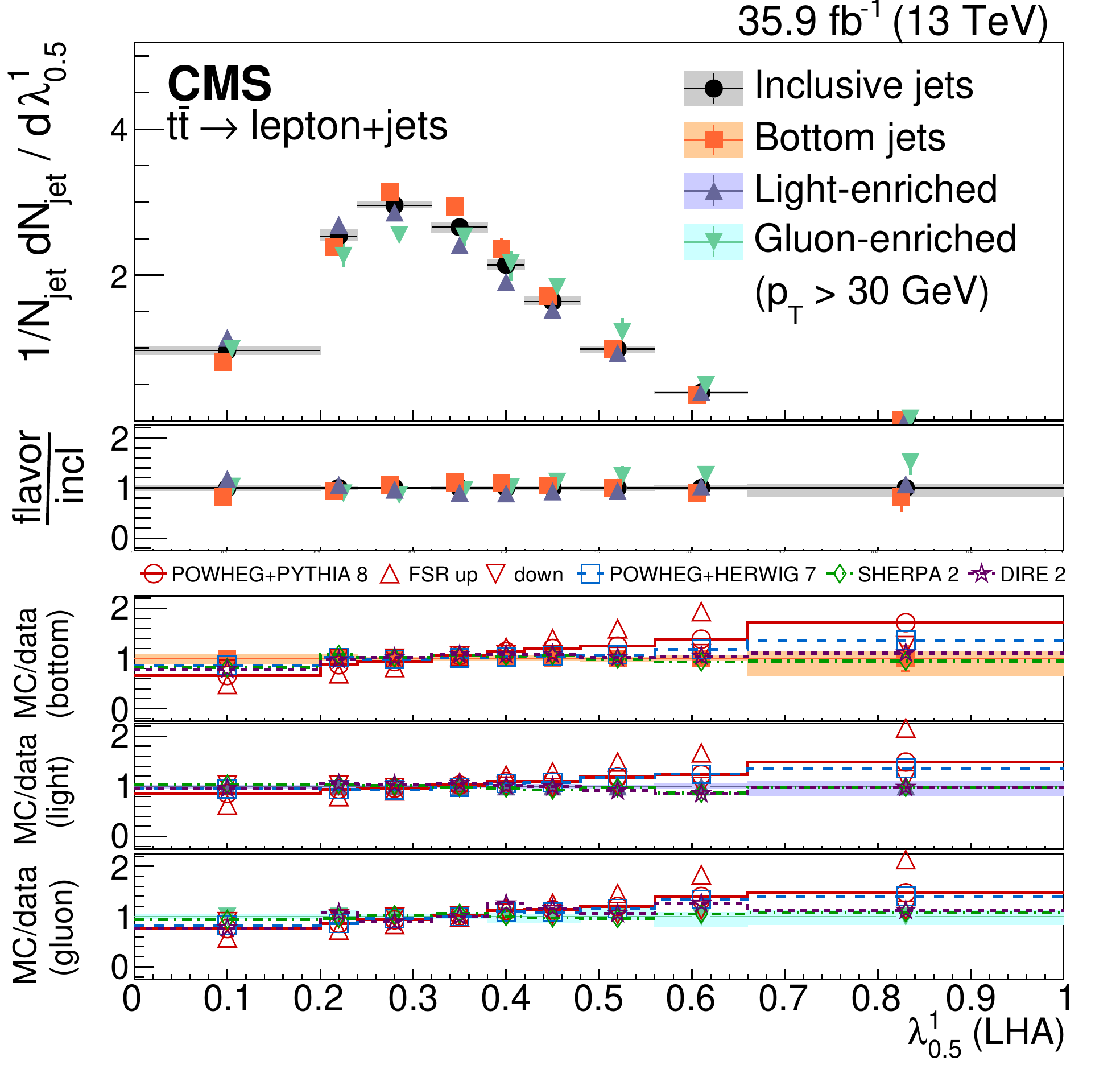}
  \includegraphics[width=0.48\textwidth]{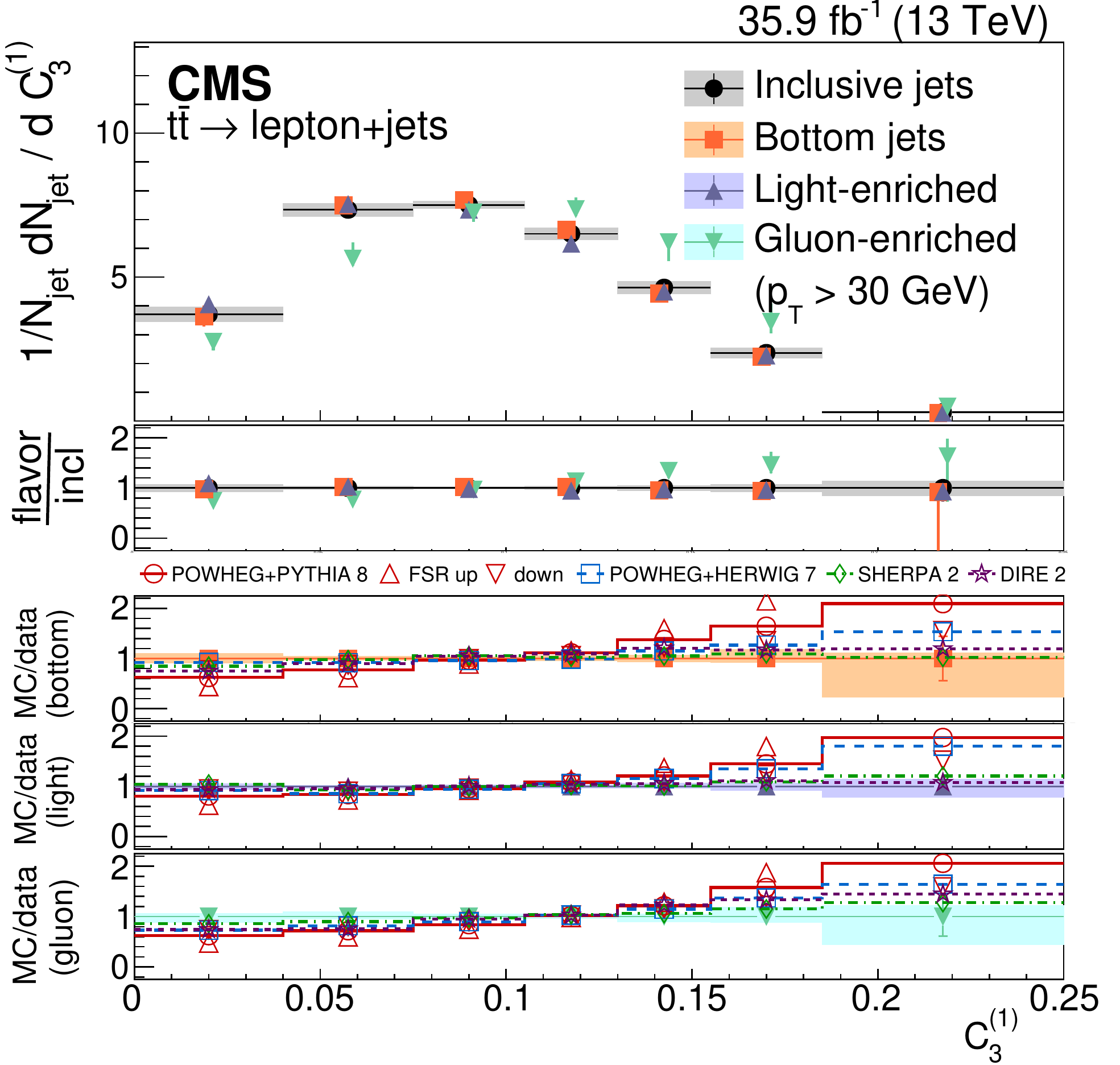}
  \caption{Distributions of the charged multiplicity (upper left), scaled \pt dispersion ($\lambda_0^{2*}$) (upper right), Les Houches angularity ($\lambda_{0.5}^1$) (lower left), and the energy correlation ratio $C_{3}^{\left(1\right)}$ (lower right), unfolded to the particle level, for jets of different flavors.
  The second panel shows the corresponding ratios of the different flavors over the inclusive jets data.
  The sub-panels show the ratios of the different MC predictions over the bottom, light-quark-enriched, and gluon-enriched jet data.
  }
  \label{fig:flavors}
\end{figure*}

Observables relevant for studies of quark/gluon discrimination, such as the charged multiplicity, scaled \pt dispersion, Les Houches angularity, and the energy correlation ratio $C_{3}^{\left(1\right)}$ are shown in Fig.~\ref{fig:flavors} for the three exclusive jet samples.
For all observables, the differences between the quark- and gluon-enriched samples do not seem to be very strong, with the energy correlation ratio $C_3^{\left(1\right)}$ providing the best separation.
This might be caused by the algorithmic definition of the samples that leads to a high contamination with other partonic flavors.
It is notable that the data/MC agreement for bottom-quark jets is significantly worse than for the light- and gluon-enriched samples, see also the $\chi^2$ tests in Section~\ref{sec:chi2}.
Therefore, an update in the MC parameter tuning and/or physics modeling may require flavor-dependent improvements to match the data.

\section{Compatibility tests with minimally correlated observables}
\label{sec:chi2}

The compatibility of the unfolded data and different MC predictions is tested by calculating $\chi^{2} = \Delta^{T} C^{-1} \Delta$, where $\Delta = \left(\vec{x}_{\text{data}} - \vec{x}_{\text{MC}}\right)$ is the vector of measurement residuals, and $C$ is the total covariance matrix of the measurement, given by $C = C_{\text{stat}} + \sum_{\text{syst}} C_{\text{syst}}$, with the vector/matrix entries for the first bin removed to make $C$ invertible.

The statistical covariance matrices $C_{\text{stat}}$ for the normalized distributions are obtained from 1000 pseudo-experiments per observable.
For uncertainties described by a single systematic shift, the systematic covariance matrix is defined as $C_{\text{syst}}\left(i,j\right) = \left(x_i^{\text{syst}} - x_i^{\text{nom}}\right) \left(x_j^{\text{syst}} - x_j^{\text{nom}}\right)$, where $x_i^{\text{nom}}$ is the vector representing the nominal result.
For uncertainties described by two opposite shifts, the systematic covariance matrix is defined as
\begin{linenomath}
\begin{align*}
C_{\text{syst}}\left(i,j\right) =& \max\left(\abs{x_i^{\text{syst}+} - x_i^{\text{nom}}}, \abs{x_i^{\text{syst}-} - x_i^{\text{nom}}}\right)\\
&\times \max\left(\abs{x_j^{\text{syst}+} - x_j^{\text{nom}}}, \abs{x_j^{\text{syst}-} - x_j^{\text{nom}}}\right)\\
&\times \sign\left(\left[x_i^{\text{syst}+} - x_i^{\text{syst}-}\right] \left[x_j^{\text{syst}+} - x_j^{\text{syst}-}\right]\right),
\end{align*}
\end{linenomath}
which corresponds to symmetrizing the largest observed shift in each bin.

\begin{figure*}[!htp]
\centering
\includegraphics[width=1.0\textwidth]{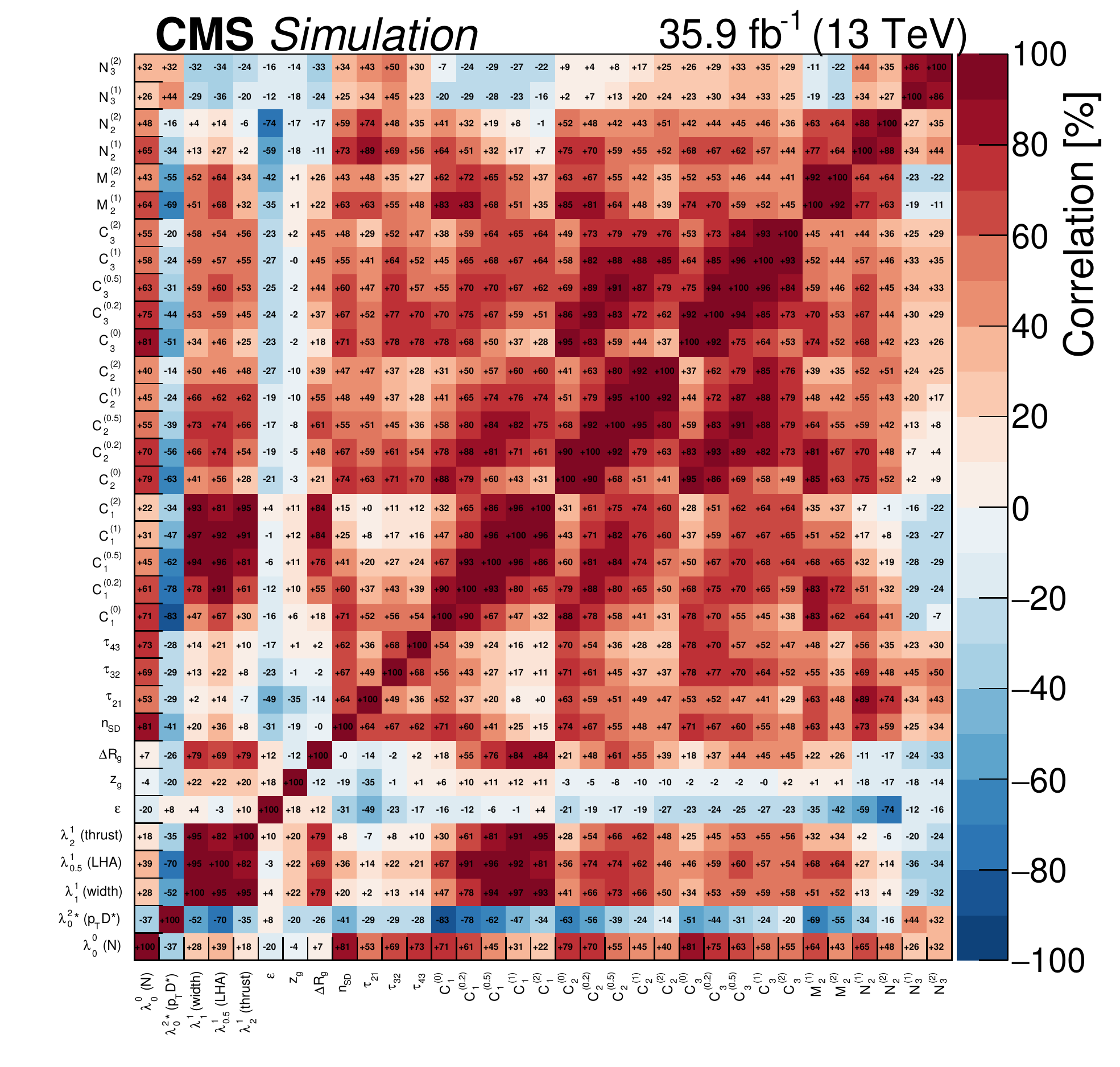}
\caption{Correlations of the jet-substructure observables used in this analysis obtained at the particle level.
}
\label{fig:obs_corr}
\end{figure*}

\begin{figure}[!htp]
\centering
\includegraphics[width=0.48\textwidth]{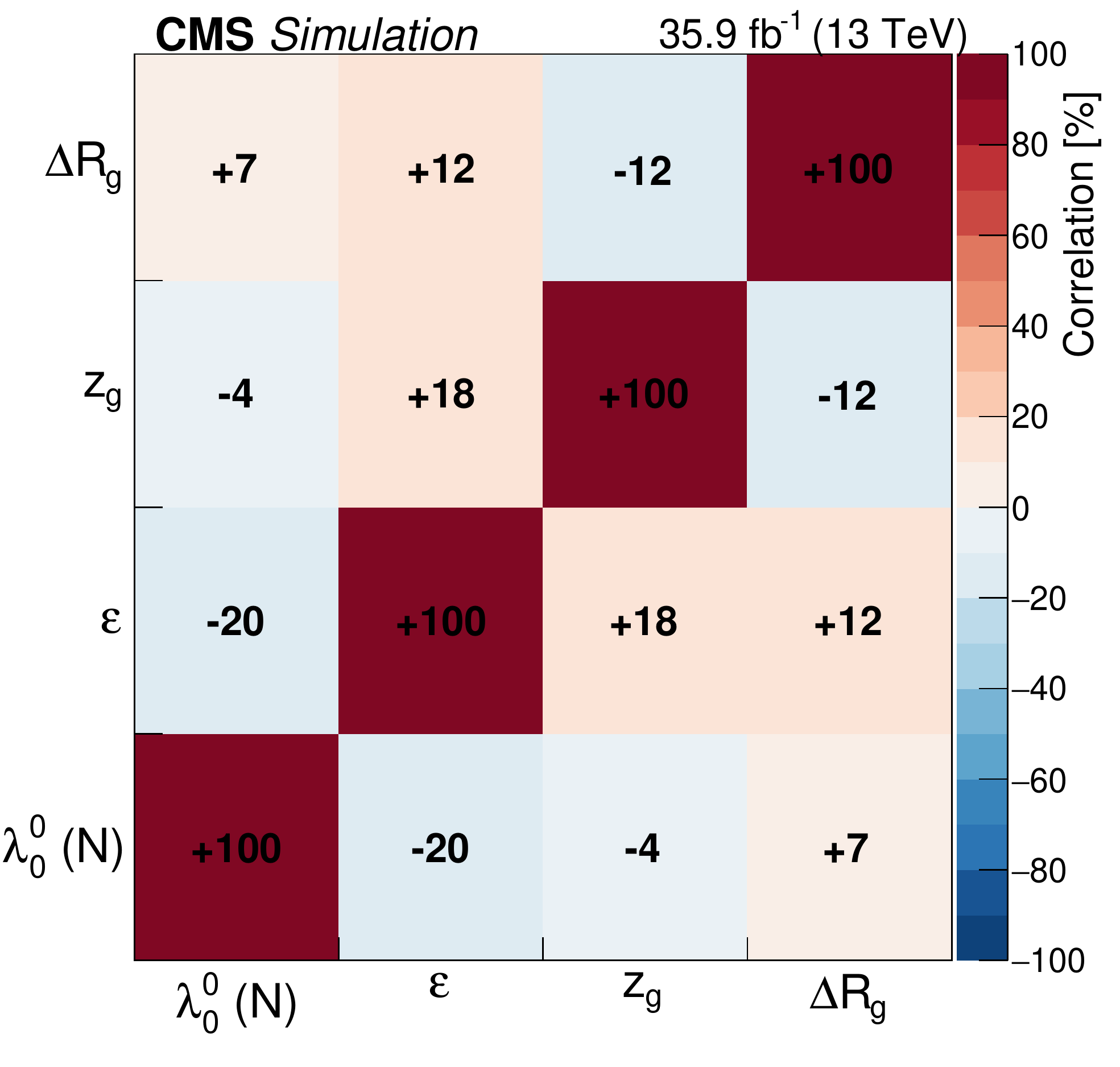}
\caption{Correlations of the jet-substructure observables used in this analysis obtained at the particle level for the set of four minimally correlated observables.
}
\label{fig:obs_corr2}
\end{figure}

By construction, the considered jet-substructure observables exhibit significant correlation with each other, as shown by the pair-wise sample Pearson correlation coefficients in Figs.~\ref{fig:obs_corr} and \ref{fig:obs_corr2}.
For further analysis, it is useful to identify a subset of observables with low correlation to each other.

A suitable subset of 4 observables is identified that have an absolute correlation of less than 30\% among each other: the charged multiplicity $\lambda_0^0$ ($N$), the eccentricity $\varepsilon$, the groomed momentum fraction $z_\mathrm{g}$, and the angle between the groomed subjets $\Delta R_\mathrm{g}$.
The associated data-to-simulation goodness-of-fit values, $\chi^{2}$, for these four low-correlation observables are listed in Tables~\ref{tab:chi2} and \ref{tab:chi2pythia}.

Among the \POWHEG + \PYTHIA~8 predictions, the FSR-down setting with $\asfsr = 0.1224$ shows improved agreement with data, except for $z_\mathrm{g}$ which does not depend on the value of \asfsr.
The agreement with data is also improved by the alternative models for CR and by the rope hadronization model~\cite{Bierlich:2014xba}.
The $\Delta R_\mathrm{g}$ observable is also shown to be sensitive to the b fragmentation function, and shows better agreement with harder fragmentation.
The agreement of the \POWHEG + \PYTHIA~8 predictions with the jet eccentricity data is poor compared to \SHERPA~2 and \POWHEG + \HERWIG~7, particularly.
The \POWHEG + \HERWIG~7 generator setup with the angular-ordered shower also provides the best description of the groomed momentum fraction $z_\mathrm{g}$.
The prediction by \SHERPA~2 has an overall good agreement with the data, but does not describe well the $\Delta R_\mathrm{g}$ of bottom-quark jets.
This might be caused by the missing ME corrections to the radiation from the b quark in the top quark decay.

\begin{table*}
{
\centering
\topcaption{$\chi^{2}$ values and the numbers of degrees of freedom (ndf) for the data-to-simulation comparison of the distributions of the four weakly-correlated jet substructure observables, $\lambda_{0}^{0}$ ($N$), $\varepsilon$, $z_\mathrm{g}$, and $\Delta R_\mathrm{g}$, for four different jet flavors and six MC generator setups.}
\label{tab:chi2}
\begin{scotch}{llrrrrrr}
 &  & \multicolumn{3}{c}{\POWHEG + \PYTHIA~8} & \POWHEG + & \SHERPA~2 & \DIRE~2 \\
 &  & FSR-down & Nominal & FSR-up & \HERWIG~7 &  &  \\ [\cmsTabSkip]
\multirow{2}{*}{\asfsr} & & 0.1224 & 0.1365 & 0.1543 & 0.1262 & 0.118 & 0.1201 \\
 &  & One-loop & One-loop & One-loop & Two-loop & Two-loop CMW & Two-loop \\ [\cmsTabSkip]
Observable & Flavor & $\chi^2$ & $\chi^2$ & $\chi^2$ & $\chi^2$ & $\chi^2$ & $\chi^2$ \\
\hline
\multirow{2}{*}{$\lambda_{0}^{0}$ ($N$)}
 & Inclusive & 23.4 & 88.0 & 390.5 & 27.4 & 16.1 & 15.1\\
\multirow{2}{*}{ndf = 8} & Bottom & 35.7 & 110.6 & 432.9 & 35.4 & 20.0 & 26.0\\
 & Light & 7.2 & 12.3 & 53.3 & 24.5 & 13.2 & 24.0\\
 & Gluon & 9.0 & 26.1 & 84.5 & 13.5 & 4.7 & 14.1\\ [\cmsTabSkip]
\multirow{2}{*}{$\varepsilon$}
 & Inclusive & 72.6 & 108.8 & 217.6 & 6.3 & 9.4 & 61.6\\
\multirow{2}{*}{ndf = 6} & Bottom & 28.2 & 48.7 & 102.9 & 2.1 & 4.8 & 21.7\\
 & Light & 27.6 & 44.6 & 89.6 & 3.9 & 2.7 & 26.3\\
 & Gluon & 57.0 & 81.3 & 133.4 & 7.5 & 19.7 & 73.6\\ [\cmsTabSkip]
\multirow{2}{*}{$z_\mathrm{g}$}
 & Inclusive & 18.9 & 20.7 & 23.2 & 1.8 & 7.7 & 16.2\\
\multirow{2}{*}{ndf = 4} & Bottom & 4.8 & 6.4 & 8.6 & 1.2 & 1.5 & 3.0\\
 & Light & 22.0 & 20.7 & 19.5 & 1.3 & 8.9 & 27.6\\
 & Gluon & 11.2 & 10.4 & 8.8 & 2.0 & 9.6 & 15.9\\ [\cmsTabSkip]
\multirow{2}{*}{$\Delta R_\mathrm{g}$}
 & Inclusive & 19.5 & 29.3 & 241.5 & 23.2 & 41.8 & 77.0\\
\multirow{2}{*}{ndf = 10} & Bottom & 23.2 & 18.4 & 227.5 & 16.6 & 79.1 & 15.8\\
 & Light & 9.3 & 29.3 & 251.0 & 120.1 & 40.2 & 221.6\\
 & Gluon & 11.7 & 8.6 & 69.5 & 19.7 & 28.3 & 33.1\\
\end{scotch}
}
\end{table*}

\begin{table*}
\centering
\topcaption{$\chi^{2}$ values and the numbers of degrees of freedom (ndf) for the data-to-simulation comparison of the distributions of the four weakly-correlated jet substructure observables, $\lambda_{0}^{0}$ ($N$), $\varepsilon$, $z_\mathrm{g}$, and $\Delta R_\mathrm{g}$, for four different jet flavors and seven \POWHEG + \PYTHIA~8 model variations.
The value of the strong coupling is $\asfsr = 0.1365$ for all predictions.}
\label{tab:chi2pythia}
\begin{scotch}{llrrrrrrr}
 &  & Nominal & \multicolumn{3}{c}{CR/hadronization} & \multicolumn{3}{c}{b fragmentation} \\
 & & & QCD & Move & Rope & Soft & Hard & Peterson \\ [\cmsTabSkip]
Observable & Flavor & $\chi^2$ & $\chi^2$ & $\chi^2$ & $\chi^2$ & $\chi^2$ & $\chi^2$ & $\chi^2$ \\
\hline
\multirow{2}{*}{$\lambda_{0}^{0}$ ($N$)}
 & Inclusive & 88.0 & 42.1 & 57.0 & 51.6 & 120.7 & 78.5 & 158.7\\
\multirow{2}{*}{ndf = 8} & Bottom & 110.6 & 80.1 & 95.7 & 65.4 & 159.3 & 96.4 & 207.6\\
 & Light & 12.3 & 9.5 & 12.3 & 10.3 & 12.6 & 12.1 & 12.6\\
 & Gluon & 26.1 & 7.4 & 13.0 & 21.5 & 27.4 & 25.5 & 27.5\\ [\cmsTabSkip]
\multirow{2}{*}{$\varepsilon$}
 & Inclusive & 108.8 & 85.3 & 89.5 & 94.6 & 118.6 & 103.3 & 108.5\\
\multirow{2}{*}{ndf = 6} & Bottom & 48.7 & 44.0 & 45.7 & 37.4 & 56.7 & 44.3 & 48.5\\
 & Light & 44.6 & 32.1 & 34.5 & 42.0 & 45.7 & 44.0 & 45.4\\
 & Gluon & 81.3 & 40.4 & 54.7 & 87.9 & 81.8 & 80.9 & 81.1\\ [\cmsTabSkip]
\multirow{2}{*}{$z_\mathrm{g}$}
 & Inclusive & 20.7 & 15.6 & 18.5 & 18.0 & 22.3 & 19.5 & 18.1\\
\multirow{2}{*}{ndf = 4} & Bottom & 6.4 & 6.0 & 5.8 & 5.2 & 7.3 & 5.7 & 4.8\\
 & Light & 20.7 & 14.8 & 18.9 & 18.8 & 20.8 & 20.7 & 20.7\\
 & Gluon & 10.4 & 6.1 & 8.6 & 9.8 & 10.5 & 10.4 & 10.4\\ [\cmsTabSkip]
\multirow{2}{*}{$\Delta R_\mathrm{g}$}
 & Inclusive & 29.3 & 24.8 & 26.1 & 23.7 & 48.2 & 23.2 & 44.7\\
\multirow{2}{*}{ndf = 10} & Bottom & 18.4 & 18.6 & 15.8 & 9.1 & 60.1 & 8.6 & 55.4\\
 & Light & 29.3 & 18.5 & 23.5 & 18.4 & 33.6 & 27.2 & 32.7\\
 & Gluon & 8.6 & 4.7 & 7.6 & 9.0 & 8.6 & 8.6 & 8.3\\
\end{scotch}
\end{table*}

\section{Extraction of the strong coupling}
\label{sec:asfsr}

\begin{figure*}[htp]
\centering
\includegraphics[width=0.48\textwidth]{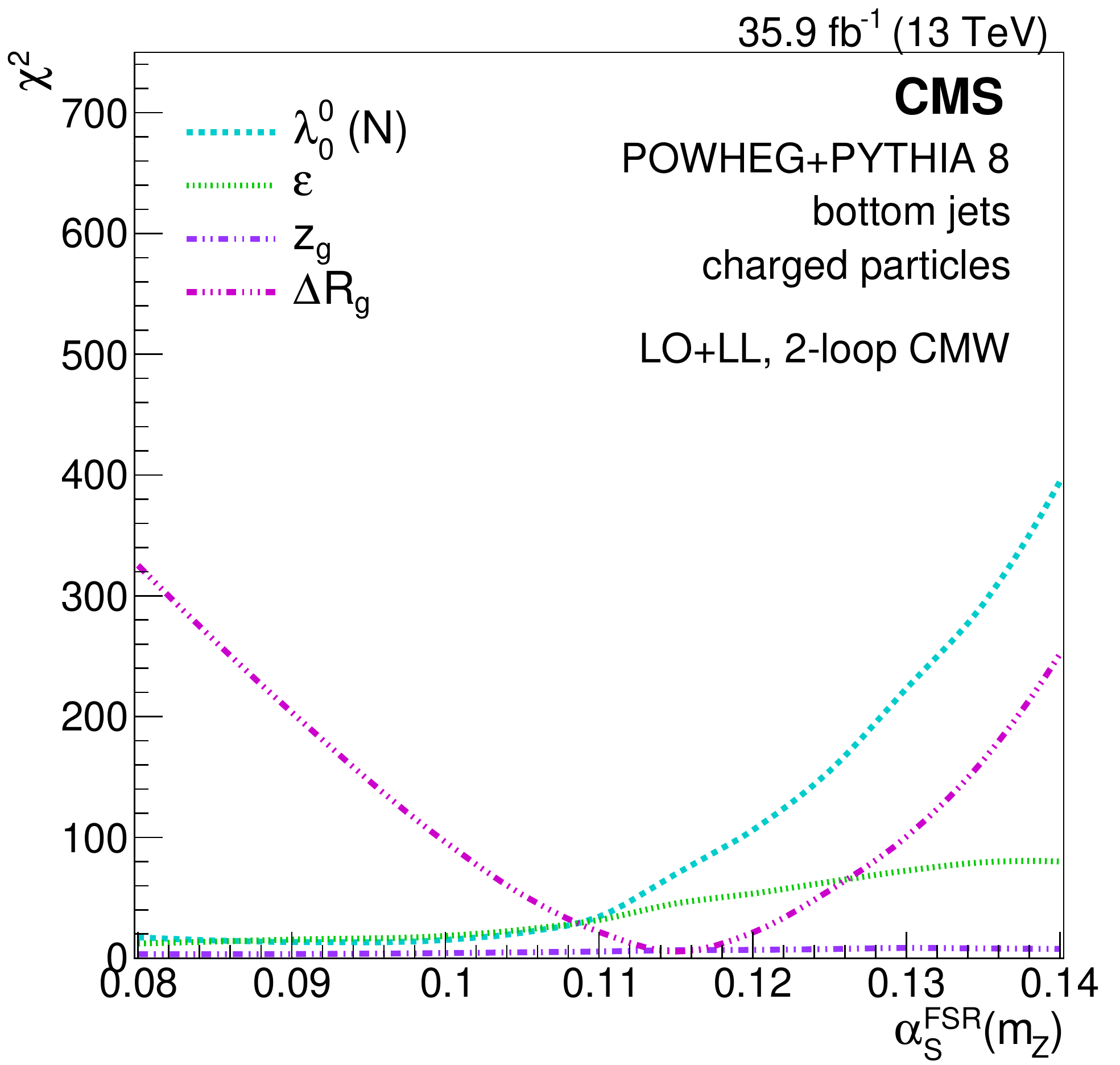}
\includegraphics[width=0.48\textwidth]{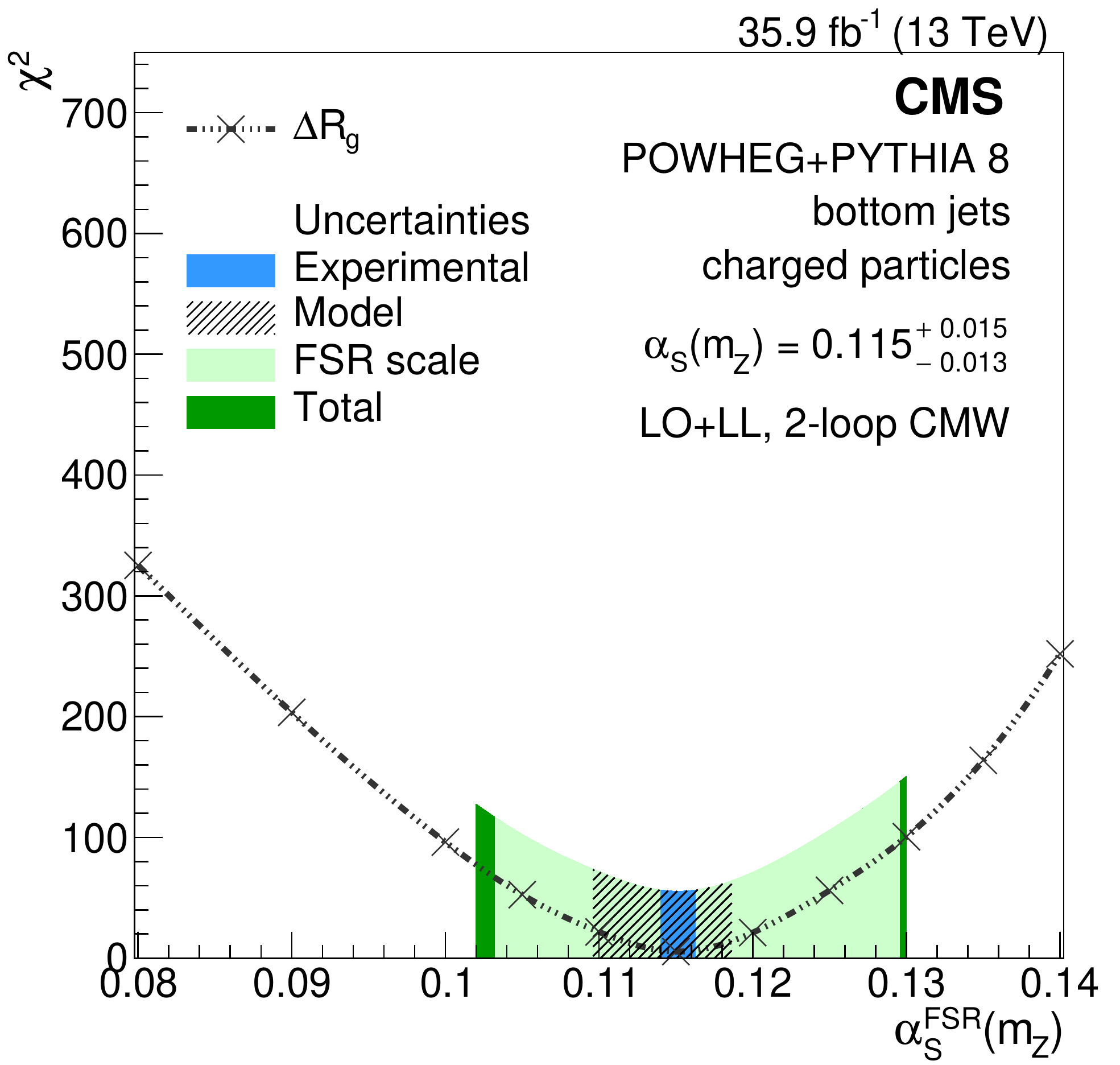}
\caption{Scans of $\chi^2$ as a function of \asfsr, derived from the bottom-quark jet sample, for the minimally-correlated observables $\lambda_0^0$ ($N$), $\varepsilon$, $z_\mathrm{g}$, and $\Delta R_\mathrm{g}$ (left), and for $\Delta R_\mathrm{g}$ alone with uncertainties indicated by the shaded areas (right).
}
\label{fig:fit_lowcor}
\end{figure*}

The value of the strong coupling preferred by the jet substructure observables can be extracted from a comparison of the measured distributions to \POWHEG + \PYTHIA~8 predictions.
Monte Carlo samples were generated with \asfsr values between 0.08 and 0.14, where higher-order corrections to soft gluon emissions are incorporated in an effective way using 2-loop running of the strong coupling and CMW rescaling~\cite{Catani:1990rr}.
The $\chi^2$ scan of \asfsr for the low-correlation observables is shown in Fig.~\ref{fig:fit_lowcor}.
The charged multiplicity and the jet eccentricity are sensitive to \asfsr but are expected to be highly affected by the modeling of nonperturbative effects, pointing to the need of tuning additional parameters.
As expected, the groomed momentum fraction $z_\mathrm{g}$ is independent of \asfsr.

The angle between the groomed subjets, $\Delta R_\mathrm{g}$, is measured with high precision and the removal of soft radiation lowers the impact of nonperturbative effects.
The value of \asmz can be extracted from this observable with an experimental uncertainty of $\pm 0.001$ using the \cPqb{} jet sample (Fig.~\ref{fig:fit_lowcor}, right).
These bottom-quark jets stem mostly from top quark decays where the \PYTHIA~8 prediction incorporates ME corrections, describing the jet substructure at LO accuracy in the hard emission limit, while also being at least LL accurate elsewhere.
The modeling uncertainties are estimated by the \POWHEG + \PYTHIA~8 variations described in Section~\ref{sec:sys}, as well as by a comparison to the results obtained with the rope hadronization model.
This extraction of \asmz is currently limited by the FSR scale uncertainties of ${}_{-0.012}^{+0.014}$.
Other relevant model uncertainties stem from the b fragmentation (${}^{+0.003}_{-0.006}$) and the alternative rope hadronization model ($+0.002$).
Taking into account all uncertainties, a value of $\asmz = 0.115^{+0.015}_{-0.013}$ is obtained from the b jet sample.
An extraction using charged+neutral particles leads to an identical result even though with a slightly larger experimental uncertainty of $\pm 0.002$.

The default \POWHEG + \PYTHIA~8 samples were generated without CMW rescaling and with first-order running of \alpS.
In this case, a value of $\asmz = 0.130^{+0.016}_{-0.020}$ is extracted from the b jet sample.
This value is in between those of the \POWHEG + \PYTHIA~8 nominal sample with $\asfsr = 0.1365$ and the ``FSR down'' sample which has an effective $\asfsr = 0.1224$ for final-state radiation.
A lower value of \asfsr also improves the data-to-simulation agrement for charged multiplicity and jet eccentricity although some discrepancy remains.

\section{Summary}

A measurement of jet substructure observables in resolved \ttbar lepton+jets events from $\Pp\Pp$ collisions at $\sqrt{s} = 13\TeV$ has been presented, including several variables relevant for quark-gluon discrimination and for heavy Lorentz-boosted object identification.
The investigated observables provide valuable insights on the perturbative and nonperturbative phases of jet evolution.
Their unfolded distributions have been derived for inclusive jets, as well as for samples enriched in jets originating from bottom quarks, light quarks, or gluons.

Data are compared to theoretical predictions either based on next-to-leading-order (NLO) matrix-element calculations (\POWHEG) interfaced with different generators for the parton shower and hadronization (either \PYTHIA~8 or \HERWIG~7), or based on \SHERPA~2 with NLO corrections, as well as on the \DIRE~2 shower model.
The correlations between all jet substructure variables have been studied.
Eliminating observables with a high level of correlation, a set of four variables is identified and used for quantifying the level of data-simulation agreement.
With the default Monte Carlo (MC) generator tunes, none of the predictions yields a good overall reproduction of the experimental distributions.
Thus, some further tuning of the models is required, with special attention to the data/MC disagreement observed in the particle multiplicity $\lambda_0^0$ and correlated observables, including those designed for quark/gluon discrimination.
The groomed momentum fraction $z_\mathrm{g}$ is directly sensitive to the parton-shower splitting functions, thereby providing a useful handle to improve their modeling in the MC generators.

The angle between the groomed subjets, $\Delta R_\mathrm{g}$, is an experimentally powerful observable for extracting the value of the strong coupling in final-state parton radiation (FSR) processes.
A value of $\asmz = 0.115^{+0.015}_{-0.013}$, including experimental as well as model uncertainties, has been extracted at leading-order plus leading-log accuracy, where the precision is limited by the FSR scale uncertainty of the \PYTHIA~8 prediction.
The data will allow for a precise determination of \asmz once predictions for top quark decays with multiple emissions at higher order combined with parton showers (ideally at approximate next-leading-log accuracy) are available.
Besides tuning and improving final-state parton showers, the present data also provide useful tests for improved quantum chromodynamics analytical calculations, including higher-order fixed and logarithmic corrections, for infrared- and/or collinear-safe observables.

\begin{acknowledgments}
We congratulate our colleagues in the CERN accelerator departments for the excellent performance of the LHC and thank the technical and administrative staffs at CERN and at other CMS institutes for their contributions to the success of the CMS effort. In addition, we gratefully acknowledge the computing centers and personnel of the Worldwide LHC Computing Grid for delivering so effectively the computing infrastructure essential to our analyses. Finally, we acknowledge the enduring support for the construction and operation of the LHC and the CMS detector provided by the following funding agencies: BMWFW and FWF (Austria); FNRS and FWO (Belgium); CNPq, CAPES, FAPERJ, FAPERGS, and FAPESP (Brazil); MES (Bulgaria); CERN; CAS, MoST, and NSFC (China); COLCIENCIAS (Colombia); MSES and CSF (Croatia); RPF (Cyprus); SENESCYT (Ecuador); MoER, ERC IUT, and ERDF (Estonia); Academy of Finland, MEC, and HIP (Finland); CEA and CNRS/IN2P3 (France); BMBF, DFG, and HGF (Germany); GSRT (Greece); NKFIA (Hungary); DAE and DST (India); IPM (Iran); SFI (Ireland); INFN (Italy); MSIP and NRF (Republic of Korea); MES (Latvia); LAS (Lithuania); MOE and UM (Malaysia); BUAP, CINVESTAV, CONACYT, LNS, SEP, and UASLP-FAI (Mexico); MOS (Montenegro); MBIE (New Zealand); PAEC (Pakistan); MSHE and NSC (Poland); FCT (Portugal); JINR (Dubna); MON, RosAtom, RAS, RFBR, and NRC KI (Russia); MESTD (Serbia); SEIDI, CPAN, PCTI, and FEDER (Spain); MOSTR (Sri Lanka); Swiss Funding Agencies (Switzerland); MST (Taipei); ThEPCenter, IPST, STAR, and NSTDA (Thailand); TUBITAK and TAEK (Turkey); NASU and SFFR (Ukraine); STFC (United Kingdom); DOE and NSF (USA).

\hyphenation{Rachada-pisek} Individuals have received support from the Marie-Curie program and the European Research Council and Horizon 2020 Grant, contract No. 675440 (European Union); the Leventis Foundation; the A. P. Sloan Foundation; the Alexander von Humboldt Foundation; the Belgian Federal Science Policy Office; the Fonds pour la Formation \`a la Recherche dans l'Industrie et dans l'Agriculture (FRIA-Belgium); the Agentschap voor Innovatie door Wetenschap en Technologie (IWT-Belgium); the F.R.S.-FNRS and FWO (Belgium) under the ``Excellence of Science - EOS" - be.h project n. 30820817; the Ministry of Education, Youth and Sports (MEYS) of the Czech Republic; the Lend\"ulet (``Momentum") Program and the J\'anos Bolyai Research Scholarship of the Hungarian Academy of Sciences, the New National Excellence Program \'UNKP, the NKFIA research grants 123842, 123959, 124845, 124850 and 125105 (Hungary); the Council of Science and Industrial Research, India; the HOMING PLUS program of the Foundation for Polish Science, cofinanced from European Union, Regional Development Fund, the Mobility Plus program of the Ministry of Science and Higher Education, the National Science Center (Poland), contracts Harmonia 2014/14/M/ST2/00428, Opus 2014/13/B/ST2/02543, 2014/15/B/ST2/03998, and 2015/19/B/ST2/02861, Sonata-bis 2012/07/E/ST2/01406; the National Priorities Research Program by Qatar National Research Fund; the Programa Estatal de Fomento de la Investigaci{\'o}n Cient{\'i}fica y T{\'e}cnica de Excelencia Mar\'{\i}a de Maeztu, grant MDM-2015-0509 and the Programa Severo Ochoa del Principado de Asturias; the Thalis and Aristeia programs cofinanced by EU-ESF and the Greek NSRF; the Rachadapisek Sompot Fund for Postdoctoral Fellowship, Chulalongkorn University and the Chulalongkorn Academic into Its 2nd Century Project Advancement Project (Thailand); the Welch Foundation, contract C-1845; and the Weston Havens Foundation (USA).
\end{acknowledgments}

\bibliography{auto_generated}
\cleardoublepage \appendix\section{The CMS Collaboration \label{app:collab}}\begin{sloppypar}\hyphenpenalty=5000\widowpenalty=500\clubpenalty=5000\vskip\cmsinstskip
\textbf{Yerevan Physics Institute, Yerevan, Armenia}\\*[0pt]
A.M.~Sirunyan, A.~Tumasyan
\vskip\cmsinstskip
\textbf{Institut f\"{u}r Hochenergiephysik, Wien, Austria}\\*[0pt]
W.~Adam, F.~Ambrogi, E.~Asilar, T.~Bergauer, J.~Brandstetter, E.~Brondolin, M.~Dragicevic, J.~Er\"{o}, A.~Escalante~Del~Valle, M.~Flechl, R.~Fr\"{u}hwirth\cmsAuthorMark{1}, V.M.~Ghete, J.~Hrubec, M.~Jeitler\cmsAuthorMark{1}, N.~Krammer, I.~Kr\"{a}tschmer, D.~Liko, T.~Madlener, I.~Mikulec, N.~Rad, H.~Rohringer, J.~Schieck\cmsAuthorMark{1}, R.~Sch\"{o}fbeck, M.~Spanring, D.~Spitzbart, A.~Taurok, W.~Waltenberger, J.~Wittmann, C.-E.~Wulz\cmsAuthorMark{1}, M.~Zarucki
% ...
\vskip\cmsinstskip
% ...
1:  Also at Vienna University of Technology, Vienna, Austria\\
% ...
\end{sloppypar}
\end{document}